\documentclass[fleqn, usenatbib]{mnras}

\usepackage{fix-cm}
\usepackage{amsmath}
\usepackage{txfonts}
\usepackage{CJKutf8}
\usepackage{amssymb}
\usepackage{mathptmx}
\usepackage{graphicx}
\usepackage{subfiles}
\usepackage{newtxtext}
\usepackage{newtxmath}
\usepackage{orcidlink}
\usepackage{subcaption}
\usepackage{fontawesome}
\usepackage[T1]{fontenc}
\usepackage[left]{lineno}
\usepackage{threeparttable}
\DeclareRobustCommand{\VAN}[3]{#2}
\let\VANthebibliography\thebibliography
\def\thebibliography{\DeclareRobustCommand{\VAN}[3]{##3}\VANthebibliography}

\title[SOM-based data augmentation]{Improved photometric redshift estimations through self-organising map-based data augmentation}

\author[Y.-H. Zhang et al.]{
Yun-Hao Zhang \begin{CJK*}{UTF8}{gbsn}(张云皓)\end{CJK*}\orcidlink{0000-0003-4916-6346},\(^{1, 2}\)\thanks{YunHao.Zhang@ed.ac.uk}
Joe Zuntz\orcidlink{0000-0001-9789-9646},\(^{1}\)
Irene Moskowitz\orcidlink{0000-0002-2206-8589},\(^{3}\)
Eric Gawiser\orcidlink{0000-0003-1530-8713},\(^{3}\)
Konrad Kuijken\orcidlink{0000-0002-3827-0175},\(^{2}\)
\newauthor
Marika Asgari\orcidlink{0000-0002-3064-083X},\(^{4}\) 
Henk Hoekstra\orcidlink{0000-0002-0641-3231},\(^{2}\)
Alex I. Malz\orcidlink{https://orcid.org/0000-0002-8676-1622},\(^{5, 6}\)
Ziang Yan \begin{CJK*}{UTF8}{gbsn}(颜子昂)\end{CJK*}\orcidlink{0000-0001-8043-5378},\(^{7}\)
Tianqing Zhang\orcidlink{https://orcid.org/0000-0002-5596-198X},\(^{8}\)
\newauthor
and The LSST Dark Energy Science Collaboration
\\
\(^{1}\) Institute for Astronomy, University of Edinburgh, Royal Observatory, Blackford Hill, Edinburgh, EH9 3HJ, The United Kingdom \\
\(^{2}\) Leiden Observatory, Leiden University, Einsteinweg 55, 2333 CC, Leiden, The Netherlands \\
\(^{3}\) Department of Physics and Astronomy, Rutgers, The State University of New Jersey, Piscataway, NJ 08854, The U.S.A. \\
\(^{4}\) School of Mathematics, Statistics and Physics, Newcastle University, Herschel Building, NE1 7RU, Newcastle-upon-Tyne, The United Kingdom \\
\(^{5}\) LSST Interdisciplinary Network for Collaboration and Computing Frameworks, 933 N. Cherry Avenue, Tucson, AZ 85721, The U.S.A. \\
\(^{6}\) McWilliams Center for Cosmology and Astrophysics, Department of Physics, Carnegie Mellon University, Pittsburgh, PA 15213, The U.S.A. \\
\(^{7}\) Ruhr University Bochum, Astronomical Institute (AIRUB), German Centre for Cosmological Lensing, 44780 Bochum, Germany \\
\(^{8}\) Department of Physics and Astronomy and PITT PACC, University of Pittsburgh, Pittsburgh, PA 15260, The U.S.A. \\
}

\date{Accepted XXX. Received YYY; in original form ZZZ}
\pubyear{2025}

\begin{document}
\label{firstpage}
\pagerange{\pageref{firstpage}--\pageref{lastpage}}
\maketitle



\begin{abstract}
We introduce a framework for the enhanced estimation of photometric redshifts using Self-Organising Maps (SOMs). Our method projects galaxy Spectral Energy Distributions (SEDs) onto a two-dimensional map, identifying regions that are sparsely sampled by existing spectroscopic observations. These under-sampled areas are then augmented with simulated galaxies, yielding a more representative spectroscopic training dataset. To assess the efficacy of this SOM-based data augmentation in the context of the forthcoming Legacy Survey of Space and Time (LSST), we employ mock galaxy catalogues from the \texttt{OpenUniverse2024} project and generate synthetic datasets that mimic the expected photometric selections of LSST after one (Y1) and ten (Y10) years of observation. We construct 501 degraded realisations of synthetic spectroscopic surveys by sampling galaxy colours, magnitudes, redshifts, and spectroscopic success rates, in order to emulate the diverse compilation of spectroscopic datasets that may exist for LSST analysis. Augmenting the degraded mock datasets with simulated galaxies from the independent \texttt{CosmoDC2} catalogues significantly improves the performance of our photometric-redshift estimates -- particularly at high redshift \((z_\mathrm{true} \gtrsim 1.5)\) -- even in the presence of differences in the underlying galaxy SED modelling between the two catalogues. This improvement is manifested in notably reduced systematic biases and a decrease in catastrophic failures by up to approximately a factor of \(2\), along with a reduction in information loss in the conditional density estimations. These results underscore the effectiveness of SOM-based augmentation in refining photometric redshift estimation, thereby enabling more robust analyses in cosmology and astrophysics for the NSF-DOE Vera C. Rubin Observatory.
\end{abstract}




\section{Introduction} \label{sec:1}
Cosmological redshift underpins myriad applications in modern astronomy. Accurate redshift estimates permit the conversion of observed magnitudes of distant galaxies into intrinsic luminosities, allowing studies of the galaxy luminosity function \citep[e.g.][]{1976ApJ...203..297S, 2015ApJ...810...71F, 2023MNRAS.523.1009B}, the evolution of galaxy size and morphology across cosmic epochs \citep[e.g.][]{2014ARA&A..52..291C, 2015ApJS..219...15S}, and scaling relations between galaxy stellar masses and properties of their interstellar medium \citep[e.g.][]{1998ApJ...498..541K, 2004ApJ...613..898T}, central supermassive black holes \citep[e.g.][]{2013ARA&A..51..511K}, and host haloes \citep[e.g.][]{2013ApJ...770...57B, 2018ARA&A..56..435W}.

In cosmology, Type Ia supernova host galaxy redshifts provided the first direct evidence of the accelerating expansion of the Universe \citep{1998AJ....116.1009R, 1999ApJ...517..565P}. Baryon Acoustic Oscillations (BAOs), measured from galaxy clustering, act as a standard ruler for cosmic distances \citep{2005ApJ...633..560E}, while large-scale structure surveys map the growth of density fluctuations \citep{1980lssu.book.....P, 2006Natur.440.1137S}. The weak gravitational lensing of distant galaxies \citep{1990ApJ...349L...1T, 1998ApJ...498...26K, 2001PhR...340..291B} enables the reconstruction of the matter distribution and serves as a sensitive probe of dark energy through cosmic-shear tomography \citep{2015RPPh...78h6901K}. Collectively, these observables demonstrate how cosmological redshift is the keystone linking local galaxy properties to the geometry and dynamics of the Universe, and a critical underpinning of the concordance \(\Lambda\)CDM model.

Recent wide-field imaging surveys offer multi-wavelength photometry for hundreds of millions of galaxies, such as the Kilo-Degree Survey \citep[KiDS\footnote{\url{https://kids.strw.leidenuniv.nl}};][]{2024A&A...686A.170W}, the Dark Energy Survey \citep[DES\footnote{\url{https://www.darkenergysurvey.org}};][]{2025arXiv250105739B}, and the Hyper Suprime-Cam Survey \citep[HSC\footnote{\url{https://hsc.mtk.nao.ac.jp/ssp/}};][]{2022PASJ...74..421L}. Nonetheless, acquiring precise redshifts using spectroscopy for such extensive samples remains exceedingly resource-demanding. As a result, studies on large-scale structures and galaxy evolution are increasingly dependent on photometric redshift methods, which estimate redshifts from broadband flux observations, a technique initially proposed by \citet{1985AJ.....90..418K, 1986ApJ...303..154L}.

Next‐generation facilities -- including the NSF-DOE Vera C. Rubin Observatory \citep[\textit{Rubin}\footnote{\url{https://rubinobservatory.org}};][]{2009arXiv0912.0201L}, the Euclid mission \citep[\textit{Euclid}\footnote{\url{https://www.cosmos.esa.int/web/euclid}};][]{2025A&A...697A...1E}, the Nancy Grace Roman Space Telescope \citep[\textit{Roman}\footnote{\url{https://roman.gsfc.nasa.gov}};][]{2015arXiv150303757S} and the China Space Station Telescope \citep[CSST\footnote{\url{http://www.bao.ac.cn/csst/}};][]{2019ApJ...883..203G} -- will deliver unprecedented imaging depth and sky coverage. Exploiting their full cosmological potential to probe dark matter and dark energy, however, demands exceptional control of systematic uncertainties, especially in photometric redshift performance \citep{2018ARA&A..56..393M}.

In recent decades, numerous techniques for estimating photometric redshifts have been developed and are generally classified into two categories based on the type of data and the priors used during inference, as reviewed in \citet{2022ARA&A..60..363N}. Comprehensive descriptions of most of these techniques are available through the Redshift Assessment Infrastructure Layers \citep[\texttt{RAIL}\footnote{\url{https://github.com/LSSTDESC/rail}};][]{2025arXiv250502928T} platform for Legacy Survey of Space and Time (LSST) Dark Energy Science Collaboration (DESC). Specifically, the most traditional approach, template fitting, matches observed broadband photometry to libraries of galaxy Spectral Energy Distribution (SED) templates, using software packages such as \texttt{LePhare} \citep{1999MNRAS.310..540A,2006A&A...457..841I}, \texttt{BPZ} \citep{2000ApJ...536..571B, 2006AJ....132..926C}, \texttt{ZEBRA} \citep{2006MNRAS.372..565F}, \texttt{EAZY} \citep{2008ApJ...686.1503B}, and \texttt{DELIGHT} \citep{2017ApJ...838....5L}. 

Modern template-based approaches incorporate sophisticated Stellar Population Synthesis (SPS) models -- such as \texttt{FSPS} \citep{2009ApJ...699..486C, 2010ApJ...708...58C, 2010ApJ...712..833C}, \texttt{BAGPIPES} \citep{2018MNRAS.480.4379C}, and \texttt{Prospector} \citep{2021ApJS..254...22J} -- that include flexible star-formation histories and improved dust and nebular-emission treatments. These developments have enabled highly competitive photometric-redshift point estimates and ensemble redshift distributions, particularly for surveys with medium- or narrow-band photometry where spectral features are more cleanly resolved \citep[e.g.][]{2009ApJ...690.1250S, 2011ApJ...742...61S, 2021MNRAS.501.6103A}. For stage-IV, broad-band imaging surveys, however, the dominant systematic uncertainties generally arise from the realism and flexibility of the adopted galaxy-population priors rather than from intrinsic limitations of the SPS models themselves. Although template fitting provides a principled route to inferring redshift posteriors from noisy broad-band data, mismatches between the assumed and true population priors can still introduce biases that ultimately limit the accuracy of broad-band photometric-redshift estimates \citep[e.g.][]{2010A&A...523A..31H,2013ApJ...775...93D}.

Conversely, a variety of machine-learning algorithms have been devised to characterise the complex relationship between galaxy broadband SEDs and redshifts, utilising galaxies with known redshifts. These include artificial neural networks \citep[\texttt{ANNz};][]{2004PASP..116..345C, 2016PASP..128j4502S}, decision trees with random forest techniques \citep[\texttt{TPZ};][]{2013MNRAS.432.1483C}, Gaussian processes \citep[\texttt{GPz};][]{2016MNRAS.462..726A}, the quasi-Newton algorithm \citep[\texttt{MLPQNA};][]{2012A&A...546A..13C}, hybrid density estimation \citep[\texttt{METAPHOR};][]{2017MNRAS.465.1959C}, boosted decision trees combined with basis function decomposition \citep[\texttt{FlexZBoost};][]{10.1214/17-EJS1302, 2020A&C....3000362D, 2020MNRAS.499.1587S}, the normalising flow algorithm \citep[\texttt{PZFlow}; ][]{2024AJ....168...80C}, and stratified learning \citep[\texttt{StratLearn-z};][]{2018MNRAS.473.3969R, 2024MNRAS.534.3808A, 2025OJAp....8E..50M}. 

These data-driven approaches rely on training samples with high completeness and precisely measured redshifts, typically drawn from wide‐area spectroscopic surveys such as the Sloan Digital Sky Survey \cite[SDSS\footnote{\url{https://www.sdss.org}};][]{2000AJ....120.1579Y}, the Two-degree Field Galaxy Redshift Survey \citep[2DFGRS\footnote{\url{http://www.2dfgrs.net}};][]{2001MNRAS.328.1039C} and the Galaxy And Mass Assembly survey \citep[GAMA\footnote{\url{https://www.gama-survey.org}};][]{2011MNRAS.413..971D}, as well as deep spectroscopic programmes (e.g.\ zCOSMOS; \citealp{2007ApJS..172...70L}, DEEP2; \citealp{2013ApJS..208....5N}, VIPERS; \citealp{2014A&A...566A.108G}). Narrow‐band imaging campaigns (e.g.\ J‐PAS\footnote{\url{https://www.j-pas.org}}; \citealp{2014arXiv1403.5237B}, PAUS\footnote{\url{https://www.paus-survey.org}}; \citealp{2025A&A...693A.102D}) extend this reach to fainter magnitudes and higher redshifts. New and forthcoming facilities -- among them the Dark Energy Spectroscopic Instrument \citep[DESI\footnote{\url{https://www.desi.lbl.gov}};][]{2024AJ....167...62D}, the Subaru Prime Focus Spectrograph \citep[PFS\footnote{\url{https://pfs.ipmu.jp}};][]{2014PASJ...66R...1T}, the 4-metre Multi-Object Spectroscopic Telescope \citep[4MOST\footnote{\url{https://www.4most.eu/cms/home/}};][]{2016SPIE.9910E..1NW} and William Herschel Telescope Enhanced Area Velocity Explorer \citep[WEAVE\footnote{\url{https://www.ing.iac.es/astronomy/instruments/weave/weaveinst.html}};][]{2024MNRAS.530.2688J} -- promise substantially improved spectroscopic completeness, thereby enhancing the training and accuracy of machine‐learning photometric‐redshift estimators.

Nevertheless, spectroscopic samples remain subject to magnitude limits and exhibit uneven coverage in redshift-colour space, resulting in non‐representative training datasets and inducing selection biases \citep{2020A&A...637A.100W, 2020MNRAS.496.4769H}. Spectroscopic misidentifications and catalogue mismatches further introduce systematic offsets in point estimates and distort the shapes of posterior redshift distributions \citep{2015ApJ...813...53M, 2020MNRAS.496.4769H}. Moreover, degeneracies in galaxy SEDs and redshift-colour ambiguities exacerbate these biases. Mitigating these challenges therefore demands robust statistical frameworks, truly representative calibration samples and explicit correction schemes that account for spectroscopic incompleteness.

To address these limitations, an alternative method introduced by \citep{2024ApJ...967L...6M} utilises synthetic mock galaxy catalogues to supplement missing galaxy SEDs, thereby providing more representative datasets for training machine learning methods to estimate photometric redshifts. Our work is therefore complementary to SPS-based methods: instead of modifying SED priors, we aim to improve the empirical completeness of the training sample in colour-magnitude space through targeted data augmentation. This requires a careful understanding of the realism of the supplemental catalogue, but was shown to provide significant improvements, particularly in high-redshift estimation where the training dataset is least representative. 

In this study, we extend this framework by using a Self-Organising Map (SOM) to identify regions of colour-magnitude space with insufficient spectroscopic-redshift coverage and preferentially augment those regions. This enables a more targeted and effective augmentation of simulated galaxies, leading to improved accuracy in photometric-redshift estimates. We also introduce two-stage controlled variations to generate a broad suite of synthetic spectroscopic samples representative of those potentially available for LSST. This allows us to quantify the impact of differing spectroscopic selection functions and to assess photometric-redshift performance under both one-year (Y1) and ten-year (Y10) LSST depths.

This work focuses on generating mock photometric and spectroscopic galaxy datasets with a realistic level of fidelity for LSST cosmological analyses, and demonstrates how SOM-based data augmentation improves photometric-redshift estimates for individual galaxies. In a forthcoming study (Zhang et al., in preparation), we extend this framework to establish tomographic binning schemes and to calibrate the ensemble redshift distributions for each tomographic bin.

The structure of the paper is as follows: Section~\ref{sec:2} provides a review of the mock galaxy catalogues; Section~\ref{sec:3} outlines our photometric and spectroscopic sample selection and augmentation process; Section~\ref{sec:4} explains our photometric redshift modelling; Section~\ref{sec:5} evaluates the precision achieved; and Section~\ref{sec:6} concludes with a summary of our findings and and presents opportunities for future research.

\section{Mock Galaxy Catalogue} \label{sec:2}
\begin{figure*}
    \includegraphics[width=\linewidth]{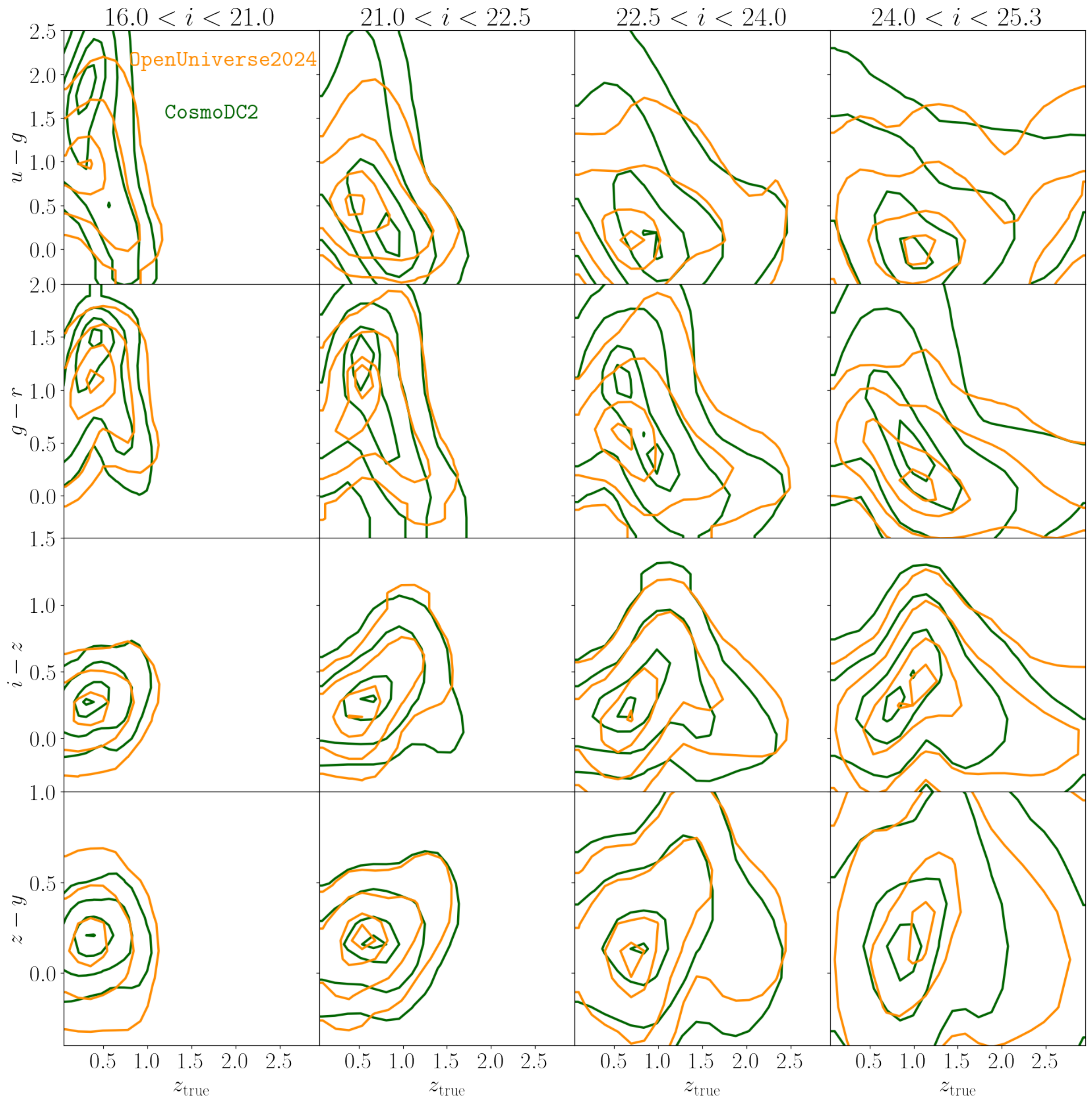}
    \caption{Comparison of the colour-redshift distributions for \texttt{OpenUniverse2024} (orange contours) and \texttt{CosmoDC2} (green contours) at LSST Y10 depth, showing confidence contours at \(0.5\sigma, 1.0\sigma, 2.0\sigma,\) and \(3.0\sigma\). From top to bottom, the rows show the \(u - g\), \(g - r\), \(i - z\), and \(z - y\) colours, while the columns correspond to bins in \(i\)-band magnitude: \(16.0 < i < 21.0\), \(21.0 < i < 22.5\), \(22.5 < i < 24.0\), and \(24.0 < i < 25.3\). The contours highlight the relative coverage of colour-redshift space: \texttt{CosmoDC2} generally spans a broader region, while there remain areas occupied only by \texttt{OpenUniverse2024}, where resampling \texttt{CosmoDC2} galaxies alone cannot fully reconcile the discrepancy.}
    \label{fig:1}
\end{figure*}

To ensure realism and utility, tests of the data‐augmentation method using simulated galaxies must employ catalogues that exhibit authentic differences. In practice, the discrepancy between the real observation and one simulation should be comparable to that between two distinct simulations employed in our study. If the mocks were identical, augmentation would trivially succeed but offer no insight into its effectiveness in real analyses; if they were unrealistically different, augmentation would fail outright. 

In this section, we outline the key features of the two input mock galaxy catalogues, \texttt{CosmoDC2} and \texttt{OpenUniverse2024}, that serve as the starting points for our data manipulation and analysis. The \texttt{OpenUniverse2024} catalogue is utilised to emulate the properties of galaxies expected in forthcoming LSST observations, whereas \texttt{CosmoDC2} is used here solely for data augmentation. Our selected catalogues closely resemble actual observations; nevertheless, they exhibit differences, offering an appropriate testing environment for the intended application of the method. This variation effectively represents the systematic discrepancies between synthetic catalogues and real observations, thereby reproducing a significant source of bias in photometric-redshift estimation through data augmentation.

\subsection{\texttt{CosmoDC2}}

The \texttt{CosmoDC2} catalogue \citep{2019ApJS..245...26K} is a comprehensive, purpose‐designed resource for \textit{Rubin} cosmology. This constitutes the foundation of the second Data Challenge \citep[DC2;][]{2021ApJS..253...31L} conducted by the LSST DESC. Derived from the Outer Rim N-body simulation \citep{2019ApJS..245...16H}, it models a cosmic volume of \( \left( 4.225\, \mathrm{Gpc} \right)^3 \) using over a trillion \(10240^3\) particles, each with a mass of \(2.6 \times 10^9 \, M_{\odot}\). The simulation suite comprises roughly \(100\) temporal snapshots, recorded from \(z \sim 10\) to \(z = 0\), and was created using the Hybrid/Hardware Accelerated Cosmology Code (HACC; \citealp{2016NewA...42...49H} on the IBM BG/Q system Mira at the Argonne Leadership Computing Facility (ALCF\footnote{\url{https://www.alcf.anl.gov}}). 

To incorporate realistic galaxy properties, \texttt{CosmoDC2} employs \texttt{GalSampler} \citep{2020MNRAS.495.5040H}, a hybrid approach that merges semi-analytic modelling using \texttt{Galacticus} \citep{2012MNRAS.419.3590B}, with empirical techniques based on the \texttt{UniverseMachine} \citep{2019MNRAS.488.3143B}. This method transfers sophisticated stellar population synthesis prescriptions to cosmological simulations via halo-to-halo correspondence, incorporating more than \(500\) galaxy attributes, including the \textit{Rubin} photometry in the \(u\), \(g\), \(r\), \(i\), \(z\), \(y\) filters, morphology, halo properties, and weak lensing shear. \texttt{CosmoDC2} has been validated against the scientific objectives of LSST DESC \citep{2022OJAp....5E...1K}, and its simulated catalogues span over \(440 \,\mathrm{deg}^2\) of sky up to a redshift of \(z = 3\), reaching a magnitude depth of \(r = 28 \) and supporting a broad range of scientific applications.

\subsection{\texttt{OpenUniverse2024}}

The \texttt{OpenUniverse2024} extragalactic catalogue \citep{2025MNRAS.544.3799O} builds upon the same Outer Rim dark‐matter simulation used in \texttt{CosmoDC2}, but uses the differentiable \texttt{Diffsky} framework to populate galaxies within the same cosmological large‐scale structure, with details introduced in \href{https://lsstdesc-diffsky.readthedocs.io/en/latest/index.html}{\texttt{lsstdesc-diffsky}}. Rather than relying on discrete merger‐tree assembly histories and star‐formation tracks, \texttt{Diffsky} utilises the auto‐differentiable modules \texttt{DiffMAH} \citep{2021OJAp....4E...7H}, \texttt{DiffStar} \citep{2023MNRAS.518..562A}, and \texttt{DSPS} \citep{2023MNRAS.521.1741H} to derive smooth, physically motivated parametrisations of mass assembly histories, star formation histories, and stellar population synthesis modelling. This approach yields robust predictions of stellar mass and star formation rates, even in poorly resolved haloes. The resulting position, lensing shear and convergence, morphological parameters, and multi‐wavelength SED of each galaxy are consolidated within \texttt{SkyCatalog}, a unified truth‐catalogue format. \texttt{OpenUniverse2024} covers approximately \(110\,\mathrm{deg}^2\) -- corresponding to both the LSST Deep Drilling Fields (DDFs; \citealp{2024ApJS..275...21G}) and the Roman High-Latitude Imaging Survey \citep{2019arXiv190205569A}.

\subsection{Comparison of catalogues}

Both \texttt{CosmoDC2} and \texttt{OpenUniverse2024} employ galaxy SED models that have been rigorously validated against deep, multi-band survey data. Owing to its fully differentiable framework, the mock colours and number counts of \texttt{OpenUniverse2024} agree with those of the COSMOS2020 survey at the per cent level \citep{2022ApJS..258...11W, 2022A&A...664A..61S}. Similarly, \texttt{CosmoDC2} demonstrates excellent consistency in its bright, low-redshift galaxy population when evaluated with the LSST DESCQA validation suite \citep{2018ApJS..234...36M, 2019ApJS..245...26K, 2021ApJS..253...31L}.

As illustrated by the green points in Figure~\ref{fig:1}, the colour–redshift distribution of \texttt{CosmoDC2} exhibits pronounced discreteness at high redshift compared to the orange points for \texttt{OpenUniverse2024}. This behaviour originates from its dependence on a restricted array of stellar population templates. We therefore utilise \texttt{CosmoDC2} to construct our augmentation catalogues, thereby compensating for gaps in spectroscopic coverage. In a practical analysis, these augmented catalogues would likewise be simulated datasets, with their SEDs mildly differing from those in the observational datasets.

Conversely, the auto-differentiable, parameterised models utilised by \texttt{OpenUniverse2024} produce smoother and more physical colour distributions, rendering it an excellent proxy for LSST-like observations. In this study, we employ \texttt{OpenUniverse2024}, a high-fidelity LSST-like mock catalogue, as a proxy for the observational dataset expected from the forthcoming survey.

At low redshift (\(z_{\mathrm{true}} \lesssim 1.5\)), both catalogues exhibit excellent agreement in the \(i - z\) and \(z - y\) colours, particularly for bright galaxies (\(i \lesssim 22.5\)). Conversely, \texttt{CosmoDC2} shows a systematic excess of redder \(u - g\) and \(g - r\) colours relative to \texttt{OpenUniverse2024} especially among low-redshift (\(z_{\mathrm{true}} \lesssim 1.5\)) and faint galaxies (\(i \gtrsim 22.5\)), reflecting differences in the treatment of stellar populations at the blue end of the SED. As detailed below, these offsets can be partly mitigated by weighted resampling of the \texttt{CosmoDC2} galaxies to match the smoother distributions of \texttt{OpenUniverse2024}. 

At fainter magnitudes (\(i \gtrsim 22.5\)) and higher redshifts (\(z_{\mathrm{true}} \gtrsim 1.5\)), where spectroscopic calibrators are scarce, the two SED models diverge markedly, reducing their overlap. This divergence persists after resampling and thus represents a significant source of systematic error in photometric-redshift modelling. Figure~\ref{fig:1} illustrates these comparisons at LSST Y10 depth -- chosen to emphasise SED differences among the faintest galaxies (\(i\gtrsim24.0\)) -- although identical trends are evident in other observing scenarios.

\section{Dataset Configuration} \label{sec:3}
\begin{figure}
    \includegraphics[width=\linewidth]{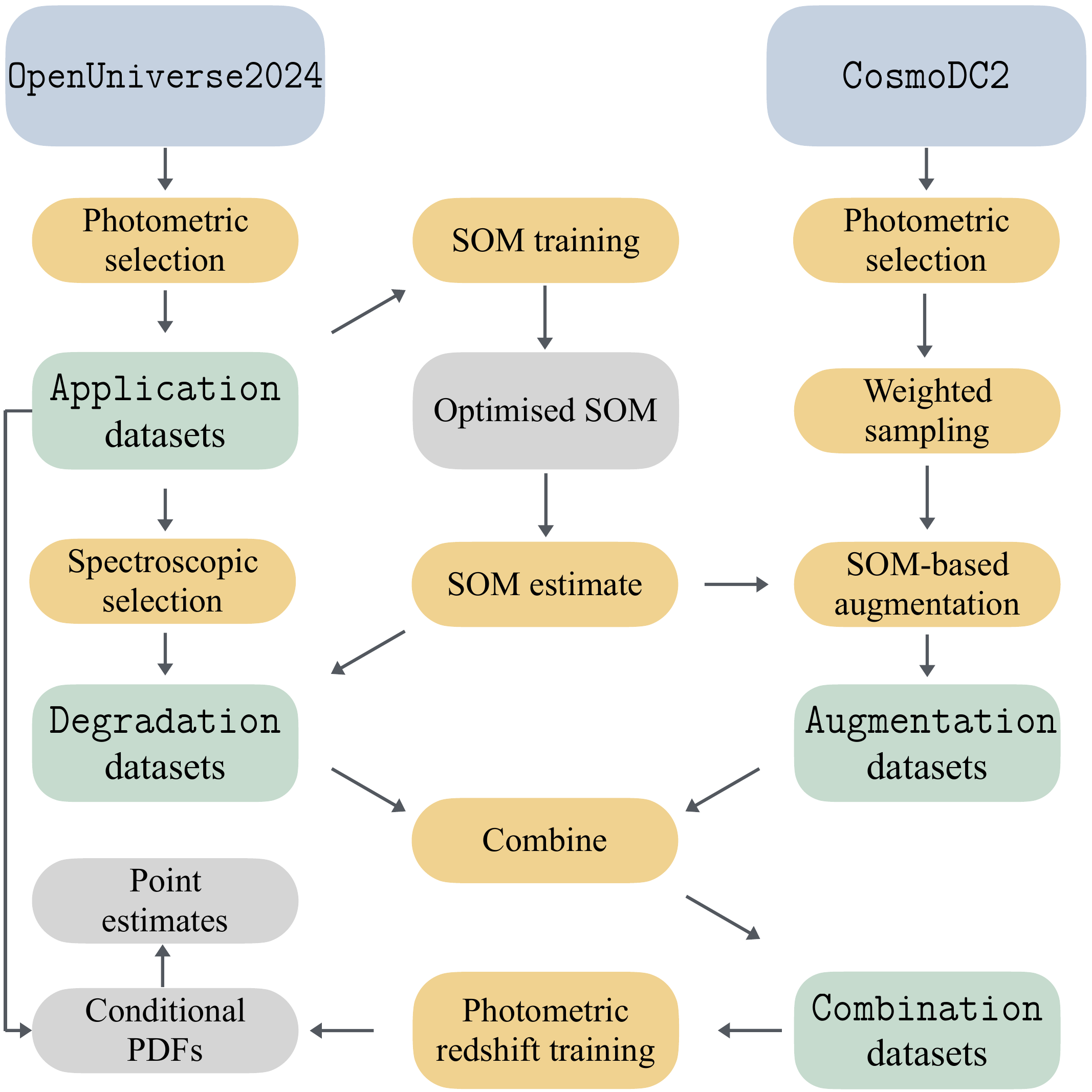}
    \caption{Schematic overview of the dataset generation pipeline. The diagram outlines the steps for constructing synthetic catalogues tailored to either photometric or spectroscopic selection criteria, using mock galaxies from \texttt{OpenUniverse2024} and \texttt{CosmoDC2}.}
    \label{fig:2}
\end{figure}

In the following, we outline our approach to utilising these advanced mock galaxy catalogues to produce several synthetic datasets. Each dataset is designed to replicate the photometric and spectroscopic selection functions of galaxy surveys for LSST photometric redshift analyses and to account for variations in selection effects. The datasets we generated are defined as follows:

\begin{itemize}
    \item \texttt{Application}: a general LSST-like photometric sample.
    \item \texttt{Degradation}: a compilation of multiple spectroscopic datasets derived from the \texttt{Application} datasets, used as part of the training sample.
    \item \texttt{Augmentation}: training datasets generated through simulations and added to the spectroscopic samples.
    \item \texttt{Combination}: complete training datasets produced through the integration of \texttt{Degradation} and \texttt{Augmentation} samples.
\end{itemize}

Figure~\ref{fig:2} presents a schematic overview of the pipeline, illustrating the principal stages from catalogue ingestion to the production of analysis-ready datasets. In order to incorporate different systematic effects into later modelling stages, we conduct \(501\) experiments that produce the four catalogues. The study involves a baseline experiment for machine learning training and \(N = 500\) supplementary experiments for statistical analysis, distinguished by differences in photometric noise realisations, spectroscopic selection functions, and samples of augmented galaxies, as outlined subsequently.

\subsection{\texttt{Application} datasets}

As mentioned earlier, the \texttt{OpenUniverse2024} catalogues serve as the foundation for replicating the photometric \texttt{Gold} selection as outlined in the LSST DESC Science Requirements Document \citep[SRD;][]{2018arXiv180901669T}. We refer to the outcome of our \texttt{Gold} selection as the \texttt{Application} datasets. We then apply an unsupervised machine‐learning technique for dimensionality reduction, the Self‐Organising Map (SOM), to map galaxy broadband SEDs onto a two‐dimensional grid, thereby facilitating subsequent analyses of spectroscopic degradation and data augmentation.

\subsubsection{\texttt{Gold} selection}

\begin{table*}
  \centering
  \begin{threeparttable}
    \caption{Statistical properties of the baseline \texttt{Application} dataset.}
    \label{tab:1}
    \begin{tabular}{lccccc}
      \hline
      Scenario & Number of galaxies & Number density & Effective number density\tnote{a} & Limiting magnitude\tnote{b} & Gold magnitude\tnote{c}  \\
      \hline
      LSST-Y1 & 2924005 & \(14.77 \, \mathrm{arcmin}^{-2}\) & \(11.14 \, \mathrm{arcmin}^{-2}\) & \(25.26 \) & \(24.05 \) \\
      LSST-Y10 & 7223012 & \(36.48 \, \mathrm{arcmin}^{-2}\) & \(27.95 \, \mathrm{arcmin}^{-2}\) & \(26.51 \) & \(25.30 \) \\
      \hline
    \end{tabular}
    \begin{tablenotes}
      \item[a] Effective number density is defined in \cite{2013MNRAS.434.2121C}.
      \item[b] \(i-\)band \(5\sigma\) point-source limiting magnitude under the prescribed observing conditions.
      \item[c] \texttt{Gold} magnitude: the \(i-\)band magnitude threshold used for \texttt{Gold} selection, as specified by LSST DESC SRD.
    \end{tablenotes}
  \end{threeparttable}
\end{table*}

Utilising the \texttt{OpenUniverse2024} catalogues, our first step is to compute realistic photometric uncertainties for observations at the LSST Y1 and Y10 depths. These uncertainties are obtained using the publicly available tool \citep[\href{https://github.com/jfcrenshaw/photerr}{\faGithub\,\texttt{PhotErr}};][]{2024AJ....168...80C}, which adopts the single-visit magnitude limits and expected numbers of visits per band from the survey specifications presented in \citet{2019ApJ...873..111I}. In our implementation, these quantities are used solely to derive uniform-depth Y1 and Y10 photometric-error models; we do not model LSST footprint variations or spatially varying depth. Only the Wide–Fast–Deep (WFD) component of the survey is considered, and the Deep Drilling Fields (DDFs) are not included. The resulting error prescriptions are applied to both point sources and extended sources using the galaxy morphological parameters described by \citet{2020A&A...642A.200V}.

In the \texttt{Application} datasets, each of the \(501\) experiments incorporates a separate random realisation of photometric noise, achieved through a unique seed to disturb the galaxy photometry. This method guarantees the creation of statistically independent synthetic datasets that mirror observational uncertainties. When calculating the perturbed magnitudes, we manually set the fluxes at the \(1 \sigma\) noise threshold for galaxies dimmer than the specified limiting magnitudes in each band, considering them as non-detections. This setting ensures that the distribution of galaxy colours remains numerically continuous, even when certain bands are not detected. 

One of the key aims of the \texttt{Gold} selection is to pinpoint galaxies with trustworthy detections. Consequently, the \texttt{Application} datasets developed subsequent to the \texttt{Gold} selection embody pragmatic photometric selection standards and serve as the groundwork for subsequent photometric-redshift modelling. Therefore, in order to eliminate very faint extended sources, we require the Signal‐to‐Noise Ratio (SNR) obtained from the combined \(r\)- and \(i\)-band photometry, (\(\mu \equiv \mathrm{SNR}_{i + r}\)), to meet the condition \(\mu > 10\). Additionally, we impose a restriction on the squared ratio of the galaxy size to the coadded point-spread function (PSF), expressed as \(\eta \equiv \left( R^{i + r} \,/\, R_{\mathrm{PSF}}^{i + r} \right)^2\), requiring \(\eta > 0.1\) in order to reduce contamination from unresolved point sources during practical analysis.

In addition, we exclude objects brighter than \(16 \) in \(i-\)band to mimic the masking of bright, saturated sources, and remove galaxies fainter than \(24.05\) for Y1 and \(25.30\) for Y10, in accordance with the \texttt{Gold} selection defined in the LSST DESC SRD. Additionally, to capture spatial sample variance, we randomly select half of the HEALPix sky pixels from \texttt{OpenUniverse2024} in each experiment.

Table~\ref{tab:1} provides a concise overview of essential information about the baseline \texttt{Application} datasets. Differences in the supplementary experiments are minimal and thus excluded for the sake of brevity.

\subsubsection{Self-Organising Map}

\begin{figure*}
    \centering
    \begin{subfigure}[b]{0.48\linewidth}
        \centering
        \includegraphics[width=\linewidth]{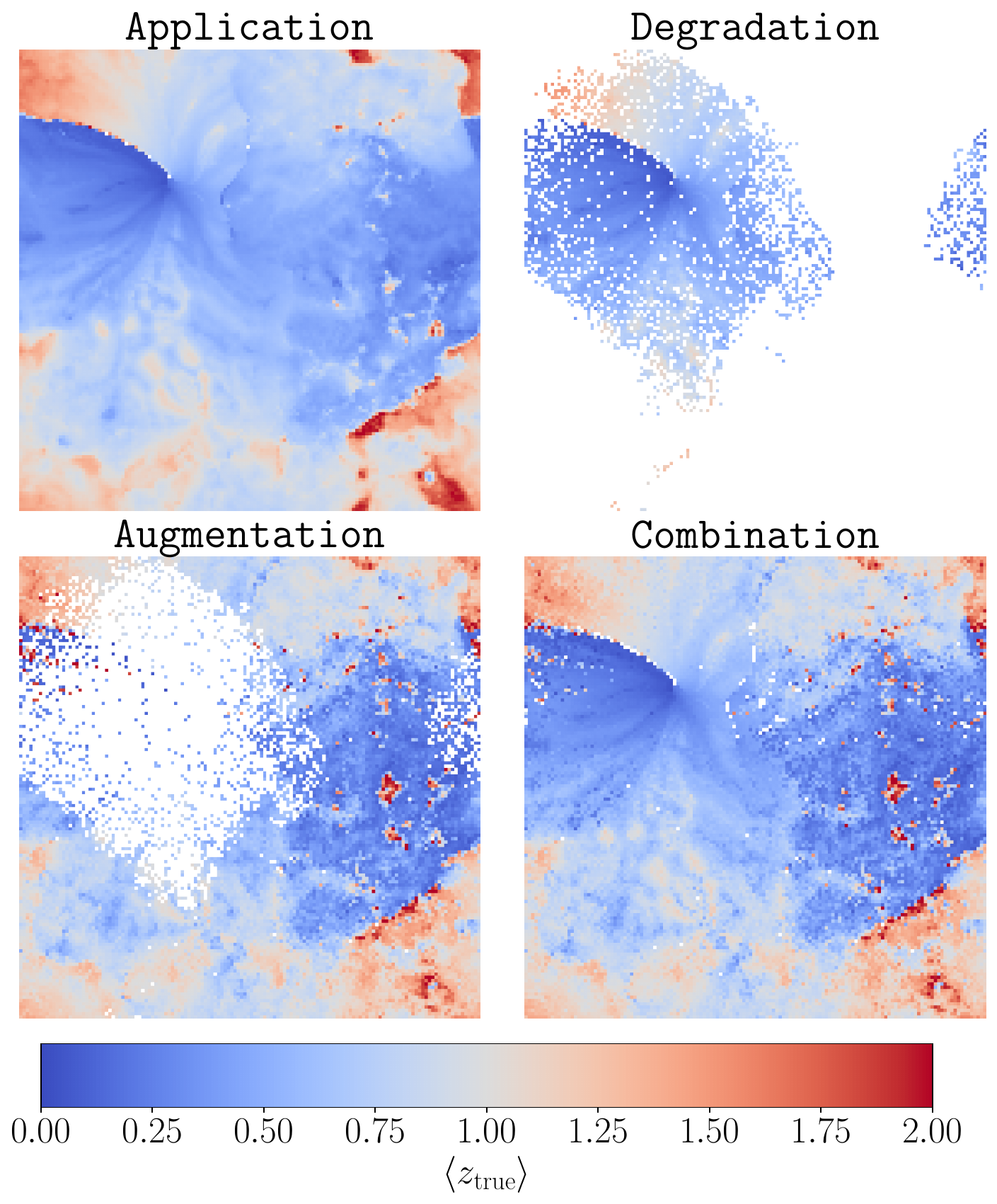}
        \caption{LSST Y1}
    \end{subfigure}
    \hspace{0.0\linewidth}
    \begin{subfigure}[b]{0.48\linewidth}
        \centering
        \includegraphics[width=\linewidth]{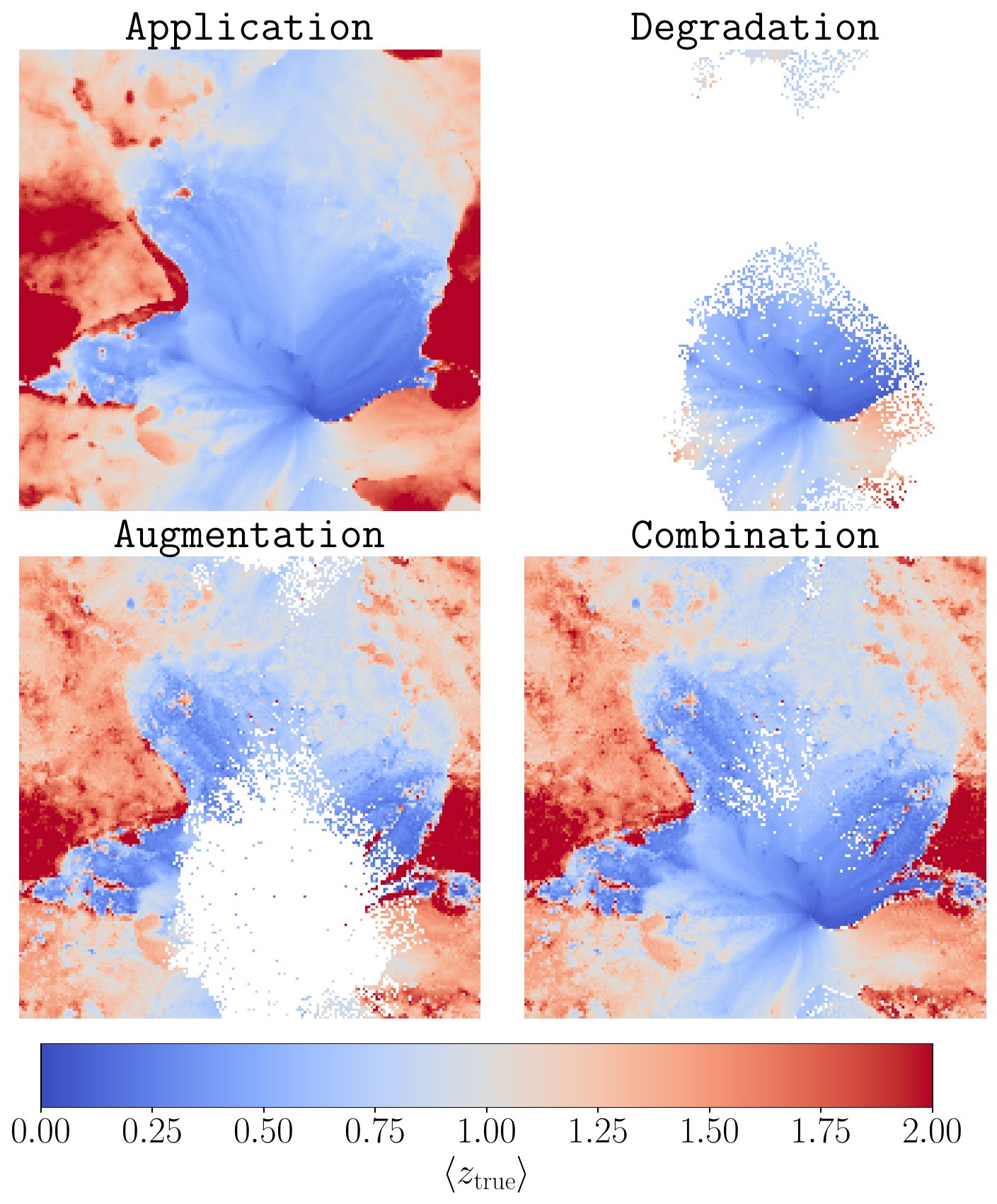}
        \caption{LSST Y10}
    \end{subfigure}
    \caption{Colour maps showing the average true redshift per SOM cell for the baseline experiment. The panels on the left correspond to SOMs trained on the LSST~Y1 photometric sample, and those on the right correspond to SOMs trained on the LSST~Y10 sample. Within each panel, the \texttt{Application}, \texttt{Degradation}, \texttt{Augmentation}, and \texttt{Combination} datasets occupy the upper-left, upper-right, lower-left, and lower-right quadrants, respectively. The \texttt{Degradation} dataset spans a noticeably smaller region of the SOM plane, reflecting the reduction in colour-magnitude coverage imposed by spectroscopic-selection effects. Meanwhile, although the \texttt{Augmentation} dataset broadly follows the structure of the \texttt{Application} sample, it departs more markedly in the Y10 case, where the underlying SED-model differences between \texttt{OpenUniverse2024} and \texttt{CosmoDC2} become increasingly evident.}
    \label{fig:3}
\end{figure*}

We employ the baseline \texttt{Application} dataset to train the SOM algorithm implemented via \texttt{Somoclu} \citep{2013arXiv1305.1422W} within the \texttt{RAIL} platform. This training compresses the high-dimensional space of a suite of representative galaxy broadband SEDs into a discrete two-dimensional map, with each cell corresponding to a sub-volume of the SED space. We choose the \(i-\)band as the reference, combining its apparent magnitudes with the colours derived relative to it. The SOM grids are configured as \(140 \times 140\) for Y1 and \(180 \times 180\) for Y10. These choices reflect a balance between achieving sufficient resolution to capture the mapping between galaxy broadband SEDs and redshift, and the practical limitations imposed by available computational resources. Although the present grid sizes are adequate for our analysis, future applications to petabyte-scale LSST datasets may require further optimisation, either through improved parallelisation and accelerated SOM training, or through carefully designed downsampling strategies for the observational catalogues.

Additionally, we adopt a Gaussian kernel with a standard deviation coefficient of \(0.5\) and a toroidal topology with hexagonal neurons for the training. The training is initialised using Principal Component Analysis (PCA) and then iterated for \(100\) epochs with linear cooling of both the neighbourhood radius and learning rate. The pre-trained SOM is subsequently applied to the remaining \(500\) statistical \texttt{Application} datasets, computing the Best Matching Unit (BMU) for each galaxy. To minimise computational costs, we refrain from retraining for each iteration and verify that the random photometric error realisation does not substantially affect the SOM mapping.

\begin{figure*}
    \centering
    \begin{subfigure}[b]{0.48\linewidth}
        \centering
        \includegraphics[width=\linewidth]{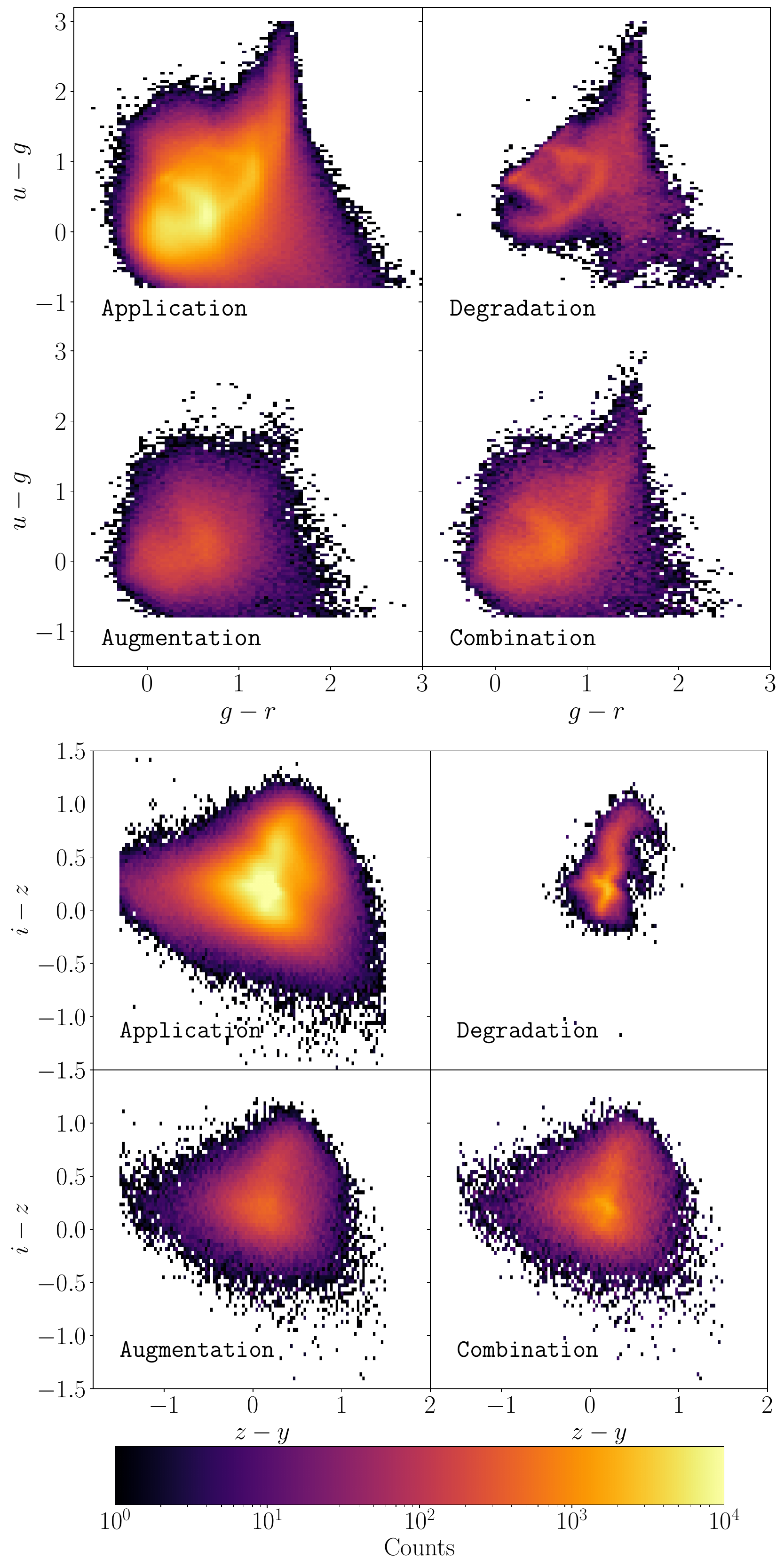}
        \caption{LSST Y1}
    \end{subfigure}
    \hspace{0.0\linewidth}
    \begin{subfigure}[b]{0.48\linewidth}
        \centering
        \includegraphics[width=\linewidth]{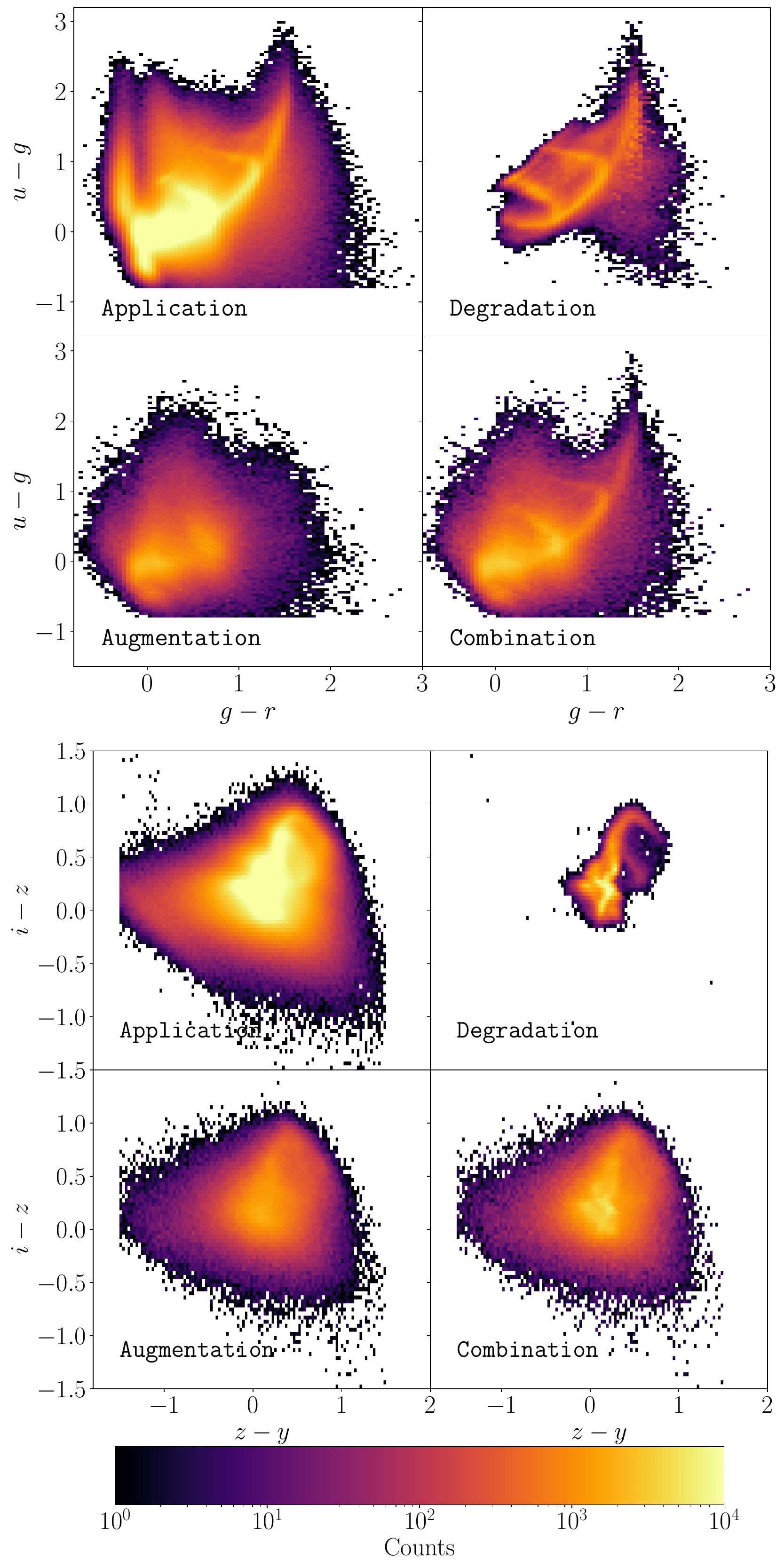}
        \caption{LSST Y10}
    \end{subfigure}
    \caption{Colour--colour diagrams for the baseline experiment, shown for LSST~Y1 (left) and LSST~Y10 (right). The upper panels display \(u - g\) versus \(g - r\), while the lower panels show \(i - z\) versus \(z - y\). Each panel is subdivided into four quadrants corresponding to the \texttt{Application} (upper left), \texttt{Degradation} (upper right), \texttt{Augmentation} (lower left), and \texttt{Combination} (lower right) datasets. A logarithmic scale highlights structures in low-density regions of colour-magnitude space while maintaining the qualitative sense of relative densities.}
    \label{fig:4}
\end{figure*}

In each panel of Figures~\ref{fig:3}–\ref{fig:6}, the left panel corresponds to the Y1 scenario and the right to Y10, both illustrating the baseline experiment as a representative example of the full set of 501 realisations. In Figure~\ref{fig:3}, the upper‐left subplot shows the SOM map of mean true redshifts derived from the baseline \texttt{Application} dataset. Figures~\ref{fig:4} displays the \(u - g\) versus \(g - r\) and \(i - z\) versus \(z - y\) colour–colour diagrams, again with the baseline \texttt{Application} data in the upper‐left quadrants of each panel. Figure~\ref{fig:5} illustrates the distribution of \(g - z\) colour against the \(i\)-band magnitude, whereas Figure~\ref{fig:6} shows the galaxy distributions in relation to the true redshifts \(z_{\mathrm{true}}\). Note that the Y10 dataset contains a higher fraction of galaxies at \(z_\mathrm{true} \gtrsim 1.0\) and fainter magnitudes (\(i > 23 \)) than Y1, reflecting its deeper survey depth.  

\subsection{\texttt{Degradation} datasets}

After constructing the \texttt{Application} datasets, we perform \(501\) realisations of a two-stage degradation procedure to mimic the selection effects that arise when compiling multiple spectroscopic surveys. This yields a suite of synthetic spectroscopic datasets representative of those plausibly accessible for LSST analysis. In the initial stage, we apply cuts in magnitude, redshift, and colour, and incorporate probabilistic sampling based on spectroscopic success rates. This produces \(501\) unique realisations of individual spectroscopic surveys, each reflecting the characteristics of different existing or planned spectroscopic programmes. The framework is flexible and can readily accommodate a wide variety of prospective spectroscopic strategies, rather than being tied to any single survey configuration.

In the second stage, we randomly sample and combine subsets of these realisations to generate \(501\) \texttt{Degradation} datasets, thereby emulating the compilation of multiple spectroscopic catalogues into a more representative training set for photometric-redshift modelling. The resulting \texttt{Degradation} datasets capture realistic variations in spectroscopic depth, colour-dependent selection biases, and redshift distributions, each of which is forward-modelled as a source of systematic uncertainty in the subsequent photometric-redshift estimation. However, our simplified prescriptions are not expected to fully capture all aspects of real spectroscopic compilations, but instead offer a forecasting tool for prospective LSST analyses.

\subsubsection{Spectroscopic selection}

For each realisation of the \texttt{Application} datasets, we draw an upper limit on the observed \(i\)-band magnitude, \(i_\ast\), uniformly from \(20  < i_\ast < 24 \) for Y1 and \(20  < i_\ast < 25 \) for Y10. These intervals mirror the depths of spectroscopic surveys likely to accompany LSST in its early and later operations.  Likewise, we assign an upper redshift cut, \(z_\ast\), by sampling uniformly within \(0.5 < z_\ast < 2.5\) for Y1 and \(0.5 < z_\ast < 3.0\) for Y10. 

This parameter configuration anticipates the increasing focus on faint and high-redshift galaxies in upcoming spectroscopic programmes. These encompass the accumulation of spectroscopy of galaxies through ongoing initiatives, as well as space-based missions like the James Webb Space Telescope \citep[JWST\footnote{\url{https://webbtelescope.org/home}};][]{2006SSRv..123..485G}, next-generation ground-based facilities -- such as the Extremely Large Telescope \citep[ELT\footnote{\url{https://elt.eso.org}};][]{2023ConPh..64...47P} and Giant Magellan Telescope \citep[GMT\footnote{\url{https://giantmagellan.org}};][]{2024SPIE13094E..17B} -- as well as preliminary Stage-V spectroscopic surveys, including DESI-II \citep{2022arXiv220903585S}, the MUltiplexed Survey Telescope (MUST\footnote{\url{https://must.astro.tsinghua.edu.cn}}) \citep{2024arXiv241107970Z}, and other potential candidates \citep{2022arXiv220904322S, 2025arXiv250307923B}.

Furthermore, deep-field spectroscopic programs frequently focus on optically red galaxies due to their robust redshift determination obtained from distinct spectral features, together with their more predictable selection functions, stronger large-scale clustering, and their detectability at greater depths. To simulate this preference, a magnitude-dependent cut in the \(g - z\) colour is applied, characterised by a linear relation with the \(i\)-band magnitude:
\begin{equation} \label{eqa:1} 
    i - i_\ast < \tan{\varphi_\ast} \left[ \left( g - z \right) - \left( g - z \right)_\ast \right].
\end{equation} 
Here, \(i_\ast\) denotes the previously defined upper limit in the \(i\)-band, \((g - z)_\ast\) the colour intercept and \(\varphi_\ast\) the slope angle. For both the LSST Y1 and Y10 scenarios, \((g - z)_\ast\) is drawn uniformly from \([1.0,\,3.0]\) and \(\varphi_\ast\) from \([0,\,\pi/2]\). 

This functional form is motivated by the colour-magnitude pre-selection strategies widely used in deep spectroscopic surveys such as DEEP2 \citep{2013ApJS..208....5N}, VVDS \citep{2005A&A...439..845L}, LEGA-C \citep{2021ApJS..256...44V}, and VUDS \citep{2015A&A...576A..79L}. These surveys often employ linear or piecewise-linear colour-magnitude boundaries to preferentially target optically red galaxies with strong continuum and absorption features, enabling robust redshift determination at faint magnitudes. Our parametrisation in Eq.~\ref{eqa:1} is therefore a simplified but flexible representation of these empirically motivated selection strategies.

In contrast, wide‐field campaigns like \textit{Euclid} and DESI also target bluer star-forming galaxies that exhibit prominent emission lines, particularly at high redshift. This selection is mainly influenced by detectability criteria regarding emission line flux and redshift window instead of colour \citep{2024A&A...689A.166C, 2023AJ....165..126R}, making them less affected by a red-galaxy exclusion. In order to simulate the influence of more intricate selection functions within the multi-dimensional colour space, we randomly choose SOM cells at a rate \(r_\ast\), which is uniformly distributed across the interval \(r \in \left[0.2, 0.8\right]\). This is applied to both the LSST Y1 and Y10 scenarios, with galaxies in the unselected cells being disregarded. Although our probabilistic prior within this parametrised framework is deliberately simplistic, it provides useful flexibility to capture variations in colour thresholds across different surveys. More sophisticated, survey‐specific colour boundaries could be incorporated in future developments.

Moreover, we account for the impact of the spectroscopic success rate \(\lambda (i)\) by modelling it as a logistic function of the galaxy \(i-\)band apparent magnitude: 
\begin{equation} \label{eqa:2}
    \lambda (i) = \frac{1}{1 + \lambda_\ast \exp{\left[ i - i_\ast \right]}}.
\end{equation} 
In this context, \(i_\ast\) represents the upper threshold previously defined in the \(i\)-band magnitude for each \texttt{Application} dataset. To introduce additional stochasticity, we draw the parameter \(\lambda_\ast\) uniformly from the range \(0.5 < \lambda_\ast < 5.0\) for both Y1 and Y10 scenarios, thereby capturing different levels of spectroscopic survey completeness. 

The logistic form in Eq.~\ref{eqa:2} produces an almost complete success rate for the brightest sources (\(i \lesssim 18 \)) and a steep decline as the SNR decreases. This behaviour reflects the well-established magnitude/SNR dependence of spectroscopic completeness observed in surveys such as zCOSMOS \citep{2009ApJS..184..218L}, VIPERS \citep{2014A&A...566A.108G}, and DESI \citep{2023AJ....165..126R}, all of which report near-unity success rates at bright magnitudes followed by a rapid fall-off beyond their survey-specific limits. Our parametrisation is intended to capture this characteristic transition while remaining survey-agnostic and flexible enough to accommodate variations in depth and targeting efficiency across realisations.

Each realisation of the first‐stage spectroscopic dataset is obtained by sampling according to Equation~\ref{eqa:2}, yielding \(100000\) to \(200000\) galaxies for Y1 and \(250000\) to \(500000\) galaxies for Y10, in line with the expected expansion of spectroscopic coverage. These sample sizes are motivated by deep‐field spectroscopic surveys, which cover relatively small areas of sky. By contrast, wide‐field campaigns observe orders of magnitude more galaxies; including them in full would cause the spectroscopic training sample to be dominated by their shallower selection functions, introducing a pronounced imbalance between bright and faint galaxies in photometric‐redshift training. Consequently, we select a subset of the wide‐field spectroscopic datasets of comparable size to our deep‐field samples, as reflected in our configuration. Moreover, given finite computational resources, much larger spectroscopic samples would substantially increase the cost of photometric‐redshift training across multiple realisations.

\subsubsection{Compilation of multiple datasets}

Finally, every \texttt{Degradation} dataset is constructed by merging \(M\) randomly chosen realisations, where \(M\) is drawn from \([4,\, 6]\) for Y1 and \([7,\, 12]\) for Y10, and then downsampling to the same total galaxy counts as the corresponding experiment of spectroscopically-selected galaxies, thereby minimising computational overhead. This procedure produces \texttt{Degradation} datasets that encapsulate multiple realistic spectroscopic selection functions while preserving sufficient variation across experiments.  

The \texttt{Degradation} datasets for pre‐trained SOM cells are shown in the upper right subplots of each panel in Figure~\ref{fig:3}, illustrating how spectroscopic selection sculpts the galaxy distribution in SOM space. Notably, many cells, particularly those at higher redshift, become empty due to these selection effects. This sparsity is also evident in the colour–colour diagrams (upper right quadrants of each panel in Figure~\ref{fig:4}), in the magnitude and colour cuts depicted in Figure~\ref{fig:5}, and in the lack of galaxies with \(z_\mathrm{true} \gtrsim 1.5\) (dark‐blue solid lines) in Figure~\ref{fig:6}. If left uncorrected, such incomplete sampling can heavily bias subsequent photometric‐redshift estimates.

\subsection{\texttt{Augmentation} datasets}

To address selection effects introduced by the degradation process, we adopt and extend an augmentation strategy originally proposed by \cite{2024ApJ...967L...6M}, adapting it for use with a SOM. Our method comprises a dual-phase procedure to construct the \texttt{Augmentation} datasets. Initially, galaxies are sampled from the \texttt{CosmoDC2} catalogues to ensure their distribution in SOM space corresponds with that of the \texttt{Application} datasets, thus aligning their overall distributions within the multi-dimensional colour-magnitude space. In the second phase, we perform targeted augmentation to populate SOM cells that are under-represented or entirely unoccupied in the corresponding \texttt{Degradation} datasets.

\subsubsection{Weighted sampling}

\begin{figure*}
    \centering
    \begin{subfigure}[b]{0.48\linewidth}
        \centering
        \includegraphics[width=\linewidth]{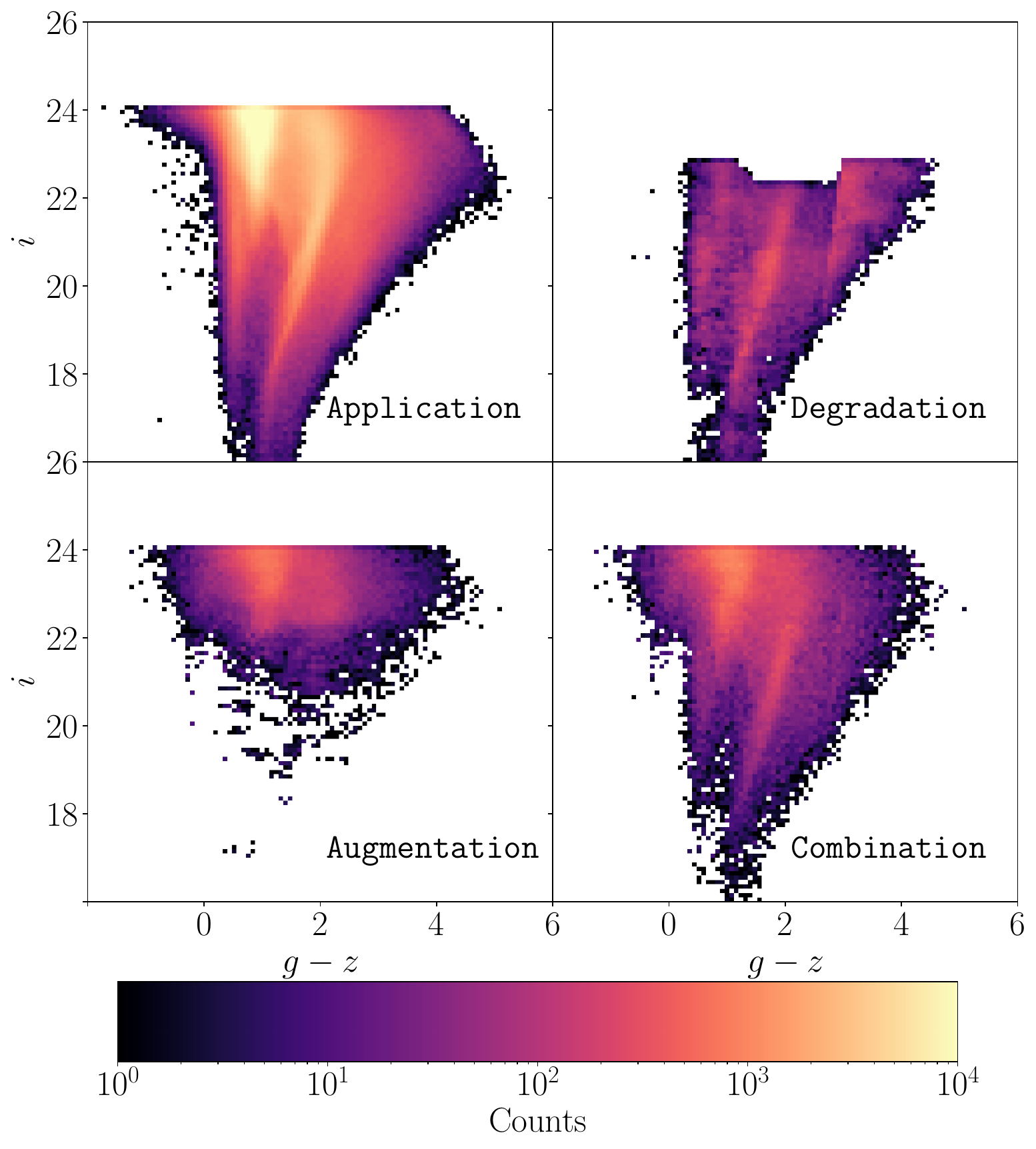}
        \caption{LSST Y1}
    \end{subfigure}
    \hspace{0.0\linewidth}
    \begin{subfigure}[b]{0.48\linewidth}
        \centering
        \includegraphics[width=\linewidth]{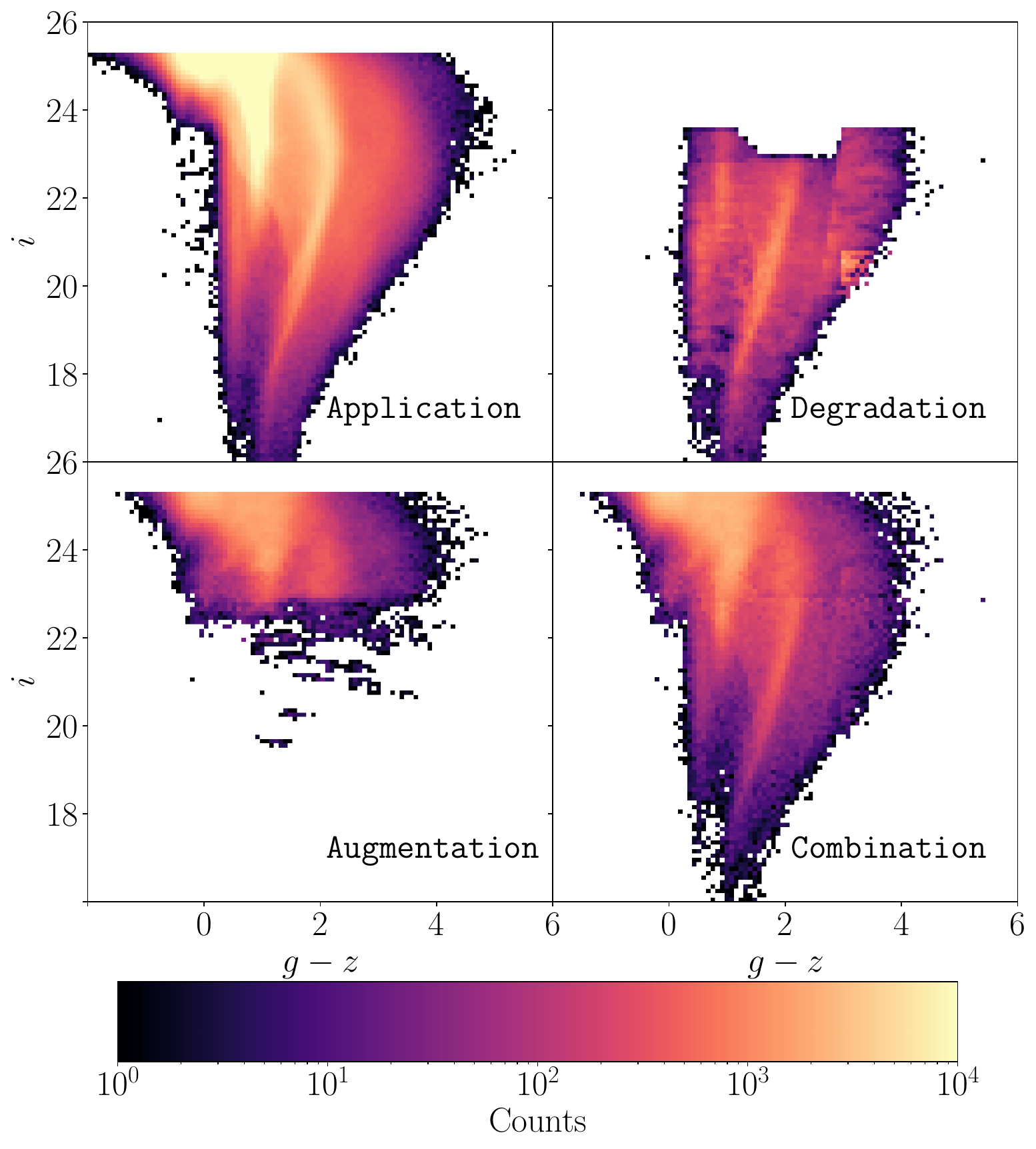}
        \caption{LSST Y10}
    \end{subfigure}
    \caption{\(g - z\) colours versus \(i\)-band apparent magnitudes for various datasets are displayed for Y1 (left panel) and Y10 (right panel), under the baseline experiment. In each panel, the upper left, upper right, lower left, and lower right quadrants correspond to the \texttt{Application}, \texttt{Degradation}, \texttt{Augmentation}, and \texttt{Combination} datasets, respectively. A logarithmic scale enhances features in low-density areas of colour-magnitude space while preserving the perceived relative densities.}
    \label{fig:5}
\end{figure*}

Initially, \texttt{CosmoDC2} galaxies are processed using the same pipeline as for the \texttt{Application} datasets. This includes applying LSST photometric errors, perturbing apparent magnitudes, enforcing the LSST DESC SRD \texttt{Gold} selection, and randomly sampling over HEALPix sky pixels to account for spatial variations, under both Y1 and Y10 observing conditions.

Then, the pre-trained SOM is used to map each selected \texttt{CosmoDC2} galaxy to its BMU in SOM space. We calculate cell-wise occupancy for both the \texttt{Application} datasets and the selected \texttt{CosmoDC2} galaxies. We then assign sampling weights to the \texttt{CosmoDC2} galaxies given by the ratio of cell occupancies, ensuring that the synthetic samples reproduce the structure of the \texttt{Application} datasets in SOM space. This selection and re-weighting procedure is repeated over \(501\) independent experiments, one baseline and \(500\) statistical runs, each matched in size to their respective \texttt{Application} datasets to minimise excessive manipulation.

\subsubsection{Adaptive data augmentation}

\begin{figure*}
    \centering
    \begin{subfigure}[b]{0.48\linewidth}
        \centering
        \includegraphics[width=\linewidth]{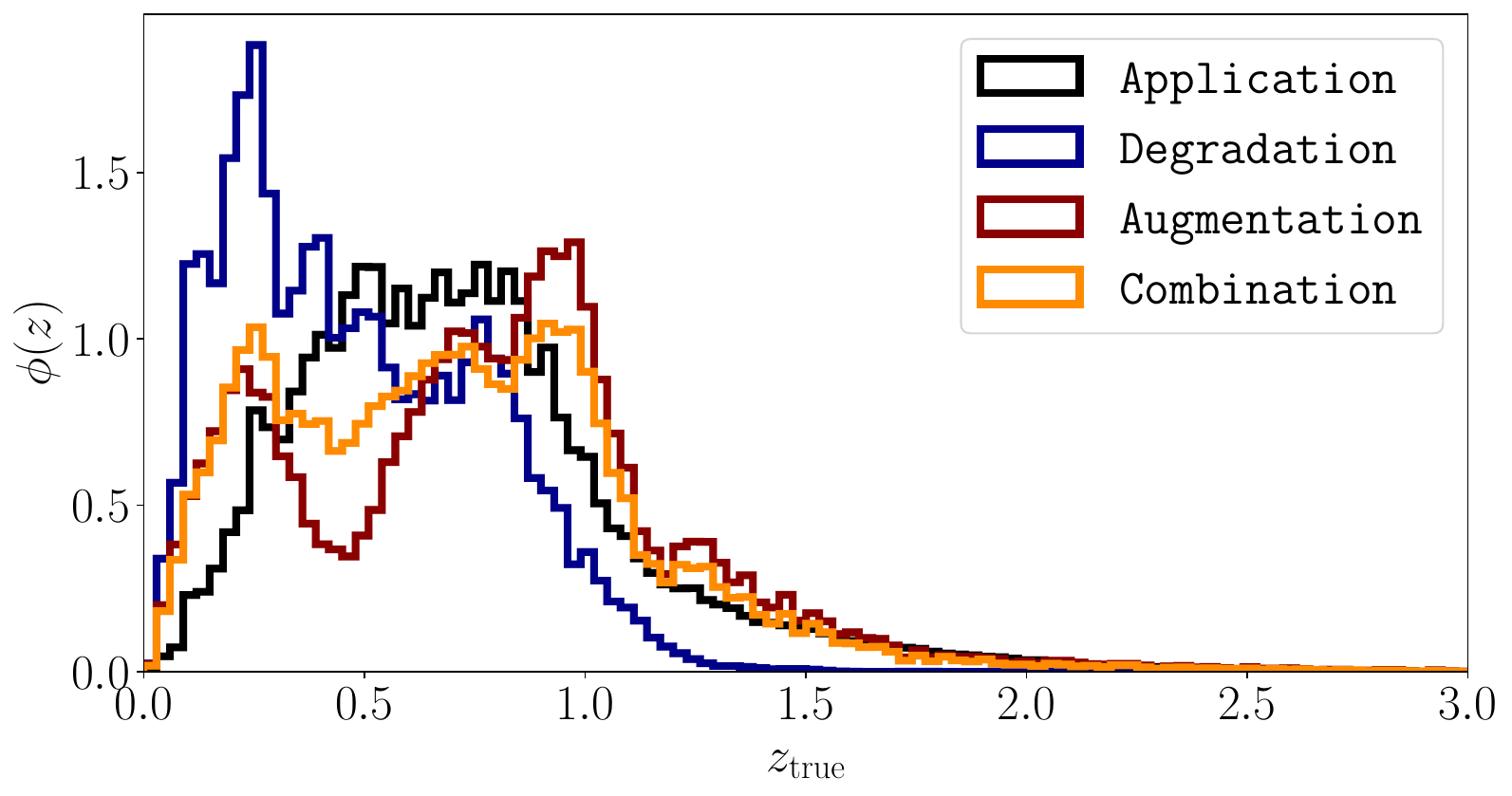}
        \caption{LSST Y1}
    \end{subfigure}
    \hspace{0.0\linewidth}
    \begin{subfigure}[b]{0.48\linewidth}
        \centering
        \includegraphics[width=\linewidth]{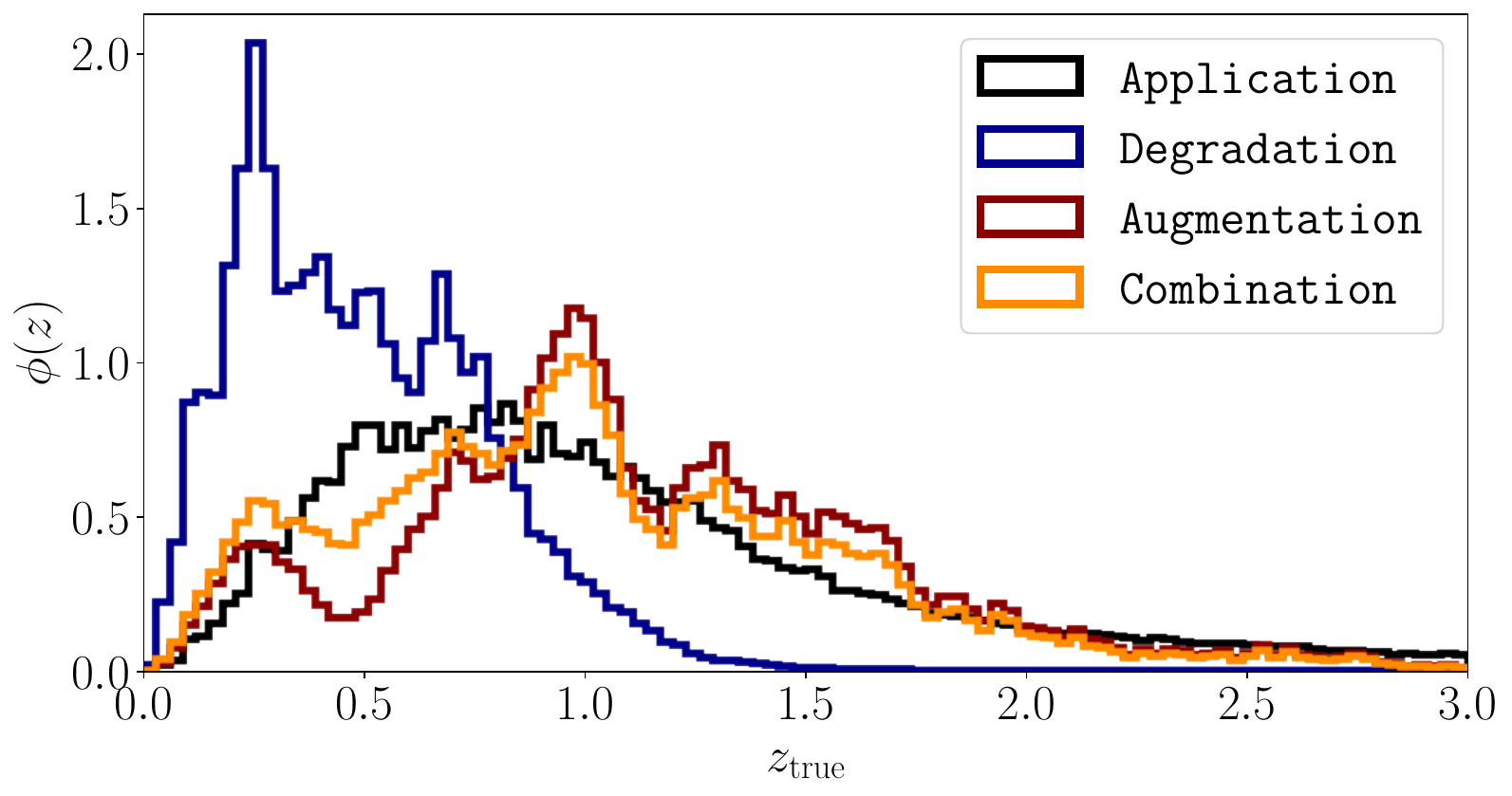}
        \caption{LSST Y10}
    \end{subfigure}
    \caption{Normalised ensemble distributions of galaxy true redshifts for various datasets are displayed for for Y1 (left panel) and Y10 (right panel), under the baseline experiment. In each panel, the black, dark-blue, dark-red, dark-orange correspond to the \texttt{Application}, \texttt{Degradation}, \texttt{Augmentation}, and \texttt{Combination} datasets, respectively.}
    \label{fig:6}
\end{figure*}

To mitigate incomplete sampling within the SOM space of the \texttt{Degradation} datasets, we build upon the matched \texttt{CosmoDC2} galaxies. First, we identify SOM cells unoccupied in the \texttt{Degradation} datasets and repopulate them using corresponding galaxies from the matched \texttt{CosmoDC2} pool. Moreover, we identify galaxies that surpass the maximum redshift and magnitude limits of the \texttt{Degradation} datasets, integrating them into the \texttt{Augmentation} datasets as well, to ensure that previously excluded populations are considered.

To adaptively regulate the number of augmented galaxies, we compute \(f_\ast\), defined as the fraction of \texttt{Application} galaxies that fall into SOM cells absent from the \texttt{Degradation} dataset. We then set the target \texttt{Augmentation}-to-\texttt{Degradation} ratio to \(f_\ast \left(1  + f_\ast\right)\), thereby smoothly scaling the size of the augmented sample to balance the two datasets. In the limit of minimal spectroscopic incompleteness \(f_\ast \ll 1\), this prescription closely approximates the ideal augmentation ratio \(\frac{f_\ast}{1 - f_\ast}\). Conversely, under the most extreme degradation scenarios where \(f_\ast \to 1\), this ratio is restricted to a maximum of \(2\). This upper bound prevents the augmented sample from becoming orders of magnitude larger than its spectroscopic counterpart, thereby avoiding situations where the photometric-redshift training would be dominated by synthetic galaxies. We find that increasing the ratio beyond this cap only adds redundant augmented galaxies, raising computational costs without altering the learned colour–redshift relations or improving photometric-redshift performance.

The distribution of galaxies within the \texttt{Augmentation} datasets is also projected onto pre-trained SOM cells, as depicted in the lower left subplots of each panel in Figure~\ref{fig:3}\({-}\)\ref{fig:5}, demonstrating how data augmentation compensates for missing information within the SOM space, colour-colour diagrams, colour-magnitude diagram, respectively. Further evidence of the effectiveness of augmentation is visible in the increased presence of galaxies at \(z_\mathrm{true} \gtrsim 1.5\) (highlighted in dark-red solid lines) in Figure~\ref{fig:6}. 

As shown in Figure~\ref{fig:1}, the colour distributions of \texttt{CosmoDC2} and \texttt{OpenUniverse2024} differ markedly in \(u - g\) and \(g - r\). In contrast, Figures~\ref{fig:4} reveals that the \texttt{Augmentation} datasets exhibit substantially improved agreement with the \texttt{Application} datasets, owing to SOM‐based weighted resampling. By matching \texttt{CosmoDC2} galaxies to occupied SOM cells, we correct misalignments introduced by the distinct SED models, particularly at the bright, low-redshift end where the two catalogues overlap.

For faint, high-redshift galaxies, the differences between the SED models become more pronounced, leading to non-overlapping colour distributions, as previously shown in Figure~\ref{fig:1}. Consequently, certain galaxy populations in the \texttt{Application} datasets remain unrepresented in the \texttt{Augmentation} datasets despite weighted resampling. For example, in the LSST Y10 scenario of Figure~\ref{fig:4}, an elongated locus of galaxies with \(g - r \approx -0.5\) and \(0.0 \lesssim u - g \lesssim 3.0\) is present in the \texttt{Application} dataset but absent in the \texttt{Augmentation} dataset. Such persistent gaps can bias inferred galaxy properties and introduce systematic errors into subsequent analyses.  

\subsection{\texttt{Combination} datasets}

By merging the \texttt{Degradation} and \texttt{Augmentation} datasets, the resulting \texttt{Combination} datasets offer enhanced representativeness of galaxy broadband broadband SEDs compared to the photometric \texttt{Application} datasets, as opposed to the unaugmented \texttt{Degradation} datasets. The lower right subplots in each panel of Figure~\ref{fig:3} display the mean true redshifts of SOM cells occupied by the \texttt{Combination} datasets, demonstrating significantly enhanced coverage of the SOM cells compared to the \texttt{Degradation} datasets alone. 

Similarly, in the colour-colour and colour-magnitude diagrams presented in the lower right subplots of each panel in Figure~\ref{fig:4}-~\ref{fig:5}, combining the sparse \texttt{Degradation} datasets, which have notable deficits at fainter magnitudes and specific colour regimes, with the targeted \texttt{Augmentation} datasets results in \texttt{Combination} datasets that more effectively span the parameter space. The resultant distributions closely resemble the overall \texttt{Application} datasets, reducing incompleteness. 

Figure~\ref{fig:6} further demonstrates these gains. The under‐densities at intermediate redshifts evident in the \texttt{Degradation} samples are filled by the extended tails of the \texttt{Augmentation} distributions, yielding a well‐balanced redshift histogram in the \texttt{Combination} datasets. Here, the augmented datasets (dark-orange solid lines) more faithfully reproduce the full \texttt{Application} redshift distributions (black solid lines). 

Compared to the original implementation of data augmentation in \citet{2024ApJ...967L...6M}, which employed simplified assumptions regarding the boundaries of \(g - z\) colour, \(i-\)band magnitude and redshift, our SOM-based approach can accommodate far more complex selection functions in the multi-dimensional colour–magnitude space. 

Figure~\ref{fig:4} demonstrates the robustness of our SOM-based augmentation in colour–colour diagrams, where no regular boundary geometry is discernible. By visualising the multi-dimensional colour–magnitude space via the SOM, we can readily identify sparsely sampled regions and apply targeted data augmentation.

Similarly, as illustrated in the left panel of Figure~\ref{fig:5}, the colour–magnitude boundary of the \texttt{Degradation} dataset no longer conforms to a simple linear relation, owing to the amalgamation of multiple mock spectroscopic surveys. Nonetheless, our SOM-based augmentation adaptively compensates for gaps in spectroscopic coverage, rendering it well suited to future analyses involving complex selection functions.

However, this compensation is not entirely flawless: a noticeable deficit of galaxies around the boundaries remains. This phenomenon arises from our current definition of the \texttt{Augmentation} dataset, which augments only wholly empty SOM cells, whereas sparsely occupied cells may also require supplementary galaxies. For truly optimal performance, a comprehensive investigation of the ideal threshold for SOM cell occupancy -- potentially with adaptive thresholds for each \texttt{Degradation} dataset -- is necessary, and constitutes an avenue for future study.  

Furthermore, some SOM cells remain unpopulated or sparsely occupied even after augmentation, reflecting persistent SED model discrepancies between \texttt{CosmoDC2} and \texttt{OpenUniverse2024}. These gaps manifest most clearly as an underestimation of the high-redshift tail (\(z_\mathrm{true} \gtrsim 2.0\)), particularly in the Y10 scenario, indicating amplified mismatches for faint, distant galaxies. Moreover, variability in the spectroscopic selection functions applied during degradation translates into differing augmentation requirements across realisations, introducing an additional systematic uncertainty that we exploit to assess the impact of incomplete coverage and selection‐induced biases on photometric-redshift estimation.

\section{Photometric Redshift Estimation} \label{sec:4}
Thanks to data augmentation, the spectroscopically informed datasets are highly representative and ideally suited for estimating photometric redshifts using machine learning, thus negating the need for predefined galaxy SED templates. However, future research into template-driven and hybrid methods would be valuable, despite being outside the scope of this study. 

In this section, we first introduce the \texttt{FlexZBoost} photometric redshift estimation method, and describe the definitions different galaxy sub-samples based on the point estimates of photometric redshifts. Although we use this specification here to characterise our photometric‐redshift quality, the SOM‐based augmentation method and the resulting datasets are applicable to many other science cases. 

\subsection{Machine learning modelling}
\begin{figure*}
    \centering
    \begin{subfigure}[b]{0.48\linewidth}
        \centering
        \includegraphics[width=\linewidth]{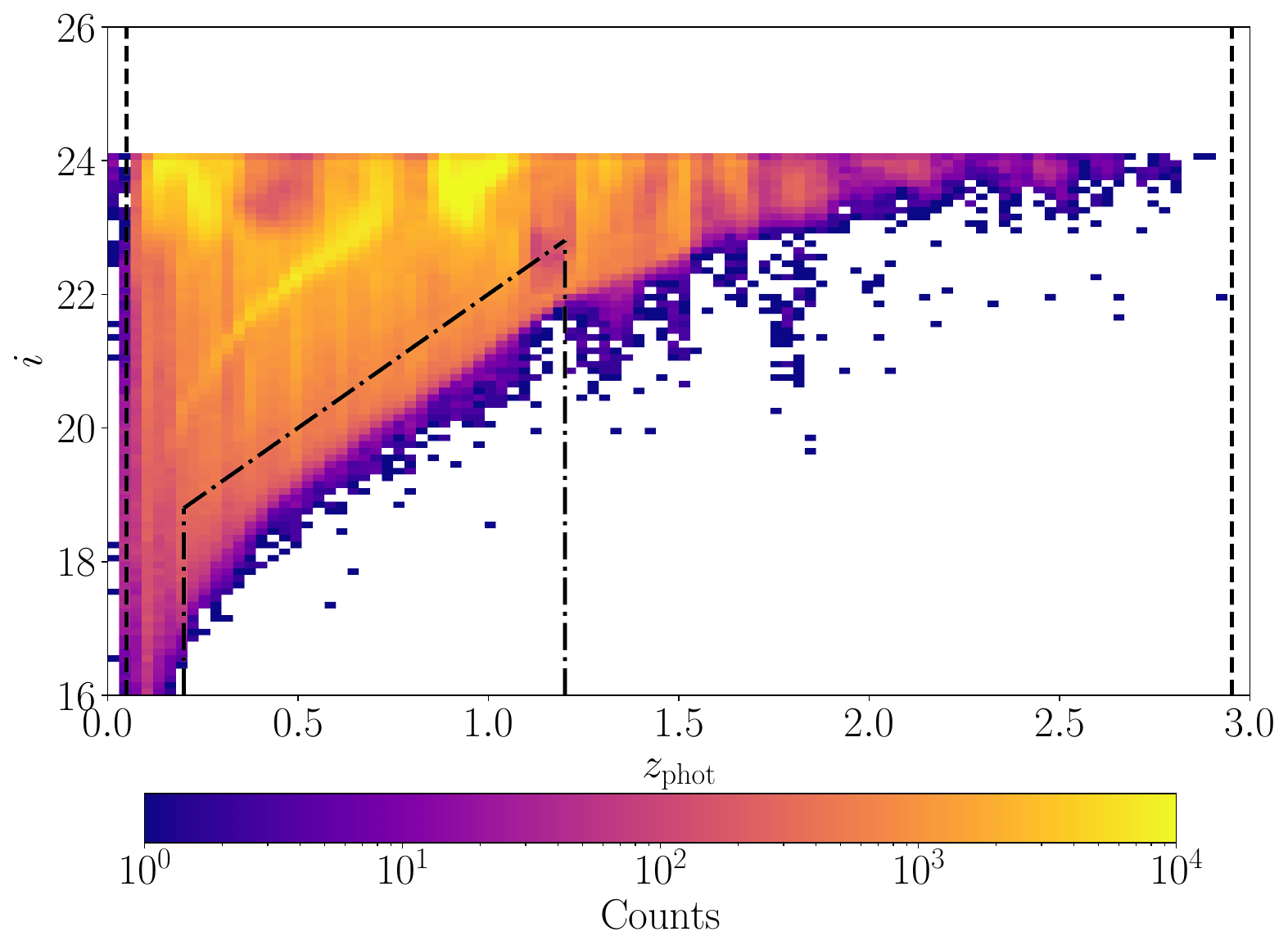}
        \caption{LSST Y1}
    \end{subfigure}
    \hspace{0.0\linewidth}
    \begin{subfigure}[b]{0.48\linewidth}
        \centering
        \includegraphics[width=\linewidth]{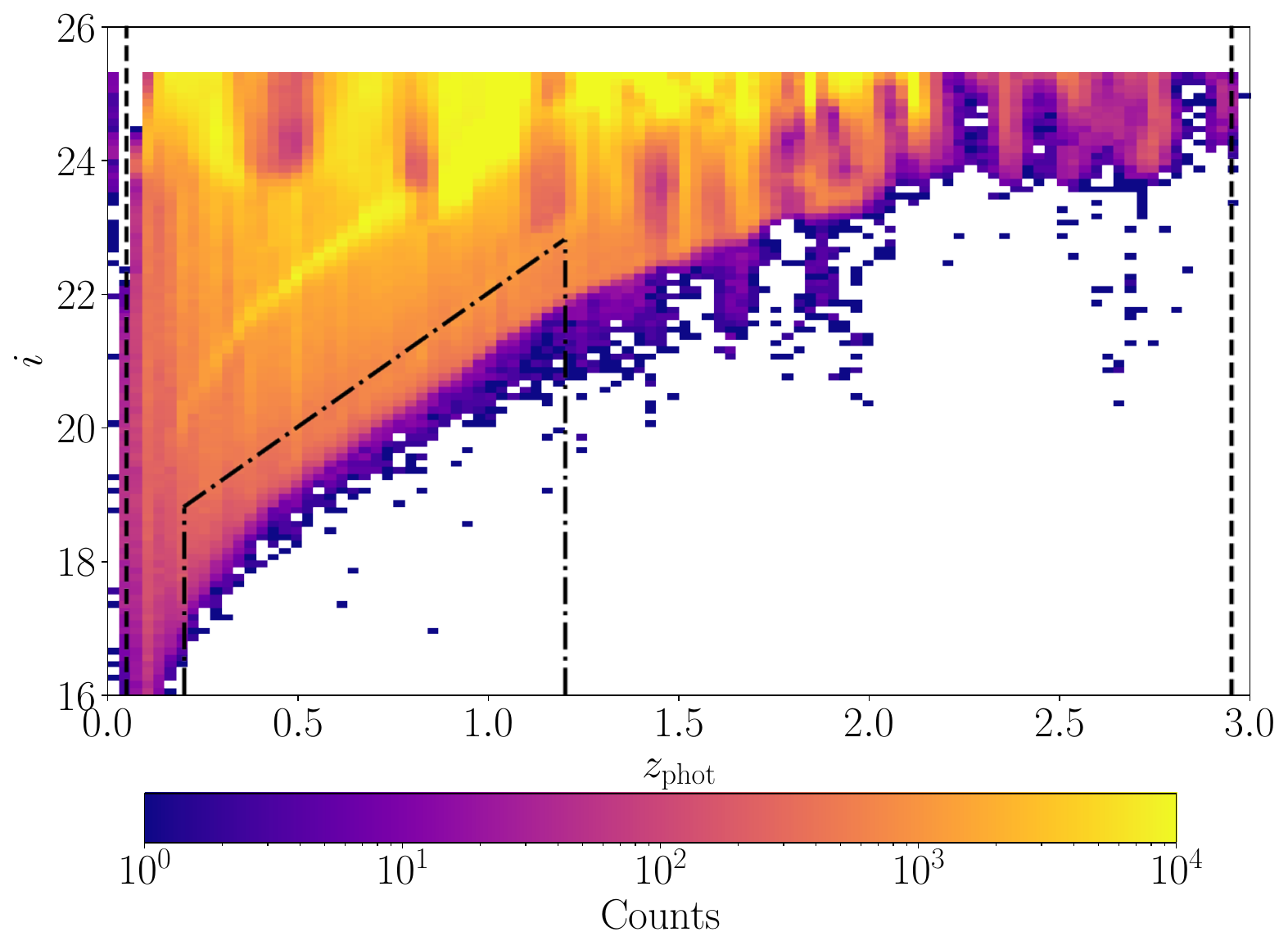}
        \caption{LSST Y10}
    \end{subfigure}
    \caption{Colour maps depicting the galaxy distribution, with the \(i\)-band magnitude on the y-axis and photometric redshift estimates, \(z_{\mathrm{phot}}\), on the x-axis, for LSST Y1 (left) and Y10 (right) under the baseline experiment. Black dashed lines indicate the selection range \(0.05 \le z_{\mathrm{phot}} \le 2.95\) used to identify source galaxies, while black dot-dashed lines represent the redshift-dependent magnitude limits imposed on lens galaxies \citep{2021PhRvD.103d3503P}.}
    \label{fig:7}
\end{figure*}

Based on a detailed comparative analysis with established methods \citep{2020MNRAS.499.1587S}, \texttt{FlexZBoost} \citep{10.1214/17-EJS1302, 2020A&C....3000362D} has demonstrated superior performance in estimating both the conditional probability density function (PDF) of photometric redshifts and the corresponding point estimates from galaxy colours. We therefore employ \texttt{FlexZBoost} in our proof‐of‐concept study to validate the efficacy of our SOM‐based data augmentation, deferring evaluation of alternative photometric‐redshift estimators to future work.  

\texttt{FlexZBoost} is built on the \texttt{Flexcode} framework, which models the conditional density \(p(z \mid \mathcal{T})\) for a given galaxy training dataset \(\mathcal{T}\) by projecting it onto an orthogonal basis of functions \(O_{k}(z)\) \(p(z \mid \mathcal{T}) \equiv \sum_{k} q_{k}(\mathcal{T}) \, O_{k}(z)\). Broadband photometry is therefore used to infer the expansion coefficients \(q_{k} (\mathcal{T}) = \int p(z \mid \mathcal{T}) \, O_{k}(z) \, \mathrm{d}z = \bigl\langle O_{k}(z) \mid \mathcal{T}\bigr\rangle\), which represent the conditional expectations of the basis functions. By employing regression in a high‐dimensional feature space, \texttt{FlexZBoost} effectively captures complex, non‐linear mappings between galaxy colours and redshift.  

Utilising the \texttt{FlexZBoost} implementation available on the \texttt{RAIL} platform, we develop our photometric redshift model using the \texttt{Combination} dataset, which integrates "observed" spectroscopic-like \texttt{Degradation} data with "simulated" \texttt{Augmentation} samples. This dataset is divided into training and validation sets in an \(80\% \,{:}\, 20\%\) proportion to perform regression on the hyperparameters of \texttt{FlexZBoost}, including the bump optimisation and sharpness parameters. The \(i\)–band apparent magnitude serves as the reference baseline against which all multi-band colours are defined.

We approximate the conditional PDF using a Fourier basis of up to \(50\) orthogonal functions uniformly covering \(0.0 \le z \le 3.0\) in steps of \(0.01\). Bump optimisation parameters are sampled over \([0.0, 0.5]\) on a \(50‐\)point grid, and sharpness parameters over \([0.5, 2.5]\) with the same grid resolution. For regression, we employ \texttt{XGBoost} \citep{2016arXiv160302754C}, minimising the Mean Squared Error (MSE) with a maximum tree depth of \(8\) and a learning rate of \(0.1\) -- choices that balance model expressivity with computational efficiency. Following hyperparameter tuning, the model is retrained on the complete \texttt{Combination} dataset. This rigorous framework provides both flexibility and robustness, effectively capturing the intrinsic complexity of galaxy populations while mitigating the risk of overfitting.

After training the \texttt{FlexZBoost} models using \(501\) experiments from the \texttt{Combination} dataset, each optimised model is applied to its respective experiment within the photometric \texttt{Application} datasets. This process generates the conditional density \(p \left( z | \mathcal{S}_n, \mathcal{P}_n \right)\) for each galaxy, with \(\mathcal{S}_n\) and \(\mathcal{P}_n\) representing the spectroscopic \texttt{Combination} and photometric \texttt{Application} datasets, respectively, for the \(n\mathrm{th}\) experiment. These conditions take into account variations in photometric noise, spatial coverage, and spectroscopic selection functions. We smoothly interpolate the conditional density over the interval \([0.0, 3.0]\) with redshift increments of \(0.01\), determining the point estimates \(z_{\mathrm{phot}} \) as the redshift that maximises the conditional PDF for each galaxy. 

\subsection{Definition of source and lens samples}
Based on point estimates, we first identify galaxies therefore serving as the background source galaxies that are gravitationally lensed by intervening mass distributions in the Universe \citep{2021A&A...645A.104A, 2022PhRvD.105b3514A, 2022PhRvD.105b3515S, 2023A&A...679A.133L, 2023PhRvD.108l3519D, 2023PhRvD.108l3518L, 2023OJAp....6E..36D, 2025arXiv250319442S, 2025arXiv250319441W}, therefore targeting galaxies with reliable measurements on weak gravitational lensing signal. Initially, we estimate the shear measurement noise, \(\sigma_\gamma\), for individual galaxies using the expression provided in \citep{2002AJ....123..583B}, written as 
\begin{equation} \label{eqa:3}
    \sigma_\gamma = \frac{a}{\mu} \left[ 1 + \left( \frac{b}{\eta} \right)^c \right].
\end{equation} The coefficients in the equation are \(\left(a,\, b,\, c\right) = \left(1.58,\, 5.03,\, 0.39\right)\) as defined by \cite{2013MNRAS.434.2121C}. In this calculation, as we described in Section~\ref{sec:2}, the variable \(\mu\) represents the \(\mathrm{SNR}\) of coadded \(r-\) and \(i-\)band photometry, while \(\eta\) signifies the squared ratio of the galaxy sizes of galaxies comparing to those of PSFs. Subsequently, we implement a selection criterion of \(\sigma_\gamma < \sigma_\epsilon\), where \(\sigma_\epsilon\) represents the intrinsic shape noise, presumed to be \(0.26\). 

On top of the constraints on the shear measurement noise \(\sigma_\gamma\), we further restrict our sample to galaxies with photometric redshifts in the range \(0.05 < z_{\mathrm{phot}} < 2.95\). This excludes objects at the boundaries of the redshift PDF, where asymmetries can lead to significant errors, and minimises contamination from higher-redshift (\(z > 3.0\)) galaxies, which contribute little to the lensing signal but complicate redshift calibration in a real analysis.  

Additionally, we define another foreground galaxy subset, referred to as lens samples, selected via precise redshift estimations to trace high‐mass haloes for large‐scale structure studies\citep{2021A&A...646A.140H, 2022PhRvD.105b3520A, 2023A&A...675A.189D, 2023PhRvD.108l3521S, 2023PhRvD.108l3517M}. We adopt the redshift‐dependent magnitude cut of \citet{2021PhRvD.103d3503P}, \( i < 4 \,\cdot\, z_{\mathrm{phot}} + 18 \), valid for \(0.2 < z_{\mathrm{phot}} < 1.2\). For both LSST Y1 and Y10, we apply the same criterion in accordance with the stringent photo-\(z\) precision requirements of the LSST DESC SRD. Future work should explore the inclusion of fainter galaxies in the Y10 lens samples -- while increased number density may enhance statistical power, it could also degrade redshift accuracy -- necessitating further investigation to optimise selection strategies \citep{2023ApJ...950...49M}.  

Figure~\ref{fig:7} shows the distribution of galaxies in \(i\)-band magnitude versus photometric redshift \(z_{\mathrm{phot}}\). Black dashed lines denote the source sample selection boundaries, which remove objects near analysis limits to prevent bias in the conditional density estimates, while black dot–dashed lines indicate the lens sample selection, isolating brighter galaxies to secure high photo-\(z\) precision. The distribution becomes increasingly disordered at faint magnitudes, highlighting the challenge of moving from training on \texttt{Degradation} datasets derived from \texttt{OpenUniverse2024} to using \texttt{Augmentation} datasets drawn from \texttt{CosmoDC2}. In the Y1 case, the spectroscopic selection limit, especially for \(20  \lesssim i_\ast \lesssim 22 \), noticeably biases redshift estimates at \(z_{\mathrm{phot}} \gtrsim 1.0\). By contrast, the deeper spectroscopic coverage in Y10 mitigates this bias, underscoring the importance of consistent lens sample definitions across survey epochs.  

\section{Results and Discussion} \label{sec:5}
In this section, we introduce the statistical metrics used to characterise the precision of our photometric redshifts. We then assess the accuracy of the point estimates and evaluate the performance of the conditional‐density modelling, thereby validating our data‐augmentation technique via comparative analysis.

\subsection{Statistical metrics}
We quantify the systematic error of each photometric‐redshift estimate \(z_\mathrm{phot}\) relative to its true redshift \(z_\mathrm{true}\) by defining the scaled bias
\begin{equation} \label{eqa:4}
    \delta_z = \frac{z_\mathrm{phot} - z_\mathrm{true}}{1 + z_\mathrm{true}}.
\end{equation}
From the distribution of \(\delta_z\), we derive two robust metrics for any galaxy subset. First, the median scaled bias, \(\bar{\delta}_z \equiv \mathrm{median} \left( \delta_z \right)\), is adopted in preference over the mean to mitigate sensitivity to outliers. Second, the Normalised Median Absolute Deviation (NMAD) of \(\delta_z\), denoted \(\sigma_z\), is defined as
\begin{equation} \label{eqa:5}
    \sigma_z = 1.4826 \times \mathrm{median} \left( \left| \delta_z - \bar{\delta}_z \right| \right).
\end{equation} 
The factor \(1.4826\) rescales the median absolute deviation into an estimator equivalent to the standard deviation for a Gaussian distribution.

Furthermore, we consider the proportion at which galaxies fall outside the outlier boundary \(\left| \delta_z \right| = 0.15\), as defined in previous studies \citep[e.g.][]{2012ApJS..198....1F, 2014ApJS..215...27Y, 2017A&A...605A..70D, 2017A&A...608A...3B, 2019A&A...622A...3D}, termed outlier fraction, 
\begin{equation} \label{eqa:6}
    f_{\mathrm{o}} = \frac{N_\mathrm{g} \left( \left| \delta_z \right| > 0.15 \right)}{N_\mathrm{g}},
\end{equation} 
where \(N_{\mathrm{g}}\) represents the total count of a specific galaxy subset. In addition, we incorporate the rates of catastrophic failures where \(\left| z_{\mathrm{phot}} - z_{\mathrm{true}} \right| > 1.0\), as outlined in \cite{2024ApJ...967L...6M}, denoted as
\begin{equation} \label{eqa:7}
    r_{\mathrm{c}} = \frac{N_\mathrm{g} \left( \left| z_{\mathrm{phot}} - z_{\mathrm{true}} \right| > 1.0 \right)}{N_\mathrm{g}}.
\end{equation}
These metrics evaluate the prevalence of notable systematic errors in the photometric-redshift determinations. In addition to point estimates, we evaluate the performance of the photometric‐redshift conditional density modelling, \(p \left(z | \mathcal{S}_n, \mathcal{P}_n \right) \). For simplicity, the explicit dependence on the \(n\)th spectroscopic dataset \(\mathcal{S}_n\) and photometric dataset \(\mathcal{P}_n\) will henceforth be omitted. We employ the cumulative distribution function,
\begin{equation} \label{eqa:8}
    q \left(z\right) \equiv \int_{0}^{z} p \left(z^\prime\right) \,\mathrm{d}z^\prime,
\end{equation}
to define summary statistics owing to its well‐behaved, bounded range. The Probability‐Integral-Transform (PIT) value -- defined as the CDF evaluated at the true redshift \(z_{\mathrm{true}}\), \(q \left( z_{\mathrm{true}} \right)\) -- has been widely used in the literature \citep[e.g.][]{2012MNRAS.421.1671B, 2017MNRAS.468.4556F, 2018PASJ...70S...9T, 2020MNRAS.499.1587S}. For an ideal conditional density model, the PIT values for any galaxy subset should follow a uniform distribution; deviations from uniformity therefore quantify information loss during the modelling procedure.

The normalised PIT histograms, \( \mathcal{H}_k \left[ q \left( z_\mathrm{true} \right) \right] \) for \(k = 1, \ldots, K\), are calculated across \(K = 10\) equally sized quantile bins, ranging from \(\left[0.0, 1.0\right]\), with intervals of \(0.1\). We calculate these histograms for both the full lens and source samples as well as for each redshift interval defined above. Thus, for each normalised PIT histogram, we then define the divergence loss, \(\mathcal{D}_q\), as the Root-Mean-Square deviation of the normalised PIT histogram from uniformity:
\begin{equation} \label{eqa:9}
    \mathcal{D}_q = \sqrt{\frac{1}{K} \sum_{k = 1}^{K} \left\{ \mathcal{H}_k \left[ q \left( z_\mathrm{true} \right) \right] - 1 \right\}^2},
\end{equation}
where unity denotes the value of the normalised histogram for an ideal uniform distribution. This metric provides a rigorous assessment of the conditional‐density modelling quality.

\begin{figure*}
    \centering
    \begin{subfigure}[b]{0.48\linewidth}
        \centering
        \includegraphics[width=\linewidth]{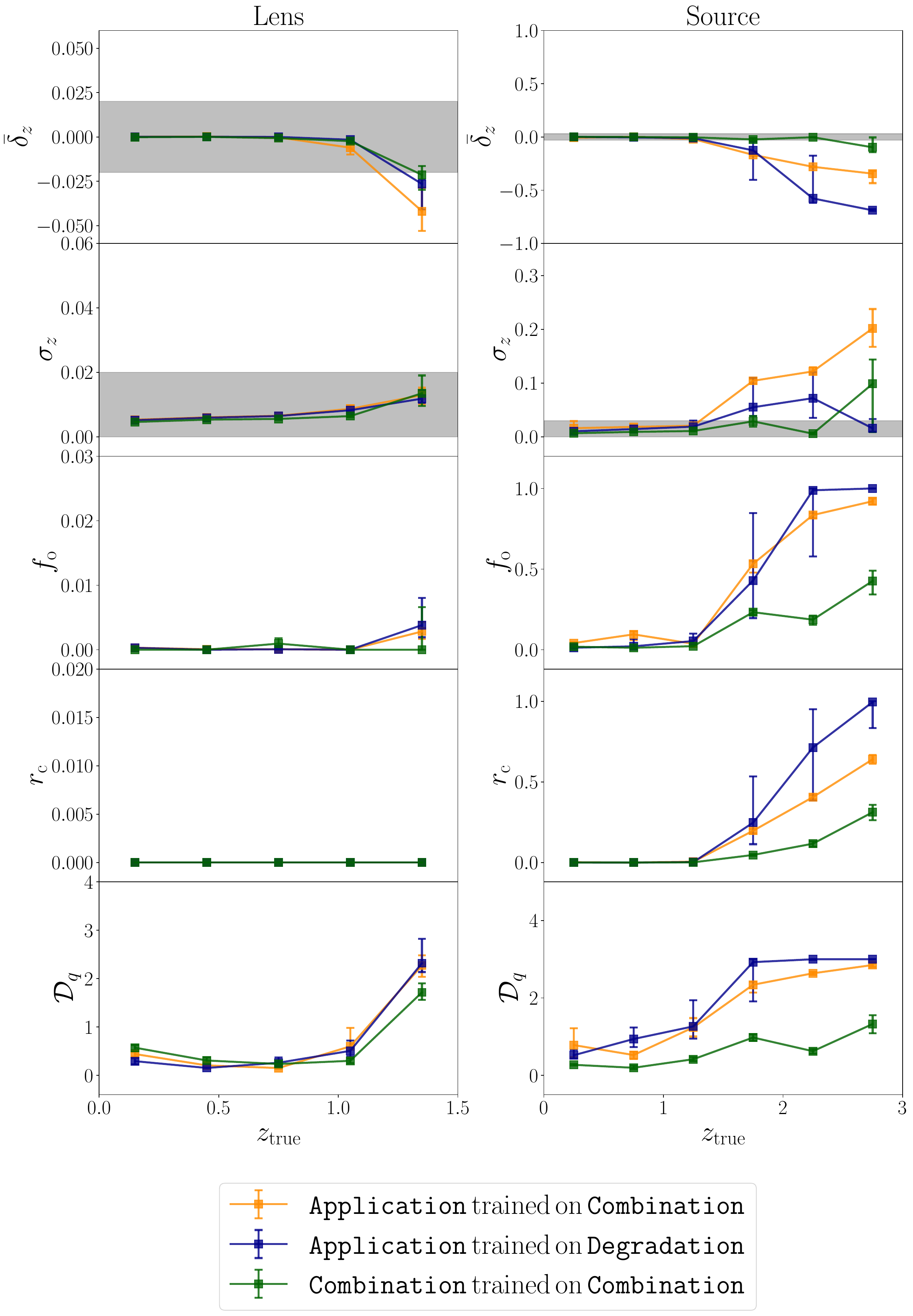}
        \caption{LSST Y1}
    \end{subfigure}
    \hspace{0.0\linewidth}
    \begin{subfigure}[b]{0.48\linewidth}
        \centering
        \includegraphics[width=\linewidth]{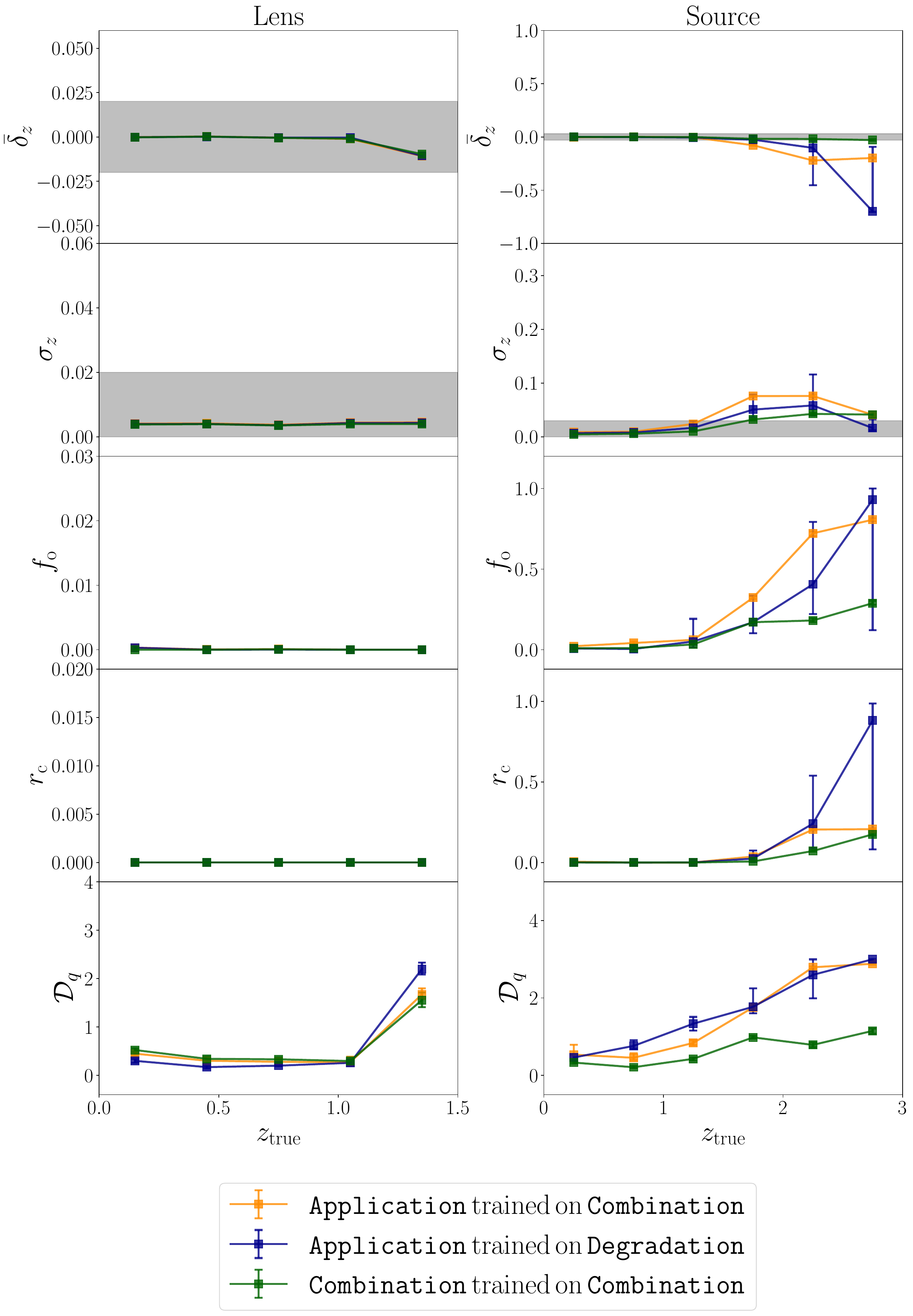}
        \caption{LSST Y10}
    \end{subfigure}
    \caption{Statistical metrics plotted as a function of true redshift, with LSST Y1 in the left panel and LSST Y10 in the right. Within each panel, the metrics are ordered from top to bottom: the median scaled bias \(\bar{\delta}_z\); the normalised median absolute deviation (NMAD) of the scaled bias \(\sigma_z\); the outlier fraction \(f_{\mathrm{o}}\); the catastrophic failure rate \(r_{\mathrm{c}}\); and the divergence loss of the normalised PIT histogram \(\mathcal{D}_q\). In each row, the left and right subplots correspond to the lens and source samples, respectively. Filled markers denote the median value of each metric across all experiments, with error bars indicating the central \(68\%\) confidence interval. Green curves show results from self-training on fully representative \texttt{Combination} datasets (ideal benchmarks); orange curves correspond to \texttt{Application} datasets trained on these augmented \texttt{Combination} datasets; and blue curves represent \texttt{Application} datasets trained on the unaugmented \texttt{Degradation} datasets. The grey bands indicate the range defined for LSST DESC SRD.}
    \label{fig:8}
\end{figure*}

For the lens sample, each metric is assessed across five redshift intervals, each with a width of \(0.3\), covering the range \(0.0 < z_\mathrm{true} < 1.5\). Similarly, for the source sample, five intervals of equal width are used, encompassing \(0.0 < z_\mathrm{true} < 3.0\). The outcomes of our statistical metrics are depicted as a function of true redshift in Figure~\ref{fig:8}, with comprehensive discussions provided thereafter.

It is important to note that this interval binning is solely used for assessing the performance of the photometric redshifts, rather than for tomographic binning in cosmological analysis. The latter is based on the derived point estimates and will be addressed in our subsequent paper.

\subsection{Evaluation of point estimate accuracy}
\begin{figure*}
    \centering
    \begin{subfigure}[b]{0.48\linewidth}
        \centering
        \includegraphics[width=\linewidth]{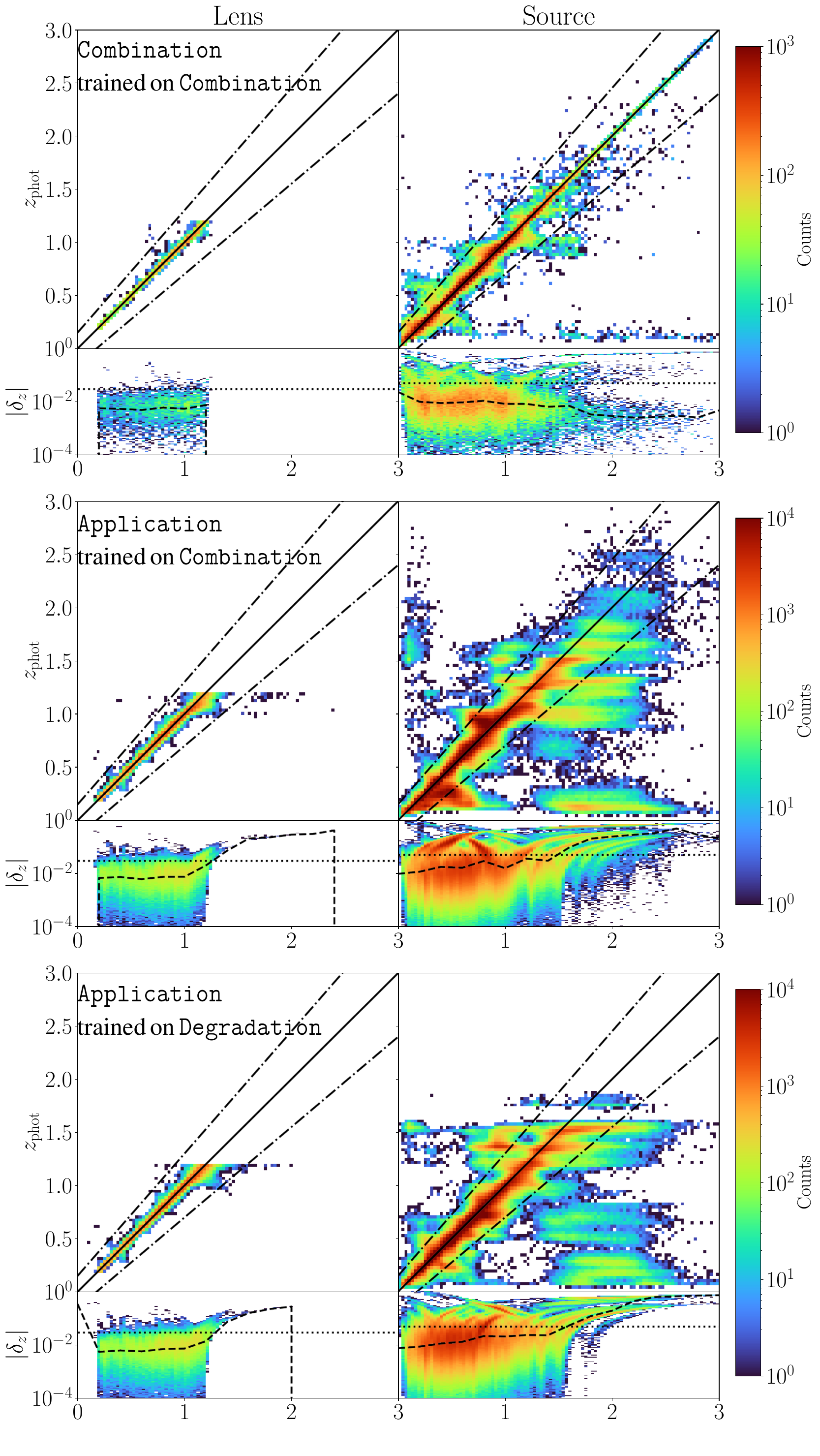}
        \caption{LSST Y1}
    \end{subfigure}
    \hspace{0.0\linewidth}
    \begin{subfigure}[b]{0.48\linewidth}
        \centering
        \includegraphics[width=\linewidth]{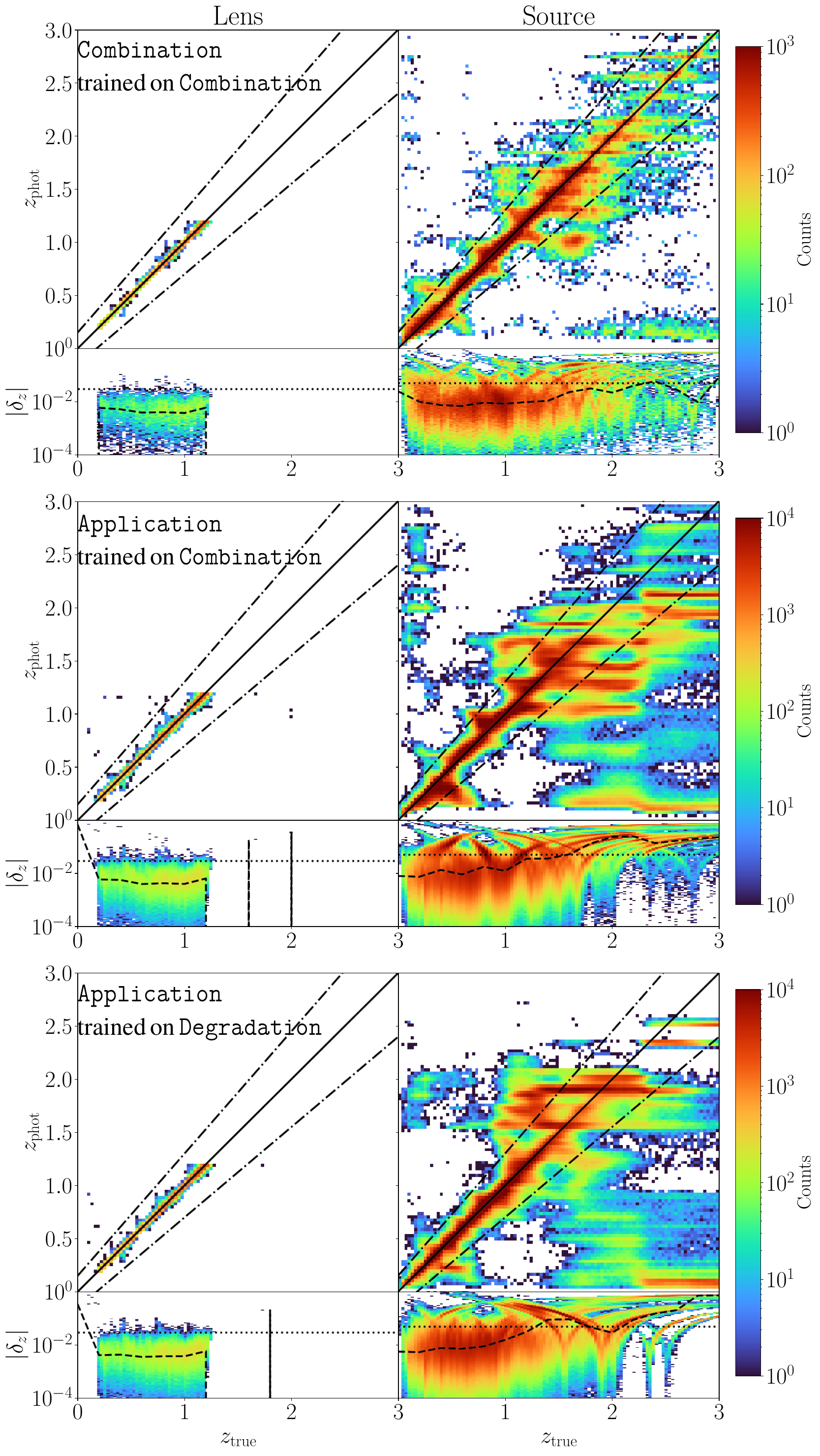}
        \caption{LSST Y10}
    \end{subfigure}
    \caption{Direct visual comparison of \(z_{\mathrm{phot}}\) versus \(z_{\mathrm{true}}\) for the baseline experiment: LSST Y1 (left) and Y10 (right). From top to bottom, each row corresponds to (i) the self-trained \texttt{Combination} dataset utilised as a fully representative  benchmark (ii) the \texttt{Application} dataset trained on the augmented \texttt{Combination} dataset; and (iii) the \texttt{Application} dataset trained the \texttt{Degradation} dataset without data augmentation. For every row, the upper panels show two-dimensional histograms for the lens (left) and source (right) samples, where the colour scale represents galaxy counts. The solid black line marks the ideal one‐to‐one relation \(z_{\mathrm{phot}} = z_{\mathrm{true}}\), and the dot‐dashed lines indicate the outlier threshold \(\delta_z = 0.15\). The lower panels show the scaled bias \(\delta_z\) as a function of \(z_{\mathrm{true}}\) for lens (left) and source (right) galaxies. The dashed black line denotes the median scaled bias \(\bar{\delta}_z\), while the dotted lines represent the LSST DESC SRD accuracy requirements of \(0.02\) for the lens sample and \(0.03\) for the source sample.}
    \label{fig:9}
\end{figure*}

Figure~\ref{fig:9} provides a direct visual assessment of the performance of our photometric redshift model under the baseline experiment. As depicted in the top panels of Figure~\ref{fig:9}, the self-training on the \texttt{Combination} dataset shows excellent agreement between \(z_{\mathrm{phot}}\) and \(z_{\mathrm{true}}\), indicating that our photometric redshift model reliably captures the colour–redshift relation. By selecting the peak of the conditional PDF, we achieve negligible bias for both the bright lens sample (left panels) and the fainter source sample (right panels), even at \(z_\mathrm{true} \gtrsim 1.5\). The resulting median absolute values of the scaled biases, \(\bar{\left| \delta \right|}_z\) consistently meet the LSST DESC SRD requirements for almost the entire redshift range. Hence, with a fully representative spectroscopic training set, \texttt{FlexZBoost} can provide photometric‐redshift estimates of outstanding accuracy.

Upon utilising the complete photometric \texttt{Application} datasets, the central panels of Figure~\ref{fig:9} demonstrate that the lens galaxies maintain a similar level of performance. In the bottom-left subplots for both the Y1 and Y10 \texttt{Application} lens samples, the median absolute values of the scaled biases, \(\bar{\left| \delta \right|}_z\), highlighted by the black dashed lines, are notably beneath the thresholds set by the LSST DESC SRD, depicted by black dotted lines. This signifies an exceptionally accurate model for this luminous population. Furthermore, practically no outliers or catastrophic failures occur in either instance, reinforcing the robustness of our photometric-redshift point estimates for bright galaxies and thus supporting precision cosmology through large-scale-structure analyses. In contrast, for the source samples, \(\bar{\left| \delta \right|}_z\) remains close to the SRD requirement up to \(z_\mathrm{true} \lesssim 1.5\). Nonetheless, a considerable residual bias becomes evident at \(z_\mathrm{true} \gtrsim 1.5\). 

Similar trends are evident in Figure~\ref{fig:8}, where the green and orange curves both correspond to models trained on the \texttt{Combination} datasets but evaluated on the \texttt{Combination} and \texttt{Application} datasets, respectively. This comparison shows that, for the lens sample, both the median scaled bias \(\bar{\delta}_z\) and its NMAD \(\sigma_z\) remain within the grey band defined by the LSST DESC SRD in all cases. For the source sample, however, we observe a significant increase in \(\bar{\delta}_z\), accompanied by a rise in its NMAD, as well as marked increases in the outlier fraction \(f_{\mathrm{o}}\), and the catastrophic failure rate \(r_{\mathrm{c}}\).

This suggests that the machine‐learning model struggles to extrapolate beyond its training domain in colour–SED space. Contributing factors include photometric noise -- which perturbs the mapping between observed SEDs and redshift, particularly for faint galaxies, thereby introducing uncertainty into \(z_{\rm phot}\) -- and systematic mismatches in SED properties. Specifically, the redshift upper limit of the \texttt{Degradation} dataset compels the model to rely on subtly discrepant galaxies in the augmented training dataset -- enhanced with high-redshift \texttt{CosmoDC2} galaxies -- compared to those in the photometric \texttt{Application} dataset from \texttt{OpenUniverse2024}. As discussed in Section~\ref{sec:3}, this SED discrepancy constitutes a primary source of bias in our photometric-redshift modelling.

The bottom two panels of Figure~\ref{fig:9} provides a qualitative comparison of photometric–redshift estimates trained exclusively on the \texttt{Degradation} datasets without augmentation. For the lens samples, incorporating the \texttt{Augmentation} datasets yields only marginal improvements: the median absolute scaled biases remain effectively unchanged whether or not augmentation is applied. Nonetheless, for the source sample in the Y1 scenario, training on the unaugmented \texttt{Degradation} dataset results in virtually no galaxies with \(z_{\mathrm{phot}} \gtrsim 1.5\). A quantitative comparison of the orange and blue curves in Figure~\ref{fig:8} reinforces these conclusions. For the lens sample, these curves coincide nearly perfectly across all statistical metrics in both the LSST Y1 and Y10 cases. This reflects that the brightest galaxies are already well represented in the spectroscopic training sets. 

Conversely, not employing data augmentation for the source sample results in significantly high outlier fractions (\(f_{\mathrm{o}} \gtrsim 0.5\)) and catastrophic failure rates (\(r_{\mathrm{c}} \gtrsim 0.5\)) at \(z_{\mathrm{true}} \gtrsim 2.0\) in the LSST Y1 scenario. Introducing SOM‐based data augmentation reduces the outlier fraction by approximately \(10 \,{-}\, 20\%\), lowers the catastrophic failure rate by a factor of \(\sim 2\), and decreases the median scaled bias by a factor of \(\sim 2 \,{-}\, 3\), thereby demonstrating the efficacy of augmentation in mitigating both bias and outliers in LSST Y1 photometric redshift point estimates. Although the NMAD of the scaled bias \(\sigma_z\) increases after data augmentation, this does not indicate a degradation in photometric performance. On the contrary, the low NMAD observed without augmentation is driven by a high fraction of catastrophic failures, which cluster galaxies around large bias values as seen in Figure~\ref{fig:9}. Consequently, the rise in NMAD post‐augmentation reflects the suppression of these extreme biases. Furthermore, the pronounced strip-like features evident in Figure~\ref{fig:9} after augmentation reflect intrinsic degeneracies in galaxy SEDs: in the absence of representative high-redshift training samples, such galaxies would otherwise be classified as catastrophic failures and, consequently, vanish from the regime around one-to-one sequence.

In the LSST Y10 scenario, the \texttt{Degradation} datasets comprise deeper spectroscopic observations than in Y1, thereby yielding improved photometric redshift performance for source galaxies even without augmentation. Nonetheless, training with the \texttt{Augmentation} datasets still confers benefits: both the median scaled bias \(\bar{\delta}_z\) and the catastrophic failure rate \(r_{\mathrm{c}}\) decrease by up to a factor of two in the highest-redshift interval (\(2.5 < z_{\mathrm{true}} < 3.0\)). However, in the Y10 case the gains are not uniform across all metrics. For instance, between \(1.5 < z_{\mathrm{true}} < 2.5\), the outlier fraction \(f_\mathrm{o}\) increases, accompanied by modest rises in the median scaled bias and its NMAD. This is likely due to residual mismatches in galaxy SEDs, reflecting limitations of our current augmentation implementation. As shown in Figure~\ref{fig:1}, the divergence between the \texttt{CosmoDC2} and \texttt{OpenUniverse2024} SEDs becomes more pronounced for the fainter galaxies probed at Y10 depths, which can introduce ambiguity into photometric-redshift training and degrade performance. This underscores the necessity for enhancing our augmented mock catalogues in preparation for future LSST Y10 photometric redshift analyses. 

Moreover, for both Y1 and Y10 scenarios, the large error bars on the blue curves denote substantial scatter in performance across unaugmented experiments -- highlighting sensitivity to the spectroscopic selection function -- whereas these variations are significantly reduced in the orange curves once augmentation is applied. Although the augmentation is not flawless, this reduced scatter across all metrics help demonstrate the efficacy of data augmentation. 

\subsection{Assessment of conditional density modelling}
\begin{figure*}
    \centering
    \begin{subfigure}[b]{0.48\linewidth}
        \centering
        \includegraphics[width=\linewidth]{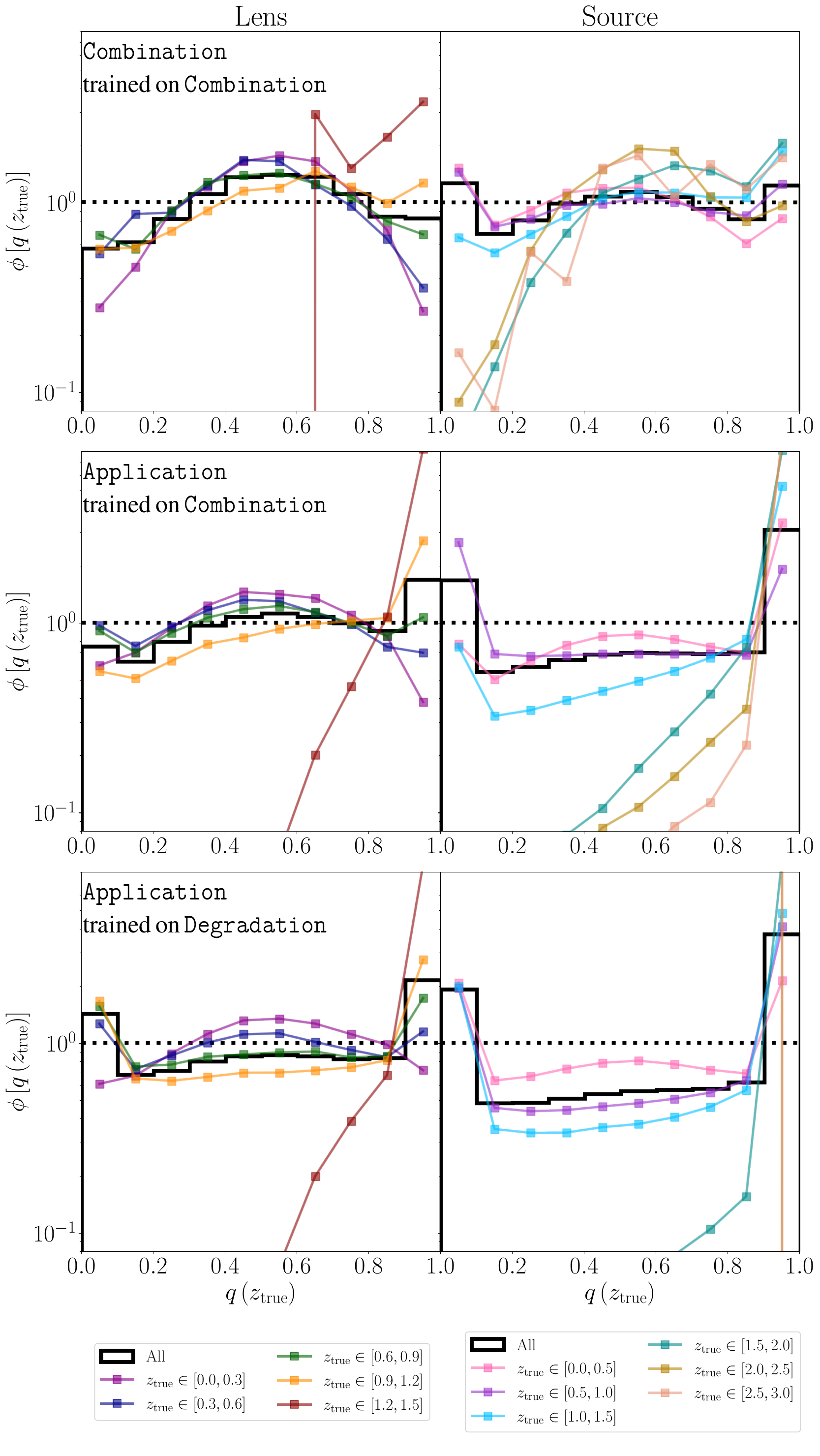}
        \caption{LSST Y1}
    \end{subfigure}
    \hspace{0.0\linewidth}
    \begin{subfigure}[b]{0.48\linewidth}
        \centering
        \includegraphics[width=\linewidth]{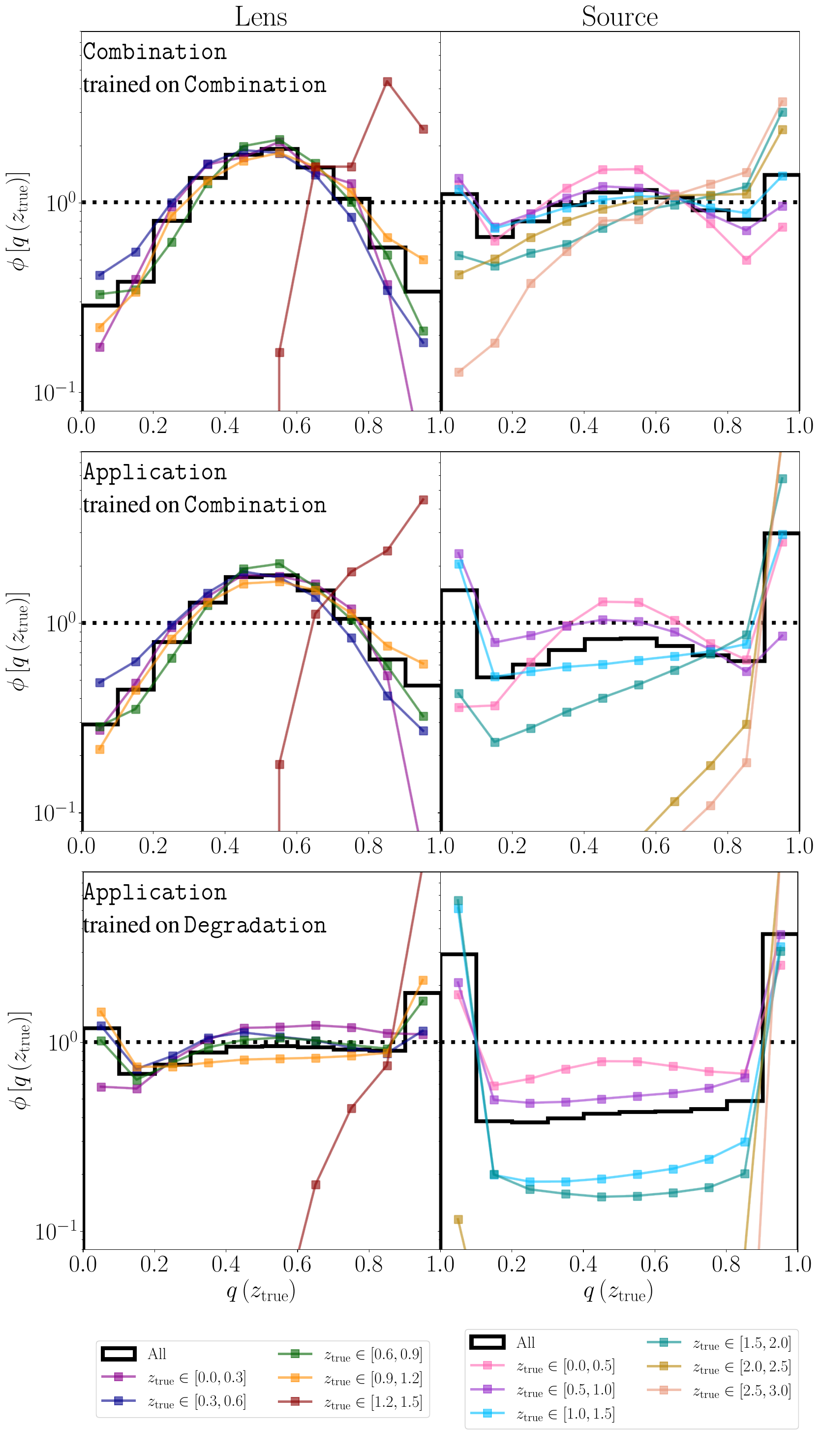}
        \caption{LSST Y10}
    \end{subfigure}
    \caption{Normalised PIT histograms for LSST Y1 (left panel) and Y10 (right panel) are depicted, under the baseline experiment. In each panel, the left subplots show lens samples, while the right subplots present source samples. The black solid lines represent the normalised PIT histogram of the whole sample, whereas the coloured solid lines illustrate the histograms for tomographic bins defined by \(z_{\mathrm{true}}\), with \(10\) bins distributed within \(\left[0.0, 3.0\right]\), spaced at intervals of \(0.3\). The upper row corresponds to the self-trained \texttt{Combination} dataset, which shows idealistic results with fully representative training; the second row relates to the \texttt{Application} dataset trained on the augmented \texttt{Combination} dataset; and the bottom row illustrates the \texttt{Application} dataset trained on the \texttt{Degradation} dataset without data augmentation.}
    \label{fig:10}
\end{figure*}

In parallel with our point‐estimate analysis, we also evaluate the quality of conditional density modelling. In the top panels of Figure~\ref{fig:10}, the normalised PIT histograms for the \texttt{Combination} datasets demonstrate the conditional‐density performance under fully representative training. For the lens samples, in both the Y1 and Y10 scenarios, a central peak exceeding unity is observed, along with dips falling below the flat line near \(q \left( z_\mathrm{true} \right) = 0\) and \(1\), across all subsets at varying redshifts. This indicates that the conditional-PDF widths for these bright populations are slightly overestimated, resulting in an excess of objects at intermediate quantiles. Such over‐smoothing stems from the basis‐function decomposition of the density model, since photometric redshift estimates for the low-redshift brightest galaxies would otherwise be expected to be sharply constrained.

The normalised PIT histogram of the all source galaxies, as shown in the solid black curve, reveal elevated values at the quantile boundaries, whilst the central peak approximates unity. Comparable patterns can be observed in histograms of low-redshift sources (\(z_\mathrm{true} \lesssim 1.5\)), which suggests their predominance in the overall distribution. This pattern indicates that the conditional-PDF of faint, low-redshift galaxies -- whose SEDs could be confused with those of higher-redshift objects -- are either underestimated regarding widths or systematically biased, unlike the bright lens samples. In the high-redshift region (\(z_\mathrm{true} \gtrsim 1.5\)), the histogram presents a marked slope even with fully representative training, suggesting a continuing systematic bias. These trends are also demonstrated in the increasing divergence loss of normalised PIT histograms in the bottom row of Figure~\ref{fig:8}, which correspond to the rise in point-estimate outliers and catastrophic failures, highlighting the difficulty of modelling precise PDFs at high redshifts. 

In the central panels of Figure~\ref{fig:10}, consistent with our point‐estimate results, applying the photometric‐redshift model to the \texttt{Application} datasets leaves the lens‐sample PIT distributions largely unchanged but precipitates a pronounced deterioration in the source‐sample performance. In particular, at \(z\gtrsim1.5\), the slope of the PIT histogram becomes markedly steeper, indicating a significant increase in average bias for faint galaxies. This trend is also reflected in the upward shift of the divergence‐loss curves from green (\texttt{Combination} datasets) to orange (\texttt{Application} datasets) in Figure~\ref{fig:8}. These findings emphasise that the inclusion of faint, high-redshift galaxies -- coupled with photometric noise and non-representative SEDs in the training data -- exacerbates the challenges of conditional-density modelling, especially for high redshift galaxies. This is one of the primary challenges in calibrating ensemble redshift distributions and will be addressed in detail in our forthcoming paper, where we will explore how various estimators can be combined to mitigate systematic biases at the population-level.

Comparison with models trained solely on the unaugmented \texttt{Degradation} dataset further emphasises the benefits of data augmentation for conditional‐density modelling. For the lens sample, augmentation induces a negligible change. In contrast, for the source sample under the LSST Y1 scenario, the divergence loss of the normalised PIT histogram decreases by approximately \(25 \,{-}\, 50\%\) even at low redshift (\(z_{\rm true} \lesssim 1.5\)), whereas improvements in point‐estimate metrics remain largely confined to high‐redshift galaxies. This effect is evident in the lower panels of Figure~\ref{fig:10}. With augmented training, the edge peaks of the normalised PIT histograms in the low‐redshift bins are moderately suppressed and the central values approach unity, yielding distributions that are closer to uniform. This improvement arises from a more faithful representation of SED degeneracies, owing to the inclusion of high‐redshift galaxies in the photometric‐redshift training.

At high redshift (\(z_{\rm true}\gtrsim1.5\)) in the LSST Y1 scenario, the unaugmented PIT histograms collapse to near zero except for a spike at \(q(z_{\rm true})=1\), indicative of pronounced systematic biases, consistent with the trends in Figures~\ref{fig:8} and \ref{fig:9}. Training on the augmented \texttt{Combination} datasets still produces a tilt relative to a uniform distribution, but a substantial fraction of galaxies exhibit markedly reduced bias, resulting in a non‐catastrophic normalised PIT histogram. This observation aligns with the reduction in catastrophic failure rates shown in Figure~\ref{fig:8}, highlighting that augmentation effectively restores missing information in the training dataset.

In the LSST Y10 scenario, the reduction in divergence loss for low-redshift galaxies remains pronounced, by up to a factor of \(\sim 2\) around \(z_\mathrm{true} \sim 1.5\), reflecting improved conditional-density modelling. At higher redshifts, however, only modest gains are seen. This is partly because the deeper spectroscopic \texttt{Degradation} datasets expected for LSST Y10 analysis already enhance the conditional-density estimates for high-redshift galaxies even without augmentation. Additionally, increasingly severe SED mismatches between the \texttt{Augmentation} and \texttt{Application} datasets for faint, high-redshift galaxies limit further improvements. Nevertheless, data augmentation still reduces the scatter in photometric-redshift performance arising from variations in the spectroscopic selection function.

\subsection{Limitations and future work}
One significant limitation of data augmentation for photometric‐redshift estimation is that the accuracy for galaxies at higher redshifts (\(z_\mathrm{true} \gtrsim 1.5\)) remains contingent upon the fidelity of the SED models in the mock catalogues even though our SOM-based weighted sample can largely align the distributions of galaxies in SED space. This shortcoming can be further alleviated by recalibrating the auto‐differentiable or semi‐analytic SED models against early LSST science observations, and by adopting a composite strategy that combines multiple simulations (e.g.\ \texttt{Cardinal}; \citealp{2024ApJ...961...59T} and \texttt{SKiLLS}; \citealp{2023A&A...670A.100L}), or even Stellar Population Synthesis (SPS) based mock galaxy SEDs using tools like \texttt{GalSBI}\citep{2025JCAP...06..007F, 2025A&A...703A.255T} or \texttt{pop-cosmos}\citep{2024ApJS..274...12A, 2024ApJ...975..145T, 2025ApJ...993..240T}, to achieve thorough coverage of the SOM space, thereby fully overlapping in the multi-dimensional colour-magnitude volume. By leveraging SOM‐based resampling, these comprehensive datasets can then be dynamically matched to observed photometry, offering a robust framework to mitigate residual discrepancies between simulations and real observations, particularly for faint, high‐redshift populations.  

Although mock catalogues such as \texttt{OpenUniverse2024} and \texttt{CosmoDC2} produce realistic galaxy properties, our analysis currently neglects pixel-level observational artefacts. Future work should incorporate forward-imaging simulations \citep{2023MNRAS.522.2801T}, for example employing \texttt{imSim} \citep{2014SPIE.9150E..14C} and \texttt{GalSim} \citep{2015A&C....10..121R}, to better capture systematics such as variable survey depth \citep{2016ApJS..226...24L, 2021A&A...648A..98J, 2025A&A...694A.259Y}, point-source contamination \citep{2015MNRAS.454.3500K}, and imperfect deblending of faint neighbouring sources \citep{2025arXiv250316680L}. These enhancements will enable a more realistic characterisation of photometric noise and selection biases, which is essential to ensure that data-augmentation techniques remain robust against the complexities inherent in real observational datasets.  

Our controlled‐degradation procedure provides only an approximation to the realistic spectroscopic samples anticipated for LSST cosmological analyses, since the distributions of magnitude and redshift limits will not be uniform across different surveys. Future work should therefore develop more comprehensive and physically motivated definitions of the spectroscopic selection function. Nonetheless, our approach furnishes a systematic framework for propagating selection-function effects into end-to-end photometric-redshift estimates. In practice, numerous spectroscopic surveys (e.g., DESI, PFS, and 4MOST), alongside deep narrow-band imaging programmes (e.g., COSMOS, PAUS, and J-PAS), will be integrated to optimise completeness and diversity within the multi-dimensional colour-magnitude space. Under this multi‐survey paradigm, controlled degradation remains a powerful tool: by selectively omitting or reweighting galaxy subsets from individual datasets, we can map the impact of incomplete spectroscopic coverage onto photometric redshifts, thereby enabling a rigorous characterisation of selection‐induced uncertainties.

In addition, as discussed in Section~\ref{sec:3}, our SOM-based data augmentation improves upon the simplified colour–magnitude–redshift boundaries used in \citet{2024ApJ...967L...6M} by providing a direct visualisation of the high-dimensional colour–magnitude manifold, thereby streamlining the identification of regions lacking spectroscopic coverage and optimising augmentation efficiency. Nonetheless, every dimensionality-reduction algorithm inevitably distorts the intrinsic topology. Future research will investigate cutting-edge techniques like Uniform Manifold Approximation and Projection \citep[UMAP;][]{2018arXiv180203426M} and \(t\)-distributed Stochastic Neighbour Embedding \citep[\(t\)-SNE;][]{vandermaatenJMLR:08a:v9}, aiming to improve augmentation fidelity. Moreover, more sophisticated and adaptable augmentation criteria can be defined within the joint colour-magnitude-redshift space, providing ample scope for further investigation.

A qualitative comparison with the approach utilised in \citet{2024ApJ...967L...6M} proves to be insightful. In that work, photometric redshifts were trained on a combination of currently available spectroscopic samples -- analogous to our LSST-Y1-like \texttt{Degradation} datasets -- but applied to an LSST-Y10 \texttt{Gold} sample, like our LSST Y10 \texttt{Application} datasets. Consequently, the level of missing spectroscopic information is considerably more significant than in our framework, where we anticipate broader spectroscopic coverage for Y10 analyses, though it remains anticipated. In \citet{2024ApJ...967L...6M}, the global photometric‐redshift metrics, such as biases, outlier fractions, and catastrophic rates, are reduced by a factor of \(\sim2\) when averaged over the full photometric sample. By contrast, in our analysis, these global statistics show only modest improvement, as they are dominated by low‐redshift galaxies with already high spectroscopic completeness. Nevertheless, our augmentation still yields clear benefits for the high‐redshift population. Taken together, the two configurations represent complementary scenarios: when the training sample is severely incomplete at the depth of the photometric data, augmentation improves overall performance; when the training sample is already broadly representative, the gains are concentrated at the faint, high‐redshift end.  

Given the differences between the mock catalogues and the varying photometric and spectroscopic selection functions adopted in this work and in \citet{2024ApJ...967L...6M}, a direct quantitative comparison is not straightforward. To enable a more rigorous assessment, we are actively analysing observational datasets from the \textit{Rubin} Data Preview~1 \citep[DP1\footnote{\url{https://dp1.lsst.io}},][]{10.71929/rubin/2570308}, which provides catalogue products over seven \(\sim 1 \,\mathrm{deg}^2\) fields for rigorous photometric-redshift validation. These analyses will benchmark our SOM-based augmentation against established methods -- both template-fitting and machine-learning -- to demonstrate its real-world efficacy. Further comparison using identical mock catalogues will enable careful evaluation of multiple augmentation strategies. The DP1 results will directly inform the integration of both colour–magnitude–redshift and SOM-based data augmentation workflows into the LSST DESC software ecosystem, with particular emphasis on the \texttt{RAIL} platform, thereby substantially enhancing photometric-redshift precision and robustness for Stage-IV cosmological studies.

Ultimately, this work has focused on constructing photometric and spectroscopic datasets and validating the efficacy of the SOM-based data-augmentation technique for photometric‐redshift estimation. In both Y1 and Y10 scenarios, the reduction of biases -- and in particular the significant decrease in catastrophic failure rates -- can substantially mitigate contamination when defining tomographic bins from photometric‐redshift estimates, thereby supporting the accurate calibration of ensemble redshift distributions. Building on this foundation, our forthcoming work will extend the framework to optimise tomographic binning strategies for both lens and source samples, following approaches discussed in the literature \cite{2021OJAp....4E..13Z, 2023ApJ...950...49M}, as well as to advance the calibration of ensemble redshift distributions. Leveraging the representative training of photometric redshifts and the improved performance achieved in both point estimates and conditional-density modelling, we will then enhance the calibration of ensemble redshift distributions by combining well‐behaved conditional PDFs with SOM‐based direct calibration \citep{2019MNRAS.489..820B, 2020A&A...637A.100W, 2025A&A...703A.144W}. This integrated approach aims to satisfy the stringent precision requirements of LSST cosmological analyses, particularly for weak gravitational lensing and large‐scale structure studies. Full details will be presented in the accompanying paper (Zhang et al., in preparation).  

\section{Conclusions} \label{sec:6}
In this study, building off the training sample augmentation technique originally introduced in \citep{2024ApJ...967L...6M}, we have developed, implemented and validated a SOM-based data-augmentation framework to improve the accuracy and robustness of photometric-redshift estimation. Starting from two independent mock catalogues -- \texttt{OpenUniverse2024} as a proxy for realistic LSST observations and \texttt{CosmoDC2} for targeted augmentation -- we conducted \(501\) experiments to generate photometric and spectroscopic datasets under both Y1 and Y10 survey depths.

We first constructed the \texttt{Application} datasets by applying LSST-like photometric errors, shear-noise and the DESC \texttt{Gold} selection to \texttt{OpenUniverse2024}, then trained a SOM on these samples to compress the high-dimensional colour–magnitude space into a two-dimensional grid of cells. Realistic spectroscopic \texttt{Degradation} datasets were produced by sampling this photometric population with magnitude, redshift and colour cuts, plus a logistic function of spectroscopic success rates, and by merging multiple independent realisations to emulate heterogeneous survey compilation.  

To compensate for under-sampling in the degraded spectroscopic datasets, we used the \texttt{CosmoDC2} catalogues with the same photometric pipeline to map each galaxy onto the pre-trained SOM. We applied cell-specific weights to harmonise their overall distributions in SOM space with the photometric \texttt{Application} datasets. We then employed adaptive augmentation by targeting SOM cells that were unoccupied in the \texttt{Degradation} datasets or where galaxies exceeded spectroscopic magnitude or redshift limits, subsequently adding suitably weighted mock galaxies until the proportion of augmented samples to degraded ones matched the deficit of occupied cells.

Using \texttt{FlexZBoost}, a conditional‐density machine‐learning estimator, we trained photometric‐redshift models on the \texttt{Combination} datasets, formed by merging the \texttt{Degradation} and \texttt{Augmentation} catalogues. We quantified point‐estimate accuracy via the median scaled bias, NMAD scatter, outlier fraction and catastrophic‐failure rate for both lens and source samples, and assessed full PDF fidelity through the divergence loss of the normalised PIT histograms. By benchmarking models trained with and without augmentation against the fully representative self‐training reference, we have demonstrated the robustness of our approach and the clear efficacy of SOM‐based data augmentation.

Across both Y1 and Y10 scenarios, the lens sample achieves median absolute scaled biases \(\bar{|\delta_z|} \sim 0.005\), comfortably within LSST DESC requirements even without augmentation—demonstrating the precision attainable for bright galaxies. For source galaxies at \(z_{\mathrm{true}}\lesssim1.5\), we likewise satisfy the science-requirement thresholds. Whilst performance declines at \(z_{\mathrm{true}}\gtrsim1.5\), training on the \texttt{Combination} datasets with SOM-based augmentation reduces the average bias and catastrophic failure rate by up to a factor of two, and enhances conditional-PDF modelling, as evidenced by markedly lower divergence losses. Importantly, the reduced scatter across independent realisations shows that augmentation renders photometric-redshift performance largely insensitive to variations in the spectroscopic selection function.

These results demonstrate that SOM‐based data augmentation robustly identify the SED space without representative spectroscopic samples and compensates for spectroscopic incompleteness and mitigates SED mismatches, yielding significantly more accurate and reliable photometric-redshift estimates. Such improvements are essential for LSST cosmological and galaxy-evolution analyses, enabling precision measurements of weak gravitational lensing, large-scale structure and the formation history of billions of galaxies across cosmic time.



\section*{Contribution Statements}

Y.-H. Z. conceptualised and directed the research, developed the methodologies, conducted the formal analysis -- which encompassed dataset construction, photometric-redshift modelling, and evaluation of the effectiveness of the SOM-based data-augmentation technique -- and also authored the majority of the manuscript.

J. Z. supervised the work, and contributed conceptually to methodology development, formal analysis, and scientific validation, and also reviewed and refined the manuscript.

I. M. established the data augmentation framework, offered insights on dataset construction and photometric redshift modelling, provided necessary scripts to facilitate the formal analysis, and also reviewed and commented on the manuscript.

E. G. contributed to designing the conceptual framework, advised on dataset construction and photometric redshift modelling, delivered crucial feedback for scientific validation, and reviewed and commented on the manuscript.

K. K. co-supervised the work, offering guidance on methodology development, formal analysis, and scientific validation, in addition to reviewing and providing feedback on the manuscript.

M. A. proposed refinements for the construction of spectroscopic datasets, advised on the application of the SOM technique, and also reviewed and commented on the manuscript.

H. H. offered insights into the discussion of the scientific findings, emphasised potential areas for future enhancements, and also reviewed and provided feedback on the draft.

Z. Y. assisted in employing the SOM technique, advising on software configuration and scientific validation, and was actively involved in implementing SOM-based data augmentation for the RAIL platform, in addition to reviewing and providing feedback on the manuscript.

A. M. and T. Z. are DESC builders and key developers of the RAIL platform, which is crucial to this work.



\section*{Acknowledgements}

This manuscript has been internally reviewed by the LSST DESC. The authors express their gratitude to Catherine Heymans and Nora Elisa Chisari for their contribution as members of the DESC publication review committee, whose insightful comments and suggestions enhanced the quality of this paper.

This study extensively utilised publicly available software, including \texttt{NumPy} \citep{2020Natur.585..357H}, \texttt{SciPy} \citep{2020NatMe..17..261V}, and \texttt{Matplotlib} \citep{2007CSE.....9...90H}, as well as the DESC softwares, particularly \texttt{RAIL} platform, including implementation of \texttt{Somoclu} and \texttt{FlexZBoost}, which were instrumental to this work.

Y.–H. Z. acknowledges support from the UK Research and Innovation (UKRI) Science and Technology Facilities Council (STFC) studentship, the Edinburgh–Leiden joint studentship awarded by the University of Edinburgh and Leiden University, and the STFC travel fund for UK participation in LSST (grant ST/X001334/1). 

J. Z. acknowledges the support by STFC funding for UK participation in LSST, through grant ST/X001334/1.

I. M. and E.G. acknowledge support from the U.S. Department of Energy, Office of Science, Office of High Energy Physics Cosmic Frontier Research program under Award Number DE-SC0010008.

M. A. receives support from the UK STFC through grant number ST/Y002652/1, as well as from the Royal Society via grant numbers RGSR2222268 and ICAR1231094.

T. Z. is supported by Schmidt Sciences.

The DESC acknowledges ongoing support from the Institut National de Physique Nucl\'eaire et de Physique des Particules in France; the Science \& Technology Facilities Council in the United Kingdom; and the Department of Energy and the LSST Discovery Alliance in the United States.  DESC uses resources of the IN2P3 Computing Center (CC-IN2P3--Lyon/Villeurbanne - France) funded by the Centre National de la Recherche Scientifique; the National Energy Research Scientific Computing Center, a DOE Office of Science User Facility supported by the Office of Science of the U.S.\ Department of Energy under Contract No.\ DE-AC02-05CH11231; STFC DiRAC HPC Facilities, funded by UK BEIS National E-infrastructure capital grants; and the UK particle physics grid, supported by the GridPP Collaboration. This work was performed in part under DOE Contract DE-AC02-76SF00515.

\section*{Data Availability}

All scripts associated with this study are publicly available in the GitHub repository \href{https://github.com/CosmoCloudZhang/SOMZCloud.git}{\faGithub \texttt{SOMZCloud}}. The generated datasets and accompanying products -- including photometric redshift estimates -- have been deposited in the LSST DESC Community File System (CFS) at the National Energy Research Scientific Computing Center (NERSC\footnote{\url{https://www.nersc.gov}}).



\bibliographystyle{mnras}
\bibliography{mnras}

@ARTICLE{2022PhRvD.105b3520A,
       author = {{Abbott}, T.~M.~C. and {Aguena}, M. and {Alarcon}, A. and {Allam}, S. and {Alves}, O. and {Amon}, A. and {Andrade-Oliveira}, F. and {Annis}, J. and {Avila}, S. and {Bacon}, D. and {Baxter}, E. and {Bechtol}, K. and {Becker}, M.~R. and {Bernstein}, G.~M. and {Bhargava}, S. and {Birrer}, S. and {Blazek}, J. and {Brandao-Souza}, A. and {Bridle}, S.~L. and {Brooks}, D. and {Buckley-Geer}, E. and {Burke}, D.~L. and {Camacho}, H. and {Campos}, A. and {Carnero Rosell}, A. and {Carrasco Kind}, M. and {Carretero}, J. and {Castander}, F.~J. and {Cawthon}, R. and {Chang}, C. and {Chen}, A. and {Chen}, R. and {Choi}, A. and {Conselice}, C. and {Cordero}, J. and {Costanzi}, M. and {Crocce}, M. and {da Costa}, L.~N. and {da Silva Pereira}, M.~E. and {Davis}, C. and {Davis}, T.~M. and {De Vicente}, J. and {DeRose}, J. and {Desai}, S. and {Di Valentino}, E. and {Diehl}, H.~T. and {Dietrich}, J.~P. and {Dodelson}, S. and {Doel}, P. and {Doux}, C. and {Drlica-Wagner}, A. and {Eckert}, K. and {Eifler}, T.~F. and {Elsner}, F. and {Elvin-Poole}, J. and {Everett}, S. and {Evrard}, A.~E. and {Fang}, X. and {Farahi}, A. and {Fernandez}, E. and {Ferrero}, I. and {Fert{\'e}}, A. and {Fosalba}, P. and {Friedrich}, O. and {Frieman}, J. and {Garc{\'\i}a-Bellido}, J. and {Gatti}, M. and {Gaztanaga}, E. and {Gerdes}, D.~W. and {Giannantonio}, T. and {Giannini}, G. and {Gruen}, D. and {Gruendl}, R.~A. and {Gschwend}, J. and {Gutierrez}, G. and {Harrison}, I. and {Hartley}, W.~G. and {Herner}, K. and {Hinton}, S.~R. and {Hollowood}, D.~L. and {Honscheid}, K. and {Hoyle}, B. and {Huff}, E.~M. and {Huterer}, D. and {Jain}, B. and {James}, D.~J. and {Jarvis}, M. and {Jeffrey}, N. and {Jeltema}, T. and {Kovacs}, A. and {Krause}, E. and {Kron}, R. and {Kuehn}, K. and {Kuropatkin}, N. and {Lahav}, O. and {Leget}, P. -F. and {Lemos}, P. and {Liddle}, A.~R. and {Lidman}, C. and {Lima}, M. and {Lin}, H. and {MacCrann}, N. and {Maia}, M.~A.~G. and {Marshall}, J.~L. and {Martini}, P. and {McCullough}, J. and {Melchior}, P. and {Mena-Fern{\'a}ndez}, J. and {Menanteau}, F. and {Miquel}, R. and {Mohr}, J.~J. and {Morgan}, R. and {Muir}, J. and {Myles}, J. and {Nadathur}, S. and {Navarro-Alsina}, A. and {Nichol}, R.~C. and {Ogando}, R.~L.~C. and {Omori}, Y. and {Palmese}, A. and {Pandey}, S. and {Park}, Y. and {Paz-Chinch{\'o}n}, F. and {Petravick}, D. and {Pieres}, A. and {Plazas Malag{\'o}n}, A.~A. and {Porredon}, A. and {Prat}, J. and {Raveri}, M. and {Rodriguez-Monroy}, M. and {Rollins}, R.~P. and {Romer}, A.~K. and {Roodman}, A. and {Rosenfeld}, R. and {Ross}, A.~J. and {Rykoff}, E.~S. and {Samuroff}, S. and {S{\'a}nchez}, C. and {Sanchez}, E. and {Sanchez}, J. and {Sanchez Cid}, D. and {Scarpine}, V. and {Schubnell}, M. and {Scolnic}, D. and {Secco}, L.~F. and {Serrano}, S. and {Sevilla-Noarbe}, I. and {Sheldon}, E. and {Shin}, T. and {Smith}, M. and {Soares-Santos}, M. and {Suchyta}, E. and {Swanson}, M.~E.~C. and {Tabbutt}, M. and {Tarle}, G. and {Thomas}, D. and {To}, C. and {Troja}, A. and {Troxel}, M.~A. and {Tucker}, D.~L. and {Tutusaus}, I. and {Varga}, T.~N. and {Walker}, A.~R. and {Weaverdyck}, N. and {Wechsler}, R. and {Weller}, J. and {Yanny}, B. and {Yin}, B. and {Zhang}, Y. and {Zuntz}, J. and {DES Collaboration}},
        title = "{Dark Energy Survey Year 3 results: Cosmological constraints from galaxy clustering and weak lensing}",
      journal = {\prd},
         year = 2022,
        month = jan,
       volume = {105},
       number = {2},
          eid = {023520},
        pages = {023520},
          doi = {10.1103/PhysRevD.105.023520},
archivePrefix = {arXiv},
       eprint = {2105.13549},
 primaryClass = {astro-ph.CO},
       adsurl = {https://ui.adsabs.harvard.edu/abs/2022PhRvD.105b3520A}
}

@ARTICLE{2019arXiv190205569A,
       author = {{Akeson}, Rachel and {Armus}, Lee and {Bachelet}, Etienne and {Bailey}, Vanessa and {Bartusek}, Lisa and {Bellini}, Andrea and {Benford}, Dominic and {Bennett}, David and {Bhattacharya}, Aparna and {Bohlin}, Ralph and {Boyer}, Martha and {Bozza}, Valerio and {Bryden}, Geoffrey and {Calchi Novati}, Sebastiano and {Carpenter}, Kenneth and {Casertano}, Stefano and {Choi}, Ami and {Content}, David and {Dayal}, Pratika and {Dressler}, Alan and {Dor{\'e}}, Olivier and {Fall}, S. Michael and {Fan}, Xiaohui and {Fang}, Xiao and {Filippenko}, Alexei and {Finkelstein}, Steven and {Foley}, Ryan and {Furlanetto}, Steven and {Kalirai}, Jason and {Gaudi}, B. Scott and {Gilbert}, Karoline and {Girard}, Julien and {Grady}, Kevin and {Greene}, Jenny and {Guhathakurta}, Puragra and {Heinrich}, Chen and {Hemmati}, Shoubaneh and {Hendel}, David and {Henderson}, Calen and {Henning}, Thomas and {Hirata}, Christopher and {Ho}, Shirley and {Huff}, Eric and {Hutter}, Anne and {Jansen}, Rolf and {Jha}, Saurabh and {Johnson}, Samson and {Jones}, David and {Kasdin}, Jeremy and {Kelly}, Patrick and {Kirshner}, Robert and {Koekemoer}, Anton and {Kruk}, Jeffrey and {Lewis}, Nikole and {Macintosh}, Bruce and {Madau}, Piero and {Malhotra}, Sangeeta and {Mandel}, Kaisey and {Massara}, Elena and {Masters}, Daniel and {McEnery}, Julie and {McQuinn}, Kristen and {Melchior}, Peter and {Melton}, Mark and {Mennesson}, Bertrand and {Peeples}, Molly and {Penny}, Matthew and {Perlmutter}, Saul and {Pisani}, Alice and {Plazas}, Andr{\'e}s and {Poleski}, Radek and {Postman}, Marc and {Ranc}, Cl{\'e}ment and {Rauscher}, Bernard and {Rest}, Armin and {Roberge}, Aki and {Robertson}, Brant and {Rodney}, Steven and {Rhoads}, James and {Rhodes}, Jason and {Ryan}, Jr., Russell and {Sahu}, Kailash and {Sand}, David and {Scolnic}, Dan and {Seth}, Anil and {Shvartzvald}, Yossi and {Siellez}, Karelle and {Smith}, Arfon and {Spergel}, David and {Stassun}, Keivan and {Street}, Rachel and {Strolger}, Louis-Gregory and {Szalay}, Alexander and {Trauger}, John and {Troxel}, M.~A. and {Turnbull}, Margaret and {van der Marel}, Roeland and {von der Linden}, Anja and {Wang}, Yun and {Weinberg}, David and {Williams}, Benjamin and {Windhorst}, Rogier and {Wollack}, Edward and {Wu}, Hao-Yi and {Yee}, Jennifer and {Zimmerman}, Neil},
        title = "{The Wide Field Infrared Survey Telescope: 100 Hubbles for the 2020s}",
      journal = {arXiv e-prints},
         year = 2019,
        month = feb,
          eid = {arXiv:1902.05569},
        pages = {arXiv:1902.05569},
          doi = {10.48550/arXiv.1902.05569},
archivePrefix = {arXiv},
       eprint = {1902.05569},
 primaryClass = {astro-ph.IM},
       adsurl = {https://ui.adsabs.harvard.edu/abs/2019arXiv190205569A}
}

@ARTICLE{2021MNRAS.501.6103A,
       author = {{Alarcon}, Alex and {Gaztanaga}, Enrique and {Eriksen}, Martin and {Baugh}, Carlton M. and {Cabayol}, Laura and {Casas}, Ricard and {Carretero}, Jorge and {Castander}, Francisco J. and {De Vicente}, Juan and {Fernandez}, Enrique and {Garcia-Bellido}, Juan and {Hildebrandt}, Hendrik and {Hoekstra}, Henk and {Joachimi}, Benjamin and {Manzoni}, Giorgio and {Miquel}, Ramon and {Norberg}, Peder and {Padilla}, Cristobal and {Renard}, Pablo and {Sanchez}, Eusebio and {Serrano}, Santiago and {Sevilla-Noarbe}, Ignacio and {Siudek}, Malgorzata and {Tallada-Cresp{\'\i}}, Pau},
        title = "{The PAU Survey: an improved photo-z sample in the COSMOS field}",
      journal = {\mnras},
         year = 2021,
        month = mar,
       volume = {501},
       number = {4},
        pages = {6103-6122},
          doi = {10.1093/mnras/staa3659},
archivePrefix = {arXiv},
       eprint = {2007.11132},
 primaryClass = {astro-ph.GA},
       adsurl = {https://ui.adsabs.harvard.edu/abs/2021MNRAS.501.6103A}
}

@ARTICLE{2023MNRAS.518..562A,
      author = {{Alarcon}, Alex and {Hearin}, Andrew P. and {Becker}, Matthew R. and {Chaves-Montero}, Jon{\'a}s},
        title = "{Diffstar: a fully parametric physical model for galaxy assembly history}",
      journal = {MNRAS},
        year = 2023,
        month = jan,
      volume = {518},
      number = {1},
        pages = {562-584},
          doi = {10.1093/mnras/stac3118},
archivePrefix = {arXiv},
      eprint = {2205.04273},
primaryClass = {astro-ph.GA},
      adsurl = {https://ui.adsabs.harvard.edu/abs/2023MNRAS.518..562A}
}

@ARTICLE{2016MNRAS.462..726A,
       author = {{Almosallam}, Ibrahim A. and {Jarvis}, Matt J. and {Roberts}, Stephen J.},
        title = "{GPZ: non-stationary sparse Gaussian processes for heteroscedastic uncertainty estimation in photometric redshifts}",
      journal = {\mnras},
         year = 2016,
        month = oct,
       volume = {462},
       number = {1},
        pages = {726-739},
          doi = {10.1093/mnras/stw1618},
archivePrefix = {arXiv},
       eprint = {1604.03593},
 primaryClass = {astro-ph.IM},
       adsurl = {https://ui.adsabs.harvard.edu/abs/2016MNRAS.462..726A}
}

@ARTICLE{2024ApJS..274...12A,
       author = {{Alsing}, Justin and {Thorp}, Stephen and {Deger}, Sinan and {Peiris}, Hiranya V. and {Leistedt}, Boris and {Mortlock}, Daniel and {Leja}, Joel},
        title = "{pop-cosmos: A Comprehensive Picture of the Galaxy Population from COSMOS Data}",
      journal = {\apjs},
         year = 2024,
        month = sep,
       volume = {274},
       number = {1},
          eid = {12},
        pages = {12},
          doi = {10.3847/1538-4365/ad5c69},
archivePrefix = {arXiv},
       eprint = {2402.00935},
 primaryClass = {astro-ph.GA},
       adsurl = {https://ui.adsabs.harvard.edu/abs/2024ApJS..274...12A}
}

@ARTICLE{2022PhRvD.105b3514A,
       author = {{Amon}, A. and {Gruen}, D. and {Troxel}, M.~A. and {MacCrann}, N. and {Dodelson}, S. and {Choi}, A. and {Doux}, C. and {Secco}, L.~F. and {Samuroff}, S. and {Krause}, E. and {Cordero}, J. and {Myles}, J. and {DeRose}, J. and {Wechsler}, R.~H. and {Gatti}, M. and {Navarro-Alsina}, A. and {Bernstein}, G.~M. and {Jain}, B. and {Blazek}, J. and {Alarcon}, A. and {Fert{\'e}}, A. and {Lemos}, P. and {Raveri}, M. and {Campos}, A. and {Prat}, J. and {S{\'a}nchez}, C. and {Jarvis}, M. and {Alves}, O. and {Andrade-Oliveira}, F. and {Baxter}, E. and {Bechtol}, K. and {Becker}, M.~R. and {Bridle}, S.~L. and {Camacho}, H. and {Carnero Rosell}, A. and {Carrasco Kind}, M. and {Cawthon}, R. and {Chang}, C. and {Chen}, R. and {Chintalapati}, P. and {Crocce}, M. and {Davis}, C. and {Diehl}, H.~T. and {Drlica-Wagner}, A. and {Eckert}, K. and {Eifler}, T.~F. and {Elvin-Poole}, J. and {Everett}, S. and {Fang}, X. and {Fosalba}, P. and {Friedrich}, O. and {Gaztanaga}, E. and {Giannini}, G. and {Gruendl}, R.~A. and {Harrison}, I. and {Hartley}, W.~G. and {Herner}, K. and {Huang}, H. and {Huff}, E.~M. and {Huterer}, D. and {Kuropatkin}, N. and {Leget}, P. and {Liddle}, A.~R. and {McCullough}, J. and {Muir}, J. and {Pandey}, S. and {Park}, Y. and {Porredon}, A. and {Refregier}, A. and {Rollins}, R.~P. and {Roodman}, A. and {Rosenfeld}, R. and {Ross}, A.~J. and {Rykoff}, E.~S. and {Sanchez}, J. and {Sevilla-Noarbe}, I. and {Sheldon}, E. and {Shin}, T. and {Troja}, A. and {Tutusaus}, I. and {Tutusaus}, I. and {Varga}, T.~N. and {Weaverdyck}, N. and {Yanny}, B. and {Yin}, B. and {Zhang}, Y. and {Zuntz}, J. and {Aguena}, M. and {Allam}, S. and {Annis}, J. and {Bacon}, D. and {Bertin}, E. and {Bhargava}, S. and {Brooks}, D. and {Buckley-Geer}, E. and {Burke}, D.~L. and {Carretero}, J. and {Costanzi}, M. and {da Costa}, L.~N. and {Pereira}, M.~E.~S. and {De Vicente}, J. and {Desai}, S. and {Dietrich}, J.~P. and {Doel}, P. and {Ferrero}, I. and {Flaugher}, B. and {Frieman}, J. and {Garc{\'\i}a-Bellido}, J. and {Gaztanaga}, E. and {Gerdes}, D.~W. and {Giannantonio}, T. and {Gschwend}, J. and {Gutierrez}, G. and {Hinton}, S.~R. and {Hollowood}, D.~L. and {Honscheid}, K. and {Hoyle}, B. and {James}, D.~J. and {Kron}, R. and {Kuehn}, K. and {Lahav}, O. and {Lima}, M. and {Lin}, H. and {Maia}, M.~A.~G. and {Marshall}, J.~L. and {Martini}, P. and {Melchior}, P. and {Menanteau}, F. and {Miquel}, R. and {Mohr}, J.~J. and {Morgan}, R. and {Ogando}, R.~L.~C. and {Palmese}, A. and {Paz-Chinch{\'o}n}, F. and {Petravick}, D. and {Pieres}, A. and {Romer}, A.~K. and {Sanchez}, E. and {Scarpine}, V. and {Schubnell}, M. and {Serrano}, S. and {Smith}, M. and {Soares-Santos}, M. and {Tarle}, G. and {Thomas}, D. and {To}, C. and {Weller}, J. and {DES Collaboration}},
        title = "{Dark Energy Survey Year 3 results: Cosmology from cosmic shear and robustness to data calibration}",
      journal = {\prd},
         year = 2022,
        month = jan,
       volume = {105},
       number = {2},
          eid = {023514},
        pages = {023514},
          doi = {10.1103/PhysRevD.105.023514},
archivePrefix = {arXiv},
       eprint = {2105.13543},
 primaryClass = {astro-ph.CO},
       adsurl = {https://ui.adsabs.harvard.edu/abs/2022PhRvD.105b3514A}
}

@ARTICLE{1999MNRAS.310..540A,
       author = {{Arnouts}, S. and {Cristiani}, S. and {Moscardini}, L. and {Matarrese}, S. and {Lucchin}, F. and {Fontana}, A. and {Giallongo}, E.},
        title = "{Measuring and modelling the redshift evolution of clustering: the Hubble Deep Field North}",
      journal = {\mnras},
         year = 1999,
        month = dec,
       volume = {310},
       number = {2},
        pages = {540-556},
          doi = {10.1046/j.1365-8711.1999.02978.x},
archivePrefix = {arXiv},
       eprint = {astro-ph/9902290},
 primaryClass = {astro-ph},
       adsurl = {https://ui.adsabs.harvard.edu/abs/1999MNRAS.310..540A}
}

@ARTICLE{2021A&A...645A.104A,
       author = {{Asgari}, Marika and {Lin}, Chieh-An and {Joachimi}, Benjamin and {Giblin}, Benjamin and {Heymans}, Catherine and {Hildebrandt}, Hendrik and {Kannawadi}, Arun and {St{\"o}lzner}, Benjamin and {Tr{\"o}ster}, Tilman and {van den Busch}, Jan Luca and {Wright}, Angus H. and {Bilicki}, Maciej and {Blake}, Chris and {de Jong}, Jelte and {Dvornik}, Andrej and {Erben}, Thomas and {Getman}, Fedor and {Hoekstra}, Henk and {K{\"o}hlinger}, Fabian and {Kuijken}, Konrad and {Miller}, Lance and {Radovich}, Mario and {Schneider}, Peter and {Shan}, HuanYuan and {Valentijn}, Edwin},
        title = "{KiDS-1000 cosmology: Cosmic shear constraints and comparison between two point statistics}",
      journal = {\aap},
         year = 2021,
        month = jan,
       volume = {645},
          eid = {A104},
        pages = {A104},
          doi = {10.1051/0004-6361/202039070},
archivePrefix = {arXiv},
       eprint = {2007.15633},
 primaryClass = {astro-ph.CO},
       adsurl = {https://ui.adsabs.harvard.edu/abs/2021A&A...645A.104A}
}

@ARTICLE{2024MNRAS.534.3808A,
       author = {{Autenrieth}, Maximilian and {Wright}, Angus H. and {Trotta}, Roberto and {van Dyk}, David A. and {Stenning}, David C. and {Joachimi}, Benjamin},
        title = "{Improved weak lensing photometric redshift calibration via StratLearn and hierarchical modelling}",
      journal = {\mnras},
         year = 2024,
        month = nov,
       volume = {534},
       number = {4},
        pages = {3808-3831},
          doi = {10.1093/mnras/stae2243},
archivePrefix = {arXiv},
       eprint = {2401.04687},
 primaryClass = {astro-ph.CO},
       adsurl = {https://ui.adsabs.harvard.edu/abs/2024MNRAS.534.3808A}
}

@ARTICLE{2001PhR...340..291B,
       author = {{Bartelmann}, M. and {Schneider}, P.},
        title = "{Weak gravitational lensing}",
      journal = {\physrep},
         year = 2001,
        month = jan,
       volume = {340},
       number = {4-5},
        pages = {291-472},
          doi = {10.1016/S0370-1573(00)00082-X},
archivePrefix = {arXiv},
       eprint = {astro-ph/9912508},
 primaryClass = {astro-ph},
       adsurl = {https://ui.adsabs.harvard.edu/abs/2001PhR...340..291B}
}

@ARTICLE{2025arXiv250105739B,
       author = {{Bechtol}, K. and {Sevilla-Noarbe}, I. and {Drlica-Wagner}, A. and {Yanny}, B. and {Gruendl}, R.~A. and {Sheldon}, E. and {Rykoff}, E.~S. and {De Vicente}, J. and {Adamow}, M. and {Anbajagane}, D. and {Becker}, M.~R. and {Bernstein}, G.~M. and {Carnero Rosell}, A. and {Gschwend}, J. and {Gorsuch}, M. and {Hartley}, W.~G. and {Jarvis}, M. and {Jeltema}, T. and {Kron}, R. and {Manning}, T.~A. and {O'Donnell}, J. and {Pieres}, A. and {Rodr{\'\i}guez-Monroy}, M. and {Sanchez Cid}, D. and {Tabbutt}, M. and {Toribio San Cipriano}, L. and {Tucker}, D.~L. and {Weaverdyck}, N. and {Yamamoto}, M. and {Abbott}, T.~M.~C. and {Aguena}, M. and {Alarc{\'o}n}, A. and {Allam}, S. and {Amon}, A. and {Andrade-Oliveira}, F. and {Avila}, S. and {Bernardinelli}, P.~H. and {Bertin}, E. and {Blazek}, J. and {Brooks}, D. and {Burke}, D.~L. and {Carretero}, J. and {Castander}, F.~J. and {Cawthon}, R. and {Chang}, C. and {Choi}, A. and {Conselice}, C. and {Costanzi}, M. and {Crocce}, M. and {da Costa}, L.~N. and {Davis}, T.~M. and {Desai}, S. and {Diehl}, H.~T. and {Dodelson}, S. and {Doel}, P. and {Doux}, C. and {Fert{\'e}}, A. and {Flaugher}, B. and {Fosalba}, P. and {Frieman}, J. and {Garc{\'\i}a-Bellido}, J. and {Gatti}, M. and {Gaztanaga}, E. and {Giannini}, G. and {Gruen}, D. and {Gutierrez}, G. and {Herner}, K. and {Hinton}, S.~R. and {Hollowood}, D.~L. and {Honscheid}, K. and {Huterer}, D. and {Jeffrey}, N. and {Krause}, E. and {Kuehn}, K. and {Lahav}, O. and {Lee}, S. and {Lidman}, C. and {Lima}, M. and {Lin}, H. and {Marshall}, J.~L. and {Mena-Fern{\'a}ndez}, J. and {Miquel}, R. and {Mohr}, J.~J. and {Muir}, J. and {Myles}, J. and {Ogando}, R.~L.~C. and {Palmese}, A. and {Plazas Malag{\'o}n}, A.~A. and {Porredon}, A. and {Prat}, J. and {Raveri}, M. and {Romer}, A.~K. and {Roodman}, A. and {Samuroff}, S. and {Sanchez}, E. and {Scarpine}, V. and {Smith}, M. and {Soares-Santos}, M. and {Suchyta}, E. and {Tarle}, G. and {Troxel}, M.~A. and {Vikram}, V. and {Walker}, A.~R. and {Weller}, J. and {Wiseman}, P. and {Zhang}, Y.},
        title = "{Dark Energy Survey Year 6 Results: Photometric Data Set for Cosmology}",
      journal = {arXiv e-prints},
         year = 2025,
        month = jan,
          eid = {arXiv:2501.05739},
        pages = {arXiv:2501.05739},
          doi = {10.48550/arXiv.2501.05739},
archivePrefix = {arXiv},
       eprint = {2501.05739},
 primaryClass = {astro-ph.CO},
       adsurl = {https://ui.adsabs.harvard.edu/abs/2025arXiv250105739B}
}

@ARTICLE{2013ApJ...770...57B,
       author = {{Behroozi}, Peter S. and {Wechsler}, Risa H. and {Conroy}, Charlie},
        title = "{The Average Star Formation Histories of Galaxies in Dark Matter Halos from z = 0-8}",
      journal = {\apj},
         year = 2013,
        month = jun,
       volume = {770},
       number = {1},
          eid = {57},
        pages = {57},
          doi = {10.1088/0004-637X/770/1/57},
archivePrefix = {arXiv},
       eprint = {1207.6105},
 primaryClass = {astro-ph.CO},
       adsurl = {https://ui.adsabs.harvard.edu/abs/2013ApJ...770...57B}
}

@ARTICLE{2019MNRAS.488.3143B,
       author = {{Behroozi}, Peter and {Wechsler}, Risa H. and {Hearin}, Andrew P. and {Conroy}, Charlie},
        title = "{UNIVERSEMACHINE: The correlation between galaxy growth and dark matter halo assembly from z = 0-10}",
      journal = {\mnras},
         year = 2019,
        month = sep,
       volume = {488},
       number = {3},
        pages = {3143-3194},
          doi = {10.1093/mnras/stz1182},
archivePrefix = {arXiv},
       eprint = {1806.07893},
 primaryClass = {astro-ph.GA},
       adsurl = {https://ui.adsabs.harvard.edu/abs/2019MNRAS.488.3143B}
}

@ARTICLE{2000ApJ...536..571B,
       author = {{Ben{\'\i}tez}, Narciso},
        title = "{Bayesian Photometric Redshift Estimation}",
      journal = {\apj},
         year = 2000,
        month = jun,
       volume = {536},
       number = {2},
        pages = {571-583},
          doi = {10.1086/308947},
archivePrefix = {arXiv},
       eprint = {astro-ph/9811189},
 primaryClass = {astro-ph},
       adsurl = {https://ui.adsabs.harvard.edu/abs/2000ApJ...536..571B}
}

@ARTICLE{2014arXiv1403.5237B,
       author = {{Benitez}, N. and {Dupke}, R. and {Moles}, M. and {Sodre}, L. and {Cenarro}, J. and {Marin-Franch}, A. and {Taylor}, K. and {Cristobal}, D. and {Fernandez-Soto}, A. and {Mendes de Oliveira}, C. and {Cepa-Nogue}, J. and {Abramo}, L.~R. and {Alcaniz}, J.~S. and {Overzier}, R. and {Hernandez-Monteagudo}, C. and {Alfaro}, E.~J. and {Kanaan}, A. and {Carvano}, J.~M. and {Reis}, R.~R.~R. and {Martinez Gonzalez}, E. and {Ascaso}, B. and {Ballesteros}, F. and {Xavier}, H.~S. and {Varela}, J. and {Ederoclite}, A. and {Vazquez Ramio}, H. and {Broadhurst}, T. and {Cypriano}, E. and {Angulo}, R. and {Diego}, J.~M. and {Zandivarez}, A. and {Diaz}, E. and {Melchior}, P. and {Umetsu}, K. and {Spinelli}, P.~F. and {Zitrin}, A. and {Coe}, D. and {Yepes}, G. and {Vielva}, P. and {Sahni}, V. and {Marcos-Caballero}, A. and {Kitaura}, F. -S. and {Maroto}, A.~L. and {Masip}, M. and {Tsujikawa}, S. and {Carneiro}, S. and {Gonzalez Nuevo}, J. and {Carvalho}, G.~C. and {Reboucas}, M.~J. and {Carvalho}, J.~C. and {Abdalla}, E. and {Bernui}, A. and {Pigozzo}, C. and {Ferreira}, E.~G.~M. and {Chandrachani Devi}, N. and {Bengaly}, Jr., C.~A.~P. and {Campista}, M. and {Amorim}, A. and {Asari}, N.~V. and {Bongiovanni}, A. and {Bonoli}, S. and {Bruzual}, G. and {Cardiel}, N. and {Cava}, A. and {Cid Fernandes}, R. and {Coelho}, P. and {Cortesi}, A. and {Delgado}, R.~G. and {Diaz Garcia}, L. and {Espinosa}, J.~M.~R. and {Galliano}, E. and {Gonzalez-Serrano}, J.~I. and {Falcon-Barroso}, J. and {Fritz}, J. and {Fernandes}, C. and {Gorgas}, J. and {Hoyos}, C. and {Jimenez-Teja}, Y. and {Lopez-Aguerri}, J.~A. and {Lopez-San Juan}, C. and {Mateus}, A. and {Molino}, A. and {Novais}, P. and {OMill}, A. and {Oteo}, I. and {Perez-Gonzalez}, P.~G. and {Poggianti}, B. and {Proctor}, R. and {Ricciardelli}, E. and {Sanchez-Blazquez}, P. and {Storchi-Bergmann}, T. and {Telles}, E. and {Schoennell}, W. and {Trujillo}, N. and {Vazdekis}, A. and {Viironen}, K. and {Daflon}, S. and {Aparicio-Villegas}, T. and {Rocha}, D. and {Ribeiro}, T. and {Borges}, M. and {Martins}, S.~L. and {Marcolino}, W. and {Martinez-Delgado}, D. and {Perez-Torres}, M.~A. and {Siffert}, B.~B. and {Calvao}, M.~O. and {Sako}, M. and {Kessler}, R. and {Alvarez-Candal}, A. and {De Pra}, M. and {Roig}, F. and {Lazzaro}, D. and {Gorosabel}, J. and {Lopes de Oliveira}, R. and {Lima-Neto}, G.~B. and {Irwin}, J. and {Liu}, J.~F. and {Alvarez}, E. and {Balmes}, I. and {Chueca}, S. and {Costa-Duarte}, M.~V. and {da Costa}, A.~A. and {Dantas}, M.~L.~L. and {Diaz}, A.~Y. and {Fabregat}, J. and {Ferrari}, F. and {Gavela}, B. and {Gracia}, S.~G. and {Gruel}, N. and {Gutierrez}, J.~L.~L. and {Guzman}, R. and {Hernandez-Fernandez}, J.~D. and {Herranz}, D. and {Hurtado-Gil}, L. and {Jablonsky}, F. and {Laporte}, R. and {Le Tiran}, L.~L. and {Licandro}, J and {Lima}, M. and {Martin}, E. and {Martinez}, V. and {Montero}, J.~J.~C. and {Penteado}, P. and {Pereira}, C.~B. and {Peris}, V. and {Quilis}, V. and {Sanchez-Portal}, M. and {Soja}, A.~C. and {Solano}, E. and {Torra}, J. and {Valdivielso}, L.},
        title = "{J-PAS: The Javalambre-Physics of the Accelerated Universe Astrophysical Survey}",
      journal = {arXiv e-prints},
         year = 2014,
        month = mar,
          eid = {arXiv:1403.5237},
        pages = {arXiv:1403.5237},
          doi = {10.48550/arXiv.1403.5237},
archivePrefix = {arXiv},
       eprint = {1403.5237},
 primaryClass = {astro-ph.CO},
       adsurl = {https://ui.adsabs.harvard.edu/abs/2014arXiv1403.5237B}
}

@ARTICLE{2012MNRAS.419.3590B,
       author = {{Benson}, Andrew J. and {Borgani}, Stefano and {De Lucia}, Gabriella and {Boylan-Kolchin}, Michael and {Monaco}, Pierluigi},
        title = "{Convergence of galaxy properties with merger tree temporal resolution}",
      journal = {\mnras},
         year = 2012,
        month = feb,
       volume = {419},
       number = {4},
        pages = {3590-3603},
          doi = {10.1111/j.1365-2966.2011.20002.x},
archivePrefix = {arXiv},
       eprint = {1107.4098},
 primaryClass = {astro-ph.CO},
       adsurl = {https://ui.adsabs.harvard.edu/abs/2012MNRAS.419.3590B}
}

@ARTICLE{2002AJ....123..583B,
       author = {{Bernstein}, G.~M. and {Jarvis}, M.},
        title = "{Shapes and Shears, Stars and Smears: Optimal Measurements for Weak Lensing}",
      journal = {\aj},
         year = 2002,
        month = feb,
       volume = {123},
       number = {2},
        pages = {583-618},
          doi = {10.1086/338085},
archivePrefix = {arXiv},
       eprint = {astro-ph/0107431},
 primaryClass = {astro-ph},
       adsurl = {https://ui.adsabs.harvard.edu/abs/2002AJ....123..583B}
}

@ARTICLE{2025arXiv250307923B,
       author = {{Besuner}, Robert and {Dey}, Arjun and {Drlica-Wagner}, Alex and {Ebina}, Haruki and {Fernandez Moroni}, Guillermo and {Ferraro}, Simone and {Forero-Romero}, Jaime and {Honscheid}, Klaus and {Jelinsky}, Pat and {Lang}, Dustin and {Levi}, Michael and {Martini}, Paul and {Myers}, Adam and {Palanque-Delabrouille}, Nathalie and {Panda}, Swayamtrupta and {Poppett}, Claire and {Sailer}, Noah and {Schlegel}, David and {Shafieloo}, Arman and {Silber}, Joseph and {White}, Martin and {Abbott}, Timothy and {Allen}, Lori and {Avila}, Santiago and {Avil{\'e}s}, Roberto and {Bailey}, Stephen and {Bault}, Abby and {Bouri}, Mohamed and {Boutsia}, Konstantina and {Burtin}, Eienne and {Chierchie}, Fernando and {Coulton}, William and {Dawson}, Kyle and {Dey}, Biprateep and {Dor{\'e}}, Olivier and {Dunlop}, Patrick and {Eisenstein}, Daniel and {Emanuele}, Castorina and {Escoffier}, Stephanie and {Estrada}, Juan and {Fagrelius}, Parker and {Fanning}, Kevin and {Fanning}, Timothy and {Font-Ribera}, Andreu and {Frieman}, Joshua and {Galal}, Malak and {Gluscevic}, Vera and {Gontcho}, Satya Gontcho A and {Green}, Daniel and {Gutierrez}, Gaston and {Guy}, Julien and {Hashemi}, Kevan and {Heathcote}, Steve and {Holland}, Steve and {Hou}, Jiamin and {Huterer}, Dragan and {Irigoyen Gimenez}, Blas and {Ivanov}, Mikhail and {Joyce}, Richard and {Jullo}, Eric and {Juneau}, Stephanie and {Juramy}, Claire and {Karcher}, Armin and {Kent}, Stephen and {Kirkby}, David and {Kneib}, Jean-Paul and {Krause}, Elisabeth and {Krolewski}, Alex and {Lahav}, Ofer and {Lapi}, Agustin and {Leauthaud}, Alexie and {Lewandowski}, Matthew and {Li}, Ting and {Lin}, Kenneth and {Loverde}, Marilena and {MacBride}, Sean and {Magneville}, Christophe and {Marshall}, Jennifer and {McDonald}, Patrick and {Miller}, Timothy and {Moustakas}, John and {M{\"u}nchmeyer}, Moritz and {Najita}, Joan and {Newman}, Jeff and {Percival}, Will and {Philcox}, Oliver and {Pires}, Priscila and {Raichoor}, Anand and {Roach}, Brandon and {Rockosi}, Constance and {Rombach}, Maxime and {Ross}, Ashley and {Sanchez}, Eusebio and {Schmidt}, Luke and {Schubnell}, Michael and {Sebok}, Rebekah and {Seljak}, Uros and {Silverstein}, Eva and {Slepian}, Zachay and {Stone}, Chris and {Stupak}, Robert and {Tarl{\'e}}, Gregory and {Li}, Ting and {Tyas}, Luke and {Vargas-Maga{\~n}a}, Mariana and {Walker}, Alistair and {Wenner}, Nicholas and {Y{\`e}che}, Christophe and {Zhang}, Yuanyuan and {Zhou}, Rongpu},
        title = "{The Spectroscopic Stage-5 Experiment}",
      journal = {arXiv e-prints},
         year = 2025,
        month = mar,
          eid = {arXiv:2503.07923},
        pages = {arXiv:2503.07923},
          doi = {10.48550/arXiv.2503.07923},
archivePrefix = {arXiv},
       eprint = {2503.07923},
 primaryClass = {astro-ph.CO},
       adsurl = {https://ui.adsabs.harvard.edu/abs/2025arXiv250307923B}
}

@ARTICLE{2012MNRAS.421.1671B,
       author = {{Bordoloi}, R. and {Lilly}, S.~J. and {Amara}, A. and {Oesch}, P.~A. and {Bardelli}, S. and {Zucca}, E. and {Vergani}, D. and {Nagao}, T. and {Murayama}, T. and {Shioya}, Y. and {Taniguchi}, Y.},
        title = "{Photo-z performance for precision cosmology - II. Empirical verification}",
      journal = {\mnras},
         year = 2012,
        month = apr,
       volume = {421},
       number = {2},
        pages = {1671-1677},
          doi = {10.1111/j.1365-2966.2012.20427.x},
archivePrefix = {arXiv},
       eprint = {1201.0995},
 primaryClass = {astro-ph.CO},
       adsurl = {https://ui.adsabs.harvard.edu/abs/2012MNRAS.421.1671B}
}

@ARTICLE{2023MNRAS.523.1009B,
       author = {{Bouwens}, Rychard and {Illingworth}, Garth and {Oesch}, Pascal and {Stefanon}, Mauro and {Naidu}, Rohan and {van Leeuwen}, Ivana and {Magee}, Dan},
        title = "{UV luminosity density results at z > 8 from the first JWST/NIRCam fields: limitations of early data sets and the need for spectroscopy}",
      journal = {\mnras},
         year = 2023,
        month = jul,
       volume = {523},
       number = {1},
        pages = {1009-1035},
          doi = {10.1093/mnras/stad1014},
archivePrefix = {arXiv},
       eprint = {2212.06683},
 primaryClass = {astro-ph.CO},
       adsurl = {https://ui.adsabs.harvard.edu/abs/2023MNRAS.523.1009B}
}

@ARTICLE{2008ApJ...686.1503B,
       author = {{Brammer}, Gabriel B. and {van Dokkum}, Pieter G. and {Coppi}, Paolo},
        title = "{EAZY: A Fast, Public Photometric Redshift Code}",
      journal = {\apj},
         year = 2008,
        month = oct,
       volume = {686},
       number = {2},
        pages = {1503-1513},
          doi = {10.1086/591786},
archivePrefix = {arXiv},
       eprint = {0807.1533},
 primaryClass = {astro-ph},
       adsurl = {https://ui.adsabs.harvard.edu/abs/2008ApJ...686.1503B}
}

@ARTICLE{2017A&A...608A...3B,
       author = {{Brinchmann}, J. and {Inami}, H. and {Bacon}, R. and {Contini}, T. and {Maseda}, M. and {Chevallard}, J. and {Bouch{\'e}}, N. and {Boogaard}, L. and {Carollo}, M. and {Charlot}, S. and {Kollatschny}, W. and {Marino}, R.~A. and {Pello}, R. and {Richard}, J. and {Schaye}, J. and {Verhamme}, A. and {Wisotzki}, L.},
        title = "{The MUSE Hubble Ultra Deep Field Survey. III. Testing photometric redshifts to 30th magnitude}",
      journal = {\aap},
         year = 2017,
        month = nov,
       volume = {608},
          eid = {A3},
        pages = {A3},
          doi = {10.1051/0004-6361/201731351},
archivePrefix = {arXiv},
       eprint = {1710.05062},
 primaryClass = {astro-ph.GA},
       adsurl = {https://ui.adsabs.harvard.edu/abs/2017A&A...608A...3B}
}

@ARTICLE{2019MNRAS.489..820B,
       author = {{Buchs}, R. and {Davis}, C. and {Gruen}, D. and {DeRose}, J. and {Alarcon}, A. and {Bernstein}, G.~M. and {S{\'a}nchez}, C. and {Myles}, J. and {Roodman}, A. and {Allen}, S. and {Amon}, A. and {Choi}, A. and {Masters}, D.~C. and {Miquel}, R. and {Troxel}, M.~A. and {Wechsler}, R.~H. and {Abbott}, T.~M.~C. and {Annis}, J. and {Avila}, S. and {Bechtol}, K. and {Bridle}, S.~L. and {Brooks}, D. and {Buckley-Geer}, E. and {Burke}, D.~L. and {Carnero Rosell}, A. and {Carrasco Kind}, M. and {Carretero}, J. and {Castander}, F.~J. and {Cawthon}, R. and {D'Andrea}, C.~B. and {da Costa}, L.~N. and {De Vicente}, J. and {Desai}, S. and {Diehl}, H.~T. and {Doel}, P. and {Drlica-Wagner}, A. and {Eifler}, T.~F. and {Evrard}, A.~E. and {Flaugher}, B. and {Fosalba}, P. and {Frieman}, J. and {Garc{\'\i}a-Bellido}, J. and {Gaztanaga}, E. and {Gruendl}, R.~A. and {Gschwend}, J. and {Gutierrez}, G. and {Hartley}, W.~G. and {Hollowood}, D.~L. and {Honscheid}, K. and {James}, D.~J. and {Kuehn}, K. and {Kuropatkin}, N. and {Lima}, M. and {Lin}, H. and {Maia}, M.~A.~G. and {March}, M. and {Marshall}, J.~L. and {Melchior}, P. and {Menanteau}, F. and {Ogando}, R.~L.~C. and {Plazas}, A.~A. and {Rykoff}, E.~S. and {Sanchez}, E. and {Scarpine}, V. and {Serrano}, S. and {Sevilla-Noarbe}, I. and {Smith}, M. and {Soares-Santos}, M. and {Sobreira}, F. and {Suchyta}, E. and {Swanson}, M.~E.~C. and {Tarle}, G. and {Thomas}, D. and {Vikram}, V. and {DES Collaboration}},
        title = "{Phenotypic redshifts with self-organizing maps: A novel method to characterize redshift distributions of source galaxies for weak lensing}",
      journal = {\mnras},
         year = 2019,
        month = oct,
       volume = {489},
       number = {1},
        pages = {820-841},
          doi = {10.1093/mnras/stz2162},
archivePrefix = {arXiv},
       eprint = {1901.05005},
 primaryClass = {astro-ph.CO},
       adsurl = {https://ui.adsabs.harvard.edu/abs/2019MNRAS.489..820B}
}

@INPROCEEDINGS{2024SPIE13094E..17B,
       author = {{Burgett}, W. and {Bernstein}, R. and {Ashby}, D. and {Bigelow}, B. and {Brossus}, G. and {Cox}, M. and {Demers}, R. and {Figueroa}, F. and {Fischer}, B. and {Groark}, F. and {Laskin}, R. and {Millan-Gabet}, R. and {Park}, S. and {Turner}, R. and {Walls}, B.},
        title = "{The Giant Magellan Telescope project in 2024: status and look ahead}",
    booktitle = {Ground-based and Airborne Telescopes X},
         year = 2024,
       editor = {{Marshall}, Heather K. and {Spyromilio}, Jason and {Usuda}, Tomonori},
       series = {Society of Photo-Optical Instrumentation Engineers (SPIE) Conference Series},
       volume = {13094},
        month = aug,
          eid = {1309417},
        pages = {1309417},
          doi = {10.1117/12.3020733},
       adsurl = {https://ui.adsabs.harvard.edu/abs/2024SPIE13094E..17B}
}

@ARTICLE{2024A&A...689A.166C,
       author = {{Cagliari}, M.~S. and {Granett}, B.~R. and {Guzzo}, L. and {Bethermin}, M. and {Bolzonella}, M. and {de la Torre}, S. and {Monaco}, P. and {Moresco}, M. and {Percival}, W.~J. and {Scarlata}, C. and {Wang}, Y. and {Ezziati}, M. and {Ilbert}, O. and {Le Brun}, V. and {Amara}, A. and {Andreon}, S. and {Auricchio}, N. and {Baldi}, M. and {Bardelli}, S. and {Bender}, R. and {Bodendorf}, C. and {Branchini}, E. and {Brescia}, M. and {Brinchmann}, J. and {Camera}, S. and {Capobianco}, V. and {Carbone}, C. and {Carretero}, J. and {Casas}, S. and {Castellano}, M. and {Cavuoti}, S. and {Cimatti}, A. and {Congedo}, G. and {Conselice}, C.~J. and {Conversi}, L. and {Copin}, Y. and {Corcione}, L. and {Courbin}, F. and {Courtois}, H.~M. and {Da Silva}, A. and {Degaudenzi}, H. and {Di Giorgio}, A.~M. and {Dinis}, J. and {Dubath}, F. and {Duncan}, C.~A.~J. and {Dupac}, X. and {Dusini}, S. and {Ealet}, A. and {Farina}, M. and {Farrens}, S. and {Ferriol}, S. and {Fotopoulou}, S. and {Frailis}, M. and {Franceschi}, E. and {Galeotta}, S. and {Gillis}, B. and {Giocoli}, C. and {Grazian}, A. and {Grupp}, F. and {Haugan}, S.~V.~H. and {Hoekstra}, H. and {Hook}, I. and {Hormuth}, F. and {Hornstrup}, A. and {Jahnke}, K. and {Keih{\"a}nen}, E. and {Kermiche}, S. and {Kiessling}, A. and {Kilbinger}, M. and {Kubik}, B. and {K{\"u}mmel}, M. and {Kunz}, M. and {Kurki-Suonio}, H. and {Ligori}, S. and {Lilje}, P.~B. and {Lindholm}, V. and {Lloro}, I. and {Maino}, D. and {Maiorano}, E. and {Mansutti}, O. and {Marggraf}, O. and {Markovic}, K. and {Martinet}, N. and {Marulli}, F. and {Massey}, R. and {Maurogordato}, S. and {McCracken}, H.~J. and {Medinaceli}, E. and {Mei}, S. and {Mellier}, Y. and {Meneghetti}, M. and {Merlin}, E. and {Meylan}, G. and {Moscardini}, L. and {Munari}, E. and {Nichol}, R.~C. and {Niemi}, S. -M. and {Padilla}, C. and {Paltani}, S. and {Pasian}, F. and {Pedersen}, K. and {Pettorino}, V. and {Pires}, S. and {Polenta}, G. and {Poncet}, M. and {Popa}, L.~A. and {Pozzetti}, L. and {Raison}, F. and {Rebolo}, R. and {Renzi}, A. and {Rhodes}, J. and {Riccio}, G. and {Romelli}, E. and {Roncarelli}, M. and {Rossetti}, E. and {Saglia}, R. and {Sapone}, D. and {Sartoris}, B. and {Schneider}, P. and {Scodeggio}, M. and {Secroun}, A. and {Seidel}, G. and {Seiffert}, M. and {Serrano}, S. and {Sirignano}, C. and {Sirri}, G. and {Skottfelt}, J. and {Stanco}, L. and {Surace}, C. and {Taylor}, A.~N. and {Teplitz}, H.~I. and {Tereno}, I. and {Toledo-Moreo}, R. and {Torradeflot}, F. and {Tutusaus}, I. and {Valentijn}, E.~A. and {Valenziano}, L. and {Vassallo}, T. and {Veropalumbo}, A. and {Weller}, J. and {Zamorani}, G. and {Zoubian}, J. and {Zucca}, E. and {Burigana}, C. and {Scottez}, V. and {Viel}, M. and {Bisigello}, L.},
        title = "{Euclid: Testing photometric selection of emission-line galaxy targets}",
      journal = {\aap},
         year = 2024,
        month = sep,
       volume = {689},
          eid = {A166},
        pages = {A166},
          doi = {10.1051/0004-6361/202449970},
archivePrefix = {arXiv},
       eprint = {2403.08726},
 primaryClass = {astro-ph.CO},
       adsurl = {https://ui.adsabs.harvard.edu/abs/2024A&A...689A.166C}
}

@ARTICLE{2018MNRAS.480.4379C,
       author = {{Carnall}, A.~C. and {McLure}, R.~J. and {Dunlop}, J.~S. and {Dav{\'e}}, R.},
        title = "{Inferring the star formation histories of massive quiescent galaxies with BAGPIPES: evidence for multiple quenching mechanisms}",
      journal = {\mnras},
         year = 2018,
        month = nov,
       volume = {480},
       number = {4},
        pages = {4379-4401},
          doi = {10.1093/mnras/sty2169},
archivePrefix = {arXiv},
       eprint = {1712.04452},
 primaryClass = {astro-ph.GA},
       adsurl = {https://ui.adsabs.harvard.edu/abs/2018MNRAS.480.4379C}
}

@ARTICLE{2013MNRAS.432.1483C,
       author = {{Carrasco Kind}, Matias and {Brunner}, Robert J.},
        title = "{TPZ: photometric redshift PDFs and ancillary information by using prediction trees and random forests}",
      journal = {\mnras},
         year = 2013,
        month = jun,
       volume = {432},
       number = {2},
        pages = {1483-1501},
          doi = {10.1093/mnras/stt574},
archivePrefix = {arXiv},
       eprint = {1303.7269},
 primaryClass = {astro-ph.CO},
       adsurl = {https://ui.adsabs.harvard.edu/abs/2013MNRAS.432.1483C}
}

@ARTICLE{2012A&A...546A..13C,
       author = {{Cavuoti}, S. and {Brescia}, M. and {Longo}, G. and {Mercurio}, A.},
        title = "{Photometric redshifts with the quasi Newton algorithm (MLPQNA) Results in the PHAT1 contest}",
      journal = {\aap},
         year = 2012,
        month = oct,
       volume = {546},
          eid = {A13},
        pages = {A13},
          doi = {10.1051/0004-6361/201219755},
archivePrefix = {arXiv},
       eprint = {1206.0876},
 primaryClass = {astro-ph.IM},
       adsurl = {https://ui.adsabs.harvard.edu/abs/2012A&A...546A..13C}
}

@ARTICLE{2017MNRAS.465.1959C,
       author = {{Cavuoti}, S. and {Amaro}, V. and {Brescia}, M. and {Vellucci}, C. and {Tortora}, C. and {Longo}, G.},
        title = "{METAPHOR: a machine-learning-based method for the probability density estimation of photometric redshifts}",
      journal = {\mnras},
         year = 2017,
        month = feb,
       volume = {465},
       number = {2},
        pages = {1959-1973},
          doi = {10.1093/mnras/stw2930},
archivePrefix = {arXiv},
       eprint = {1611.02162},
 primaryClass = {astro-ph.IM},
       adsurl = {https://ui.adsabs.harvard.edu/abs/2017MNRAS.465.1959C}
}

@ARTICLE{2013MNRAS.434.2121C,
       author = {{Chang}, C. and {Jarvis}, M. and {Jain}, B. and {Kahn}, S.~M. and {Kirkby}, D. and {Connolly}, A. and {Krughoff}, S. and {Peng}, E. -H. and {Peterson}, J.~R.},
        title = "{The effective number density of galaxies for weak lensing measurements in the LSST project}",
      journal = {\mnras},
         year = 2013,
        month = sep,
       volume = {434},
       number = {3},
        pages = {2121-2135},
          doi = {10.1093/mnras/stt1156},
archivePrefix = {arXiv},
       eprint = {1305.0793},
 primaryClass = {astro-ph.CO},
       adsurl = {https://ui.adsabs.harvard.edu/abs/2013MNRAS.434.2121C}
}

@ARTICLE{2016arXiv160302754C,
       author = {{Chen}, Tianqi and {Guestrin}, Carlos},
        title = "{XGBoost: A Scalable Tree Boosting System}",
      journal = {arXiv e-prints},
         year = 2016,
        month = mar,
          eid = {arXiv:1603.02754},
        pages = {arXiv:1603.02754},
          doi = {10.48550/arXiv.1603.02754},
archivePrefix = {arXiv},
       eprint = {1603.02754},
 primaryClass = {cs.LG},
       adsurl = {https://ui.adsabs.harvard.edu/abs/2016arXiv160302754C}
}

@ARTICLE{2006AJ....132..926C,
       author = {{Coe}, Dan and {Ben{\'\i}tez}, Narciso and {S{\'a}nchez}, Sebasti{\'a}n F. and {Jee}, Myungkook and {Bouwens}, Rychard and {Ford}, Holland},
        title = "{Galaxies in the Hubble Ultra Deep Field. I. Detection, Multiband Photometry, Photometric Redshifts, and Morphology}",
      journal = {\aj},
         year = 2006,
        month = aug,
       volume = {132},
       number = {2},
        pages = {926-959},
          doi = {10.1086/505530},
archivePrefix = {arXiv},
       eprint = {astro-ph/0605262},
 primaryClass = {astro-ph},
       adsurl = {https://ui.adsabs.harvard.edu/abs/2006AJ....132..926C}
}

@ARTICLE{2001MNRAS.328.1039C,
       author = {{Colless}, Matthew and {Dalton}, Gavin and {Maddox}, Steve and {Sutherland}, Will and {Norberg}, Peder and {Cole}, Shaun and {Bland-Hawthorn}, Joss and {Bridges}, Terry and {Cannon}, Russell and {Collins}, Chris and {Couch}, Warrick and {Cross}, Nicholas and {Deeley}, Kathryn and {De Propris}, Roberto and {Driver}, Simon P. and {Efstathiou}, George and {Ellis}, Richard S. and {Frenk}, Carlos S. and {Glazebrook}, Karl and {Jackson}, Carole and {Lahav}, Ofer and {Lewis}, Ian and {Lumsden}, Stuart and {Madgwick}, Darren and {Peacock}, John A. and {Peterson}, Bruce A. and {Price}, Ian and {Seaborne}, Mark and {Taylor}, Keith},
        title = "{The 2dF Galaxy Redshift Survey: spectra and redshifts}",
      journal = {\mnras},
         year = 2001,
        month = dec,
       volume = {328},
       number = {4},
        pages = {1039-1063},
          doi = {10.1046/j.1365-8711.2001.04902.x},
archivePrefix = {arXiv},
       eprint = {astro-ph/0106498},
 primaryClass = {astro-ph},
       adsurl = {https://ui.adsabs.harvard.edu/abs/2001MNRAS.328.1039C}
}

@ARTICLE{2004PASP..116..345C,
       author = {{Collister}, Adrian A. and {Lahav}, Ofer},
        title = "{ANNz: Estimating Photometric Redshifts Using Artificial Neural Networks}",
      journal = {\pasp},
         year = 2004,
        month = apr,
       volume = {116},
       number = {818},
        pages = {345-351},
          doi = {10.1086/383254},
archivePrefix = {arXiv},
       eprint = {astro-ph/0311058},
 primaryClass = {astro-ph},
       adsurl = {https://ui.adsabs.harvard.edu/abs/2004PASP..116..345C}
}

@INPROCEEDINGS{2014SPIE.9150E..14C,
       author = {{Connolly}, Andrew J. and {Angeli}, George Z. and {Chandrasekharan}, Srinivasan and {Claver}, Charles F. and {Cook}, Kem and {Ivezic}, Zeljko and {Jones}, R. Lynne and {Krughoff}, K. Simon and {Peng}, En-Hsin and {Peterson}, John and {Petry}, Catherine and {Rasmussen}, Andrew P. and {Ridgway}, Stephen T. and {Saha}, Abhijit and {Sembroski}, Glenn and {vanderPlas}, Jacob and {Yoachim}, Peter},
        title = "{An end-to-end simulation framework for the Large Synoptic Survey Telescope}",
    booktitle = {Modeling, Systems Engineering, and Project Management for Astronomy VI},
         year = 2014,
       editor = {{Angeli}, George Z. and {Dierickx}, Philippe},
       series = {Society of Photo-Optical Instrumentation Engineers (SPIE) Conference Series},
       volume = {9150},
        month = aug,
          eid = {915014},
        pages = {915014},
          doi = {10.1117/12.2054953},
       adsurl = {https://ui.adsabs.harvard.edu/abs/2014SPIE.9150E..14C}
}

@ARTICLE{2009ApJ...699..486C,
       author = {{Conroy}, Charlie and {Gunn}, James E. and {White}, Martin},
        title = "{The Propagation of Uncertainties in Stellar Population Synthesis Modeling. I. The Relevance of Uncertain Aspects of Stellar Evolution and the Initial Mass Function to the Derived Physical Properties of Galaxies}",
      journal = {\apj},
         year = 2009,
        month = jul,
       volume = {699},
       number = {1},
        pages = {486-506},
          doi = {10.1088/0004-637X/699/1/486},
archivePrefix = {arXiv},
       eprint = {0809.4261},
 primaryClass = {astro-ph},
       adsurl = {https://ui.adsabs.harvard.edu/abs/2009ApJ...699..486C}
}

@ARTICLE{2010ApJ...708...58C,
       author = {{Conroy}, Charlie and {White}, Martin and {Gunn}, James E.},
        title = "{The Propagation of Uncertainties in Stellar Population Synthesis Modeling. II. The Challenge of Comparing Galaxy Evolution Models to Observations}",
      journal = {\apj},
         year = 2010,
        month = jan,
       volume = {708},
       number = {1},
        pages = {58-70},
          doi = {10.1088/0004-637X/708/1/58},
archivePrefix = {arXiv},
       eprint = {0904.0002},
 primaryClass = {astro-ph.CO},
       adsurl = {https://ui.adsabs.harvard.edu/abs/2010ApJ...708...58C}
}

@ARTICLE{2010ApJ...712..833C,
       author = {{Conroy}, Charlie and {Gunn}, James E.},
        title = "{The Propagation of Uncertainties in Stellar Population Synthesis Modeling. III. Model Calibration, Comparison, and Evaluation}",
      journal = {\apj},
         year = 2010,
        month = apr,
       volume = {712},
       number = {2},
        pages = {833-857},
          doi = {10.1088/0004-637X/712/2/833},
archivePrefix = {arXiv},
       eprint = {0911.3151},
 primaryClass = {astro-ph.CO},
       adsurl = {https://ui.adsabs.harvard.edu/abs/2010ApJ...712..833C}
}

@ARTICLE{2014ARA&A..52..291C,
       author = {{Conselice}, Christopher J.},
        title = "{The Evolution of Galaxy Structure Over Cosmic Time}",
      journal = {\araa},
         year = 2014,
        month = aug,
       volume = {52},
        pages = {291-337},
          doi = {10.1146/annurev-astro-081913-040037},
archivePrefix = {arXiv},
       eprint = {1403.2783},
 primaryClass = {astro-ph.GA},
       adsurl = {https://ui.adsabs.harvard.edu/abs/2014ARA&A..52..291C}
}

@ARTICLE{2024AJ....168...80C,
       author = {{Crenshaw}, John Franklin and {Kalmbach}, J. Bryce and {Gagliano}, Alexander and {Yan}, Ziang and {Connolly}, Andrew J. and {Malz}, Alex I. and {Schmidt}, Samuel J. and {The LSST Dark Energy Science Collaboration}},
        title = "{Probabilistic Forward Modeling of Galaxy Catalogs with Normalizing Flows}",
      journal = {\aj},
         year = 2024,
        month = aug,
       volume = {168},
       number = {2},
          eid = {80},
        pages = {80},
          doi = {10.3847/1538-3881/ad54bf},
archivePrefix = {arXiv},
       eprint = {2405.04740},
 primaryClass = {astro-ph.IM},
       adsurl = {https://ui.adsabs.harvard.edu/abs/2024AJ....168...80C}
}

@ARTICLE{2013ApJ...775...93D,
       author = {{Dahlen}, Tomas and {Mobasher}, Bahram and {Faber}, Sandra M. and {Ferguson}, Henry C. and {Barro}, Guillermo and {Finkelstein}, Steven L. and {Finlator}, Kristian and {Fontana}, Adriano and {Gruetzbauch}, Ruth and {Johnson}, Seth and {Pforr}, Janine and {Salvato}, Mara and {Wiklind}, Tommy and {Wuyts}, Stijn and {Acquaviva}, Viviana and {Dickinson}, Mark E. and {Guo}, Yicheng and {Huang}, Jiasheng and {Huang}, Kuang-Han and {Newman}, Jeffrey A. and {Bell}, Eric F. and {Conselice}, Christopher J. and {Galametz}, Audrey and {Gawiser}, Eric and {Giavalisco}, Mauro and {Grogin}, Norman A. and {Hathi}, Nimish and {Kocevski}, Dale and {Koekemoer}, Anton M. and {Koo}, David C. and {Lee}, Kyoung-Soo and {McGrath}, Elizabeth J. and {Papovich}, Casey and {Peth}, Michael and {Ryan}, Russell and {Somerville}, Rachel and {Weiner}, Benjamin and {Wilson}, Grant},
        title = "{A Critical Assessment of Photometric Redshift Methods: A CANDELS Investigation}",
      journal = {\apj},
         year = 2013,
        month = oct,
       volume = {775},
       number = {2},
          eid = {93},
        pages = {93},
          doi = {10.1088/0004-637X/775/2/93},
archivePrefix = {arXiv},
       eprint = {1308.5353},
 primaryClass = {astro-ph.CO},
       adsurl = {https://ui.adsabs.harvard.edu/abs/2013ApJ...775...93D}
}

@ARTICLE{2023PhRvD.108l3519D,
       author = {{Dalal}, Roohi and {Li}, Xiangchong and {Nicola}, Andrina and {Zuntz}, Joe and {Strauss}, Michael A. and {Sugiyama}, Sunao and {Zhang}, Tianqing and {Rau}, Markus M. and {Mandelbaum}, Rachel and {Takada}, Masahiro and {More}, Surhud and {Miyatake}, Hironao and {Kannawadi}, Arun and {Shirasaki}, Masato and {Taniguchi}, Takanori and {Takahashi}, Ryuichi and {Osato}, Ken and {Hamana}, Takashi and {Oguri}, Masamune and {Nishizawa}, Atsushi J. and {Malag{\'o}n}, Andr{\'e}s A. Plazas and {Sunayama}, Tomomi and {Alonso}, David and {Slosar}, An{\v{z}}e and {Luo}, Wentao and {Armstrong}, Robert and {Bosch}, James and {Hsieh}, Bau-Ching and {Komiyama}, Yutaka and {Lupton}, Robert H. and {Lust}, Nate B. and {MacArthur}, Lauren A. and {Miyazaki}, Satoshi and {Murayama}, Hitoshi and {Nishimichi}, Takahiro and {Okura}, Yuki and {Price}, Paul A. and {Tait}, Philip J. and {Tanaka}, Masayuki and {Wang}, Shiang-Yu},
        title = "{Hyper Suprime-Cam Year 3 results: Cosmology from cosmic shear power spectra}",
      journal = {\prd},
         year = 2023,
        month = dec,
       volume = {108},
       number = {12},
          eid = {123519},
        pages = {123519},
          doi = {10.1103/PhysRevD.108.123519},
archivePrefix = {arXiv},
       eprint = {2304.00701},
 primaryClass = {astro-ph.CO},
       adsurl = {https://ui.adsabs.harvard.edu/abs/2023PhRvD.108l3519D}
}

@ARTICLE{2020A&C....3000362D,
       author = {{Dalmasso}, N. and {Pospisil}, T. and {Lee}, A.~B. and {Izbicki}, R. and {Freeman}, P.~E. and {Malz}, A.~I.},
        title = "{Conditional density estimation tools in python and R with applications to photometric redshifts and likelihood-free cosmological inference}",
      journal = {Astronomy and Computing},
         year = 2020,
        month = jan,
       volume = {30},
          eid = {100362},
        pages = {100362},
          doi = {10.1016/j.ascom.2019.100362},
archivePrefix = {arXiv},
       eprint = {1908.11523},
 primaryClass = {astro-ph.IM},
       adsurl = {https://ui.adsabs.harvard.edu/abs/2020A&C....3000362D}
}

@ARTICLE{2023OJAp....6E..36D,
       author = {{Dark Energy Survey and Kilo-Degree Survey Collaboration} and {Abbott}, T.~M.~C. and {Aguena}, M. and {Alarcon}, A. and {Alves}, O. and {Amon}, A. and {Andrade-Oliveira}, F. and {Asgari}, M. and {Avila}, S. and {Bacon}, D. and {Bechtol}, K. and {Becker}, M.~R. and {Bernstein}, G.~M. and {Bertin}, E. and {Bilicki}, M. and {Blazek}, J. and {Bocquet}, S. and {Brooks}, D. and {Burger}, P. and {Burke}, D.~L. and {Camacho}, H. and {Campos}, A. and {Carnero Rosell}, A. and {Carrasco Kind}, M. and {Carretero}, J. and {Castander}, F.~J. and {Cawthon}, R. and {Chang}, C. and {Chen}, R. and {Choi}, A. and {Conselice}, C. and {Cordero}, J. and {Crocce}, M. and {da Costa}, L.~N. and {da Silva Pereira}, M.~E. and {Dalal}, R. and {Davis}, C. and {de Jong}, J.~T.~A. and {DeRose}, J. and {Desai}, S. and {Diehl}, H.~T. and {Dodelson}, S. and {Doel}, P. and {Doux}, C. and {Drlica-Wagner}, A. and {Dvornik}, A. and {Eckert}, K. and {Eifler}, T.~F. and {Elvin-Poole}, J. and {Everett}, S. and {Fang}, X. and {Ferrero}, I. and {Fert{\'e}}, A. and {Flaugher}, B. and {Friedrich}, O. and {Frieman}, J. and {Garc{\'\i}a-Bellido}, J. and {Gatti}, M. and {Giannini}, G. and {Giblin}, B. and {Gruen}, D. and {Gruendl}, R.~A. and {Gutierrez}, G. and {Harrison}, I. and {Hartley}, W.~G. and {Herner}, K. and {Heymans}, C. and {Hildebrandt}, H. and {Hinton}, S.~R. and {Hoekstra}, H. and {Hollowood}, D.~L. and {Honscheid}, K. and {Huang}, H. and {Huff}, E.~M. and {Huterer}, D. and {James}, D.~J. and {Jarvis}, M. and {Jeffrey}, N. and {Jeltema}, T. and {Joachimi}, B. and {Joudaki}, S. and {Kannawadi}, A. and {Krause}, E. and {Kuehn}, K. and {Kuijken}, K. and {Kuropatkin}, N. and {Lahav}, O. and {Leget}, P. -F. and {Lemos}, P. and {Li}, S. -S. and {Li}, X. and {Liddle}, A.~R. and {Lima}, M. and {Lin}, C. -A. and {Lin}, H. and {MacCrann}, N. and {Mahony}, C. and {Marshall}, J.~L. and {McCullough}, J. and {Mena-Fern{\'a}ndez}, J. and {Menanteau}, F. and {Miquel}, R. and {Mohr}, J.~J. and {Muir}, J. and {Myles}, J. and {Napolitano}, N. and {Navarro-Alsina}, A. and {Ogando}, R.~L.~C. and {Palmese}, A. and {Pandey}, S. and {Park}, Y. and {Paterno}, M. and {Peacock}, J.~A. and {Petravick}, D. and {Pieres}, A. and {Plazas Malag{\'o}n}, A.~A. and {Porredon}, A. and {Prat}, J. and {Radovich}, M. and {Raveri}, M. and {Reischke}, R. and {Robertson}, N.~C. and {Rollins}, R.~P. and {Romer}, A.~K. and {Roodman}, A. and {Rykoff}, E.~S. and {Samuroff}, S. and {S{\'a}nchez}, C. and {Sanchez}, E. and {Sanchez}, J. and {Schneider}, P. and {Secco}, L.~F. and {Sevilla-Noarbe}, I. and {Shan}, H. -Y. and {Sheldon}, E. and {Shin}, T. and {Sif{\'o}n}, C. and {Smith}, M. and {Soares-Santos}, M. and {St{\"o}lzner}, B. and {Suchyta}, E. and {Swanson}, M.~E.~C. and {Tarle}, G. and {Thomas}, D. and {To}, C. and {Troxel}, M.~A. and {Tr{\"o}ster}, T. and {Tutusaus}, I. and {van den Busch}, J.~L. and {Varga}, T.~N. and {Walker}, A.~R. and {Weaverdyck}, N. and {Wechsler}, R.~H. and {Weller}, J. and {Wiseman}, P. and {Wright}, A.~H. and {Yanny}, B. and {Yin}, B. and {Yoon}, M. and {Zhang}, Y. and {Zuntz}, J.},
        title = "{DES Y3 + KiDS-1000: Consistent cosmology combining cosmic shear surveys}",
      journal = {The Open Journal of Astrophysics},
         year = 2023,
        month = oct,
       volume = {6},
          eid = {36},
        pages = {36},
          doi = {10.21105/astro.2305.17173},
archivePrefix = {arXiv},
       eprint = {2305.17173},
 primaryClass = {astro-ph.CO},
       adsurl = {https://ui.adsabs.harvard.edu/abs/2023OJAp....6E..36D}
}

@ARTICLE{2017A&A...605A..70D,
       author = {{Davidzon}, I. and {Ilbert}, O. and {Laigle}, C. and {Coupon}, J. and {McCracken}, H.~J. and {Delvecchio}, I. and {Masters}, D. and {Capak}, P. and {Hsieh}, B.~C. and {Le F{\`e}vre}, O. and {Tresse}, L. and {Bethermin}, M. and {Chang}, Y. -Y. and {Faisst}, A.~L. and {Le Floc'h}, E. and {Steinhardt}, C. and {Toft}, S. and {Aussel}, H. and {Dubois}, C. and {Hasinger}, G. and {Salvato}, M. and {Sanders}, D.~B. and {Scoville}, N. and {Silverman}, J.~D.},
        title = "{The COSMOS2015 galaxy stellar mass function . Thirteen billion years of stellar mass assembly in ten snapshots}",
      journal = {\aap},
         year = 2017,
        month = sep,
       volume = {605},
          eid = {A70},
        pages = {A70},
          doi = {10.1051/0004-6361/201730419},
archivePrefix = {arXiv},
       eprint = {1701.02734},
 primaryClass = {astro-ph.GA},
       adsurl = {https://ui.adsabs.harvard.edu/abs/2017A&A...605A..70D}
}

@ARTICLE{2025A&A...693A.102D,
       author = {{Daza-Perilla}, I.~V. and {Eriksen}, M. and {Navarro-Giron{\'e}s}, D. and {Gonzalez}, E.~J. and {Rodriguez}, F. and {Gazta{\~n}aga}, E. and {Baugh}, C.~M. and {Lares}, M. and {Cabayol-Garcia}, L. and {Castander}, F.~J. and {Siudek}, M. and {Wittje}, A. and {Hildebrandt}, H. and {Casas}, R. and {Tallada-Cresp{\'\i}}, P. and {Garcia-Bellido}, J. and {Sanchez}, E. and {Sevilla-Noarbe}, I. and {Miquel}, R. and {Padilla}, C. and {Renard}, P. and {Carretero}, J. and {De Vicente}, J.},
        title = "{The PAU survey: Enhancing photometric redshift estimation using DEEPz}",
      journal = {\aap},
         year = 2025,
        month = jan,
       volume = {693},
          eid = {A102},
        pages = {A102},
          doi = {10.1051/0004-6361/202452053},
archivePrefix = {arXiv},
       eprint = {2408.16864},
 primaryClass = {astro-ph.GA},
       adsurl = {https://ui.adsabs.harvard.edu/abs/2025A&A...693A.102D}
}

@ARTICLE{2024AJ....167...62D,
       author = {{DESI Collaboration} and {Adame}, A.~G. and {Aguilar}, J. and {Ahlen}, S. and {Alam}, S. and {Aldering}, G. and {Alexander}, D.~M. and {Alfarsy}, R. and {Allende Prieto}, C. and {Alvarez}, M. and {Alves}, O. and {Anand}, A. and {Andrade-Oliveira}, F. and {Armengaud}, E. and {Asorey}, J. and {Avila}, S. and {Aviles}, A. and {Bailey}, S. and {Balaguera-Antol{\'\i}nez}, A. and {Ballester}, O. and {Baltay}, C. and {Bault}, A. and {Bautista}, J. and {Behera}, J. and {Beltran}, S.~F. and {BenZvi}, S. and {Beraldo e Silva}, L. and {Bermejo-Climent}, J.~R. and {Berti}, A. and {Besuner}, R. and {Beutler}, F. and {Bianchi}, D. and {Blake}, C. and {Blum}, R. and {Bolton}, A.~S. and {Brieden}, S. and {Brodzeller}, A. and {Brooks}, D. and {Brown}, Z. and {Buckley-Geer}, E. and {Burtin}, E. and {Cabayol-Garcia}, L. and {Cai}, Z. and {Canning}, R. and {Cardiel-Sas}, L. and {Carnero Rosell}, A. and {Castander}, F.~J. and {Cervantes-Cota}, J.~L. and {Chabanier}, S. and {Chaussidon}, E. and {Chaves-Montero}, J. and {Chen}, S. and {Chen}, X. and {Chuang}, C. and {Claybaugh}, T. and {Cole}, S. and {Cooper}, A.~P. and {Cuceu}, A. and {Davis}, T.~M. and {Dawson}, K. and {de Belsunce}, R. and {de la Cruz}, R. and {de la Macorra}, A. and {de Mattia}, A. and {Demina}, R. and {Demirbozan}, U. and {DeRose}, J. and {Dey}, A. and {Dey}, B. and {Dhungana}, G. and {Ding}, J. and {Ding}, Z. and {Doel}, P. and {Doshi}, R. and {Douglass}, K. and {Edge}, A. and {Eftekharzadeh}, S. and {Eisenstein}, D.~J. and {Elliott}, A. and {Escoffier}, S. and {Fagrelius}, P. and {Fan}, X. and {Fanning}, K. and {Fawcett}, V.~A. and {Ferraro}, S. and {Ereza}, J. and {Flaugher}, B. and {Font-Ribera}, A. and {Forero-S{\'a}nchez}, D. and {Forero-Romero}, J.~E. and {Frenk}, C.~S. and {G{\"a}nsicke}, B.~T. and {Garc{\'\i}a}, L. {\'A}. and {Garc{\'\i}a-Bellido}, J. and {Garcia-Quintero}, C. and {Garrison}, L.~H. and {Gil-Mar{\'\i}n}, H. and {Golden-Marx}, J. and {Gontcho A Gontcho}, S. and {Gonzalez-Morales}, A.~X. and {Gonzalez-Perez}, V. and {Gordon}, C. and {Graur}, O. and {Green}, D. and {Gruen}, D. and {Guy}, J. and {Hadzhiyska}, B. and {Hahn}, C. and {Han}, J.~J. and {Hanif}, M.~M.~S. and {Herrera-Alcantar}, H.~K. and {Honscheid}, K. and {Hou}, J. and {Howlett}, C. and {Huterer}, D. and {Ir{\v{s}}i{\v{c}}}, V. and {Ishak}, M. and {Jana}, A. and {Jiang}, L. and {Jimenez}, J. and {Jing}, Y.~P. and {Joudaki}, S. and {Jullo}, E. and {Joyce}, R. and {Juneau}, S. and {Kizhuprakkat}, N. and {Kara{\c{c}}ayl{\i}}, N.~G. and {Karim}, T. and {Kehoe}, R. and {Kent}, S. and {Khederlarian}, A. and {Kim}, S. and {Kirkby}, D. and {Kisner}, T. and {Kitaura}, F. and {Kneib}, J. and {Koposov}, S.~E. and {Kov{\'a}cs}, A. and {Kremin}, A. and {Krolewski}, A. and {L'Huillier}, B. and {Lahav}, O. and {Lambert}, A. and {Lamman}, C. and {Lan}, T. -W. and {Landriau}, M. and {Lang}, D. and {Lange}, J.~U. and {Lasker}, J. and {Le Guillou}, L. and {Leauthaud}, A. and {Levi}, M.~E. and {Li}, T.~S. and {Linder}, E. and {Lyons}, A. and {Magneville}, C. and {Manera}, M. and {Manser}, C.~J. and {Margala}, D. and {Martini}, P. and {McDonald}, P. and {Medina}, G.~E. and {Medina-Varela}, L. and {Meisner}, A. and {Mena-Fern{\'a}ndez}, J. and {Meneses-Rizo}, J. and {Mezcua}, M. and {Miquel}, R. and {Montero-Camacho}, P. and {Moon}, J. and {Moore}, S. and {Moustakas}, J. and {Mueller}, E. and {Mundet}, J. and {Mu{\~n}oz-Guti{\'e}rrez}, A. and {Myers}, A.~D. and {Nadathur}, S. and {Napolitano}, L. and {Neveux}, R. and {Newman}, J.~A. and {Nie}, J. and {Niz}, G. and {Norberg}, P. and {Noriega}, H.~E. and {Paillas}, E. and {Palanque-Delabrouille}, N. and {Palmese}, A. and {Zhiwei}, P. and {Parkinson}, D. and {Penmetsa}, S. and {Percival}, W.~J. and {P{\'e}rez-Fern{\'a}ndez}, A. and {P{\'e}rez-R{\`a}fols}, I. and {Pieri}, M. and {Poppett}, C. and {Porredon}, A. and {Prada}, F. and {Pucha}, R. and {Raichoor}, A. and {Ram{\'\i}rez-P{\'e}rez}, C.},
        title = "{Validation of the Scientific Program for the Dark Energy Spectroscopic Instrument}",
      journal = {\aj},
         year = 2024,
        month = feb,
       volume = {167},
       number = {2},
          eid = {62},
        pages = {62},
          doi = {10.3847/1538-3881/ad0b08},
archivePrefix = {arXiv},
       eprint = {2306.06307},
 primaryClass = {astro-ph.CO},
       adsurl = {https://ui.adsabs.harvard.edu/abs/2024AJ....167...62D}
}

@ARTICLE{2019A&A...622A...3D,
       author = {{Duncan}, K.~J. and {Sabater}, J. and {R{\"o}ttgering}, H.~J.~A. and {Jarvis}, M.~J. and {Smith}, D.~J.~B. and {Best}, P.~N. and {Callingham}, J.~R. and {Cochrane}, R. and {Croston}, J.~H. and {Hardcastle}, M.~J. and {Mingo}, B. and {Morabito}, L. and {Nisbet}, D. and {Prandoni}, I. and {Shimwell}, T.~W. and {Tasse}, C. and {White}, G.~J. and {Williams}, W.~L. and {Alegre}, L. and {Chy{\.z}y}, K.~T. and {G{\"u}rkan}, G. and {Hoeft}, M. and {Kondapally}, R. and {Mechev}, A.~P. and {Miley}, G.~K. and {Schwarz}, D.~J. and {van Weeren}, R.~J.},
        title = "{The LOFAR Two-metre Sky Survey. IV. First Data Release: Photometric redshifts and rest-frame magnitudes}",
      journal = {\aap},
         year = 2019,
        month = feb,
       volume = {622},
          eid = {A3},
        pages = {A3},
          doi = {10.1051/0004-6361/201833562},
archivePrefix = {arXiv},
       eprint = {1811.07928},
 primaryClass = {astro-ph.GA},
       adsurl = {https://ui.adsabs.harvard.edu/abs/2019A&A...622A...3D}
}

@ARTICLE{2023A&A...675A.189D,
       author = {{Dvornik}, Andrej and {Heymans}, Catherine and {Asgari}, Marika and {Mahony}, Constance and {Joachimi}, Benjamin and {Bilicki}, Maciej and {Chisari}, Elisa and {Hildebrandt}, Hendrik and {Hoekstra}, Henk and {Johnston}, Harry and {Kuijken}, Konrad and {Mead}, Alexander and {Miyatake}, Hironao and {Nishimichi}, Takahiro and {Reischke}, Robert and {Unruh}, Sandra and {Wright}, Angus H.},
        title = "{KiDS-1000: Combined halo-model cosmology constraints from galaxy abundance, galaxy clustering, and galaxy-galaxy lensing}",
      journal = {\aap},
         year = 2023,
        month = jul,
       volume = {675},
          eid = {A189},
        pages = {A189},
          doi = {10.1051/0004-6361/202245158},
archivePrefix = {arXiv},
       eprint = {2210.03110},
 primaryClass = {astro-ph.CO},
       adsurl = {https://ui.adsabs.harvard.edu/abs/2023A&A...675A.189D}
}

@ARTICLE{2011MNRAS.413..971D,
       author = {{Driver}, S.~P. and {Hill}, D.~T. and {Kelvin}, L.~S. and {Robotham}, A.~S.~G. and {Liske}, J. and {Norberg}, P. and {Baldry}, I.~K. and {Bamford}, S.~P. and {Hopkins}, A.~M. and {Loveday}, J. and {Peacock}, J.~A. and {Andrae}, E. and {Bland-Hawthorn}, J. and {Brough}, S. and {Brown}, M.~J.~I. and {Cameron}, E. and {Ching}, J.~H.~Y. and {Colless}, M. and {Conselice}, C.~J. and {Croom}, S.~M. and {Cross}, N.~J.~G. and {de Propris}, R. and {Dye}, S. and {Drinkwater}, M.~J. and {Ellis}, S. and {Graham}, Alister W. and {Grootes}, M.~W. and {Gunawardhana}, M. and {Jones}, D.~H. and {van Kampen}, E. and {Maraston}, C. and {Nichol}, R.~C. and {Parkinson}, H.~R. and {Phillipps}, S. and {Pimbblet}, K. and {Popescu}, C.~C. and {Prescott}, M. and {Roseboom}, I.~G. and {Sadler}, E.~M. and {Sansom}, A.~E. and {Sharp}, R.~G. and {Smith}, D.~J.~B. and {Taylor}, E. and {Thomas}, D. and {Tuffs}, R.~J. and {Wijesinghe}, D. and {Dunne}, L. and {Frenk}, C.~S. and {Jarvis}, M.~J. and {Madore}, B.~F. and {Meyer}, M.~J. and {Seibert}, M. and {Staveley-Smith}, L. and {Sutherland}, W.~J. and {Warren}, S.~J.},
        title = "{Galaxy and Mass Assembly (GAMA): survey diagnostics and core data release}",
      journal = {\mnras},
         year = 2011,
        month = may,
       volume = {413},
       number = {2},
        pages = {971-995},
          doi = {10.1111/j.1365-2966.2010.18188.x},
archivePrefix = {arXiv},
       eprint = {1009.0614},
 primaryClass = {astro-ph.CO},
       adsurl = {https://ui.adsabs.harvard.edu/abs/2011MNRAS.413..971D}
}

@ARTICLE{2005ApJ...633..560E,
       author = {{Eisenstein}, Daniel J. and {Zehavi}, Idit and {Hogg}, David W. and {Scoccimarro}, Roman and {Blanton}, Michael R. and {Nichol}, Robert C. and {Scranton}, Ryan and {Seo}, Hee-Jong and {Tegmark}, Max and {Zheng}, Zheng and {Anderson}, Scott F. and {Annis}, Jim and {Bahcall}, Neta and {Brinkmann}, Jon and {Burles}, Scott and {Castander}, Francisco J. and {Connolly}, Andrew and {Csabai}, Istvan and {Doi}, Mamoru and {Fukugita}, Masataka and {Frieman}, Joshua A. and {Glazebrook}, Karl and {Gunn}, James E. and {Hendry}, John S. and {Hennessy}, Gregory and {Ivezi{\'c}}, Zeljko and {Kent}, Stephen and {Knapp}, Gillian R. and {Lin}, Huan and {Loh}, Yeong-Shang and {Lupton}, Robert H. and {Margon}, Bruce and {McKay}, Timothy A. and {Meiksin}, Avery and {Munn}, Jeffery A. and {Pope}, Adrian and {Richmond}, Michael W. and {Schlegel}, David and {Schneider}, Donald P. and {Shimasaku}, Kazuhiro and {Stoughton}, Christopher and {Strauss}, Michael A. and {SubbaRao}, Mark and {Szalay}, Alexander S. and {Szapudi}, Istv{\'a}n and {Tucker}, Douglas L. and {Yanny}, Brian and {York}, Donald G.},
        title = "{Detection of the Baryon Acoustic Peak in the Large-Scale Correlation Function of SDSS Luminous Red Galaxies}",
      journal = {\apj},
         year = 2005,
        month = nov,
       volume = {633},
       number = {2},
        pages = {560-574},
          doi = {10.1086/466512},
archivePrefix = {arXiv},
       eprint = {astro-ph/0501171},
 primaryClass = {astro-ph},
       adsurl = {https://ui.adsabs.harvard.edu/abs/2005ApJ...633..560E}
}

@ARTICLE{2025A&A...697A...1E,
       author = {{Euclid Collaboration} and {Mellier}, Y. and {Abdurro'uf} and {Acevedo Barroso}, J.~A. and {Ach{\'u}carro}, A. and {Adamek}, J. and {Adam}, R. and {Addison}, G.~E. and {Aghanim}, N. and {Aguena}, M. and {Ajani}, V. and {Akrami}, Y. and {Al-Bahlawan}, A. and {Alavi}, A. and {Albuquerque}, I.~S. and {Alestas}, G. and {Alguero}, G. and {Allaoui}, A. and {Allen}, S.~W. and {Allevato}, V. and {Alonso-Tetilla}, A.~V. and {Altieri}, B. and {Alvarez-Candal}, A. and {Alvi}, S. and {Amara}, A. and {Amendola}, L. and {Amiaux}, J. and {Andika}, I.~T. and {Andreon}, S. and {Andrews}, A. and {Angora}, G. and {Angulo}, R.~E. and {Annibali}, F. and {Anselmi}, A. and {Anselmi}, S. and {Arcari}, S. and {Archidiacono}, M. and {Aric{\`o}}, G. and {Arnaud}, M. and {Arnouts}, S. and {Asgari}, M. and {Asorey}, J. and {Atayde}, L. and {Atek}, H. and {Atrio-Barandela}, F. and {Aubert}, M. and {Aubourg}, E. and {Auphan}, T. and {Auricchio}, N. and {Aussel}, B. and {Aussel}, H. and {Avelino}, P.~P. and {Avgoustidis}, A. and {Avila}, S. and {Awan}, S. and {Azzollini}, R. and {Baccigalupi}, C. and {Bachelet}, E. and {Bacon}, D. and {Baes}, M. and {Bagley}, M.~B. and {Bahr-Kalus}, B. and {Balaguera-Antolinez}, A. and {Balbinot}, E. and {Balcells}, M. and {Baldi}, M. and {Baldry}, I. and {Balestra}, A. and {Ballardini}, M. and {Ballester}, O. and {Balogh}, M. and {Ba{\~n}ados}, E. and {Barbier}, R. and {Bardelli}, S. and {Baron}, M. and {Barreiro}, T. and {Barrena}, R. and {Barriere}, J. -C. and {Barros}, B.~J. and {Barthelemy}, A. and {Bartolo}, N. and {Basset}, A. and {Battaglia}, P. and {Battisti}, A.~J. and {Baugh}, C.~M. and {Baumont}, L. and {Bazzanini}, L. and {Beaulieu}, J. -P. and {Beckmann}, V. and {Belikov}, A.~N. and {Bel}, J. and {Bellagamba}, F. and {Bella}, M. and {Bellini}, E. and {Benabed}, K. and {Bender}, R. and {Benevento}, G. and {Bennett}, C.~L. and {Benson}, K. and {Bergamini}, P. and {Bermejo-Climent}, J.~R. and {Bernardeau}, F. and {Bertacca}, D. and {Berthe}, M. and {Berthier}, J. and {Bethermin}, M. and {Beutler}, F. and {Bevillon}, C. and {Bhargava}, S. and {Bhatawdekar}, R. and {Bianchi}, D. and {Bisigello}, L. and {Biviano}, A. and {Blake}, R.~P. and {Blanchard}, A. and {Blazek}, J. and {Blot}, L. and {Bosco}, A. and {Bodendorf}, C. and {Boenke}, T. and {B{\"o}hringer}, H. and {Boldrini}, P. and {Bolzonella}, M. and {Bonchi}, A. and {Bonici}, M. and {Bonino}, D. and {Bonino}, L. and {Bonvin}, C. and {Bon}, W. and {Booth}, J.~T. and {Borgani}, S. and {Borlaff}, A.~S. and {Borsato}, E. and {Bose}, B. and {Botticella}, M.~T. and {Boucaud}, A. and {Bouche}, F. and {Boucher}, J.~S. and {Boutigny}, D. and {Bouvard}, T. and {Bouwens}, R. and {Bouy}, H. and {Bowler}, R.~A.~A. and {Bozza}, V. and {Bozzo}, E. and {Branchini}, E. and {Brando}, G. and {Brau-Nogue}, S. and {Brekke}, P. and {Bremer}, M.~N. and {Brescia}, M. and {Breton}, M. -A. and {Brinchmann}, J. and {Brinckmann}, T. and {Brockley-Blatt}, C. and {Brodwin}, M. and {Brouard}, L. and {Brown}, M.~L. and {Bruton}, S. and {Bucko}, J. and {Buddelmeijer}, H. and {Buenadicha}, G. and {Buitrago}, F. and {Burger}, P. and {Burigana}, C. and {Busillo}, V. and {Busonero}, D. and {Cabanac}, R. and {Cabayol-Garcia}, L. and {Cagliari}, M.~S. and {Caillat}, A. and {Caillat}, L. and {Calabrese}, M. and {Calabro}, A. and {Calderone}, G. and {Calura}, F. and {Camacho Quevedo}, B. and {Camera}, S. and {Campos}, L. and {Ca{\~n}as-Herrera}, G. and {Candini}, G.~P. and {Cantiello}, M. and {Capobianco}, V. and {Cappellaro}, E. and {Cappelluti}, N. and {Cappi}, A. and {Caputi}, K.~I. and {Cara}, C. and {Carbone}, C. and {Cardone}, V.~F. and {Carella}, E. and {Carlberg}, R.~G. and {Carle}, M. and {Carminati}, L. and {Caro}, F. and {Carrasco}, J.~M. and {Carretero}, J. and {Carrilho}, P. and {Carron Duque}, J. and {Carry}, B.},
        title = "{Euclid: I. Overview of the Euclid mission}",
      journal = {\aap},
         year = 2025,
        month = may,
       volume = {697},
          eid = {A1},
        pages = {A1},
          doi = {10.1051/0004-6361/202450810},
archivePrefix = {arXiv},
       eprint = {2405.13491},
 primaryClass = {astro-ph.CO},
       adsurl = {https://ui.adsabs.harvard.edu/abs/2025A&A...697A...1E}
}

@ARTICLE{2006MNRAS.372..565F,
       author = {{Feldmann}, R. and {Carollo}, C.~M. and {Porciani}, C. and {Lilly}, S.~J. and {Capak}, P. and {Taniguchi}, Y. and {Le F{\`e}vre}, O. and {Renzini}, A. and {Scoville}, N. and {Ajiki}, M. and {Aussel}, H. and {Contini}, T. and {McCracken}, H. and {Mobasher}, B. and {Murayama}, T. and {Sanders}, D. and {Sasaki}, S. and {Scarlata}, C. and {Scodeggio}, M. and {Shioya}, Y. and {Silverman}, J. and {Takahashi}, M. and {Thompson}, D. and {Zamorani}, G.},
        title = "{The Zurich Extragalactic Bayesian Redshift Analyzer and its first application: COSMOS}",
      journal = {\mnras},
         year = 2006,
        month = oct,
       volume = {372},
       number = {2},
        pages = {565-577},
          doi = {10.1111/j.1365-2966.2006.10930.x},
archivePrefix = {arXiv},
       eprint = {astro-ph/0609044},
 primaryClass = {astro-ph},
       adsurl = {https://ui.adsabs.harvard.edu/abs/2006MNRAS.372..565F}
}

@ARTICLE{2015ApJ...810...71F,
       author = {{Finkelstein}, Steven L. and {Ryan}, Jr., Russell E. and {Papovich}, Casey and {Dickinson}, Mark and {Song}, Mimi and {Somerville}, Rachel S. and {Ferguson}, Henry C. and {Salmon}, Brett and {Giavalisco}, Mauro and {Koekemoer}, Anton M. and {Ashby}, Matthew L.~N. and {Behroozi}, Peter and {Castellano}, Marco and {Dunlop}, James S. and {Faber}, Sandy M. and {Fazio}, Giovanni G. and {Fontana}, Adriano and {Grogin}, Norman A. and {Hathi}, Nimish and {Jaacks}, Jason and {Kocevski}, Dale D. and {Livermore}, Rachael and {McLure}, Ross J. and {Merlin}, Emiliano and {Mobasher}, Bahram and {Newman}, Jeffrey A. and {Rafelski}, Marc and {Tilvi}, Vithal and {Willner}, S.~P.},
        title = "{The Evolution of the Galaxy Rest-frame Ultraviolet Luminosity Function over the First Two Billion Years}",
      journal = {\apj},
         year = 2015,
        month = sep,
       volume = {810},
       number = {1},
          eid = {71},
        pages = {71},
          doi = {10.1088/0004-637X/810/1/71},
archivePrefix = {arXiv},
       eprint = {1410.5439},
 primaryClass = {astro-ph.GA},
       adsurl = {https://ui.adsabs.harvard.edu/abs/2015ApJ...810...71F}
}

@ARTICLE{2025JCAP...06..007F,
       author = {{Fischbacher}, Silvan and {Kacprzak}, Tomasz and {Tortorelli}, Luca and {Moser}, Beatrice and {Refregier}, Alexandre and {Gebhardt}, Patrick and {Gruen}, Daniel},
        title = "{GalSBI: phenomenological galaxy population model for cosmology using simulation-based inference}",
      journal = {\jcap},
         year = 2025,
        month = jun,
       volume = {2025},
       number = {6},
          eid = {007},
        pages = {007},
          doi = {10.1088/1475-7516/2025/06/007},
archivePrefix = {arXiv},
       eprint = {2412.08701},
 primaryClass = {astro-ph.CO},
       adsurl = {https://ui.adsabs.harvard.edu/abs/2025JCAP...06..007F}
}

@ARTICLE{2012ApJS..198....1F,
       author = {{Fotopoulou}, S. and {Salvato}, M. and {Hasinger}, G. and {Rovilos}, E. and {Brusa}, M. and {Egami}, E. and {Lutz}, D. and {Burwitz}, V. and {Henry}, J.~P. and {Huang}, J.~H. and {Rigopoulou}, D. and {Vaccari}, M.},
        title = "{Photometry and Photometric Redshift Catalogs for the Lockman Hole Deep Field}",
      journal = {\apjs},
         year = 2012,
        month = jan,
       volume = {198},
       number = {1},
          eid = {1},
        pages = {1},
          doi = {10.1088/0067-0049/198/1/1},
archivePrefix = {arXiv},
       eprint = {1110.0960},
 primaryClass = {astro-ph.CO},
       adsurl = {https://ui.adsabs.harvard.edu/abs/2012ApJS..198....1F}
}

@ARTICLE{2017MNRAS.468.4556F,
       author = {{Freeman}, P.~E. and {Izbicki}, R. and {Lee}, A.~B.},
        title = "{A unified framework for constructing, tuning and assessing photometric redshift density estimates in a selection bias setting}",
      journal = {\mnras},
         year = 2017,
        month = jul,
       volume = {468},
       number = {4},
        pages = {4556-4565},
          doi = {10.1093/mnras/stx764},
archivePrefix = {arXiv},
       eprint = {1703.09242},
 primaryClass = {astro-ph.IM},
       adsurl = {https://ui.adsabs.harvard.edu/abs/2017MNRAS.468.4556F}
}

@ARTICLE{2006SSRv..123..485G,
       author = {{Gardner}, Jonathan P. and {Mather}, John C. and {Clampin}, Mark and {Doyon}, Rene and {Greenhouse}, Matthew A. and {Hammel}, Heidi B. and {Hutchings}, John B. and {Jakobsen}, Peter and {Lilly}, Simon J. and {Long}, Knox S. and {Lunine}, Jonathan I. and {McCaughrean}, Mark J. and {Mountain}, Matt and {Nella}, John and {Rieke}, George H. and {Rieke}, Marcia J. and {Rix}, Hans-Walter and {Smith}, Eric P. and {Sonneborn}, George and {Stiavelli}, Massimo and {Stockman}, H.~S. and {Windhorst}, Rogier A. and {Wright}, Gillian S.},
        title = "{The James Webb Space Telescope}",
      journal = {\ssr},
         year = 2006,
        month = apr,
       volume = {123},
       number = {4},
        pages = {485-606},
          doi = {10.1007/s11214-006-8315-7},
archivePrefix = {arXiv},
       eprint = {astro-ph/0606175},
 primaryClass = {astro-ph},
       adsurl = {https://ui.adsabs.harvard.edu/abs/2006SSRv..123..485G}
}

@ARTICLE{2019ApJ...883..203G,
       author = {{Gong}, Yan and {Liu}, Xiangkun and {Cao}, Ye and {Chen}, Xuelei and {Fan}, Zuhui and {Li}, Ran and {Li}, Xiao-Dong and {Li}, Zhigang and {Zhang}, Xin and {Zhan}, Hu},
        title = "{Cosmology from the Chinese Space Station Optical Survey (CSS-OS)}",
      journal = {\apj},
         year = 2019,
        month = oct,
       volume = {883},
       number = {2},
          eid = {203},
        pages = {203},
          doi = {10.3847/1538-4357/ab391e},
archivePrefix = {arXiv},
       eprint = {1901.04634},
 primaryClass = {astro-ph.CO},
       adsurl = {https://ui.adsabs.harvard.edu/abs/2019ApJ...883..203G}
}

@ARTICLE{2024ApJS..275...21G,
       author = {{Gris}, Philippe and {Awan}, Humna and {Becker}, Matthew R. and {Lin}, Huan and {Gawiser}, Eric and {Jha}, Saurabh W. and {The LSST Dark Energy Science Collaboration}},
        title = "{A Cohesive Deep Drilling Field Strategy for LSST Cosmology}",
      journal = {\apjs},
         year = 2024,
        month = dec,
       volume = {275},
       number = {2},
          eid = {21},
        pages = {21},
          doi = {10.3847/1538-4365/ad79f5},
archivePrefix = {arXiv},
       eprint = {2405.10781},
 primaryClass = {astro-ph.CO},
       adsurl = {https://ui.adsabs.harvard.edu/abs/2024ApJS..275...21G}
}

@ARTICLE{2014A&A...566A.108G,
       author = {{Guzzo}, L. and {Scodeggio}, M. and {Garilli}, B. and {Granett}, B.~R. and {Fritz}, A. and {Abbas}, U. and {Adami}, C. and {Arnouts}, S. and {Bel}, J. and {Bolzonella}, M. and {Bottini}, D. and {Branchini}, E. and {Cappi}, A. and {Coupon}, J. and {Cucciati}, O. and {Davidzon}, I. and {De Lucia}, G. and {de la Torre}, S. and {Franzetti}, P. and {Fumana}, M. and {Hudelot}, P. and {Ilbert}, O. and {Iovino}, A. and {Krywult}, J. and {Le Brun}, V. and {Le F{\`e}vre}, O. and {Maccagni}, D. and {Ma{\l}ek}, K. and {Marulli}, F. and {McCracken}, H.~J. and {Paioro}, L. and {Peacock}, J.~A. and {Polletta}, M. and {Pollo}, A. and {Schlagenhaufer}, H. and {Tasca}, L.~A.~M. and {Tojeiro}, R. and {Vergani}, D. and {Zamorani}, G. and {Zanichelli}, A. and {Burden}, A. and {Di Porto}, C. and {Marchetti}, A. and {Marinoni}, C. and {Mellier}, Y. and {Moscardini}, L. and {Nichol}, R.~C. and {Percival}, W.~J. and {Phleps}, S. and {Wolk}, M.},
        title = "{The VIMOS Public Extragalactic Redshift Survey (VIPERS). An unprecedented view of galaxies and large-scale structure at 0.5 < z < 1.2}",
      journal = {\aap},
         year = 2014,
        month = jun,
       volume = {566},
          eid = {A108},
        pages = {A108},
          doi = {10.1051/0004-6361/201321489},
archivePrefix = {arXiv},
       eprint = {1303.2623},
 primaryClass = {astro-ph.CO},
       adsurl = {https://ui.adsabs.harvard.edu/abs/2014A&A...566A.108G}
}

@ARTICLE{2016NewA...42...49H,
       author = {{Habib}, Salman and {Pope}, Adrian and {Finkel}, Hal and {Frontiere}, Nicholas and {Heitmann}, Katrin and {Daniel}, David and {Fasel}, Patricia and {Morozov}, Vitali and {Zagaris}, George and {Peterka}, Tom and {Vishwanath}, Venkatram and {Luki{\'c}}, Zarija and {Sehrish}, Saba and {Liao}, Wei-keng},
        title = "{HACC: Simulating sky surveys on state-of-the-art supercomputing architectures}",
      journal = {\na},
         year = 2016,
        month = jan,
       volume = {42},
        pages = {49-65},
          doi = {10.1016/j.newast.2015.06.003},
archivePrefix = {arXiv},
       eprint = {1410.2805},
 primaryClass = {astro-ph.IM},
       adsurl = {https://ui.adsabs.harvard.edu/abs/2016NewA...42...49H}
}

@ARTICLE{2020Natur.585..357H,
       author = {{Harris}, Charles R. and {Millman}, K. Jarrod and {van der Walt}, St{\'e}fan J. and {Gommers}, Ralf and {Virtanen}, Pauli and {Cournapeau}, David and {Wieser}, Eric and {Taylor}, Julian and {Berg}, Sebastian and {Smith}, Nathaniel J. and {Kern}, Robert and {Picus}, Matti and {Hoyer}, Stephan and {van Kerkwijk}, Marten H. and {Brett}, Matthew and {Haldane}, Allan and {del R{\'\i}o}, Jaime Fern{\'a}ndez and {Wiebe}, Mark and {Peterson}, Pearu and {G{\'e}rard-Marchant}, Pierre and {Sheppard}, Kevin and {Reddy}, Tyler and {Weckesser}, Warren and {Abbasi}, Hameer and {Gohlke}, Christoph and {Oliphant}, Travis E.},
        title = "{Array programming with NumPy}",
      journal = {\nat},
         year = 2020,
        month = sep,
       volume = {585},
       number = {7825},
        pages = {357-362},
          doi = {10.1038/s41586-020-2649-2},
archivePrefix = {arXiv},
       eprint = {2006.10256},
 primaryClass = {cs.MS},
       adsurl = {https://ui.adsabs.harvard.edu/abs/2020Natur.585..357H}
}

@ARTICLE{2020MNRAS.496.4769H,
       author = {{Hartley}, W.~G. and {Chang}, C. and {Samani}, S. and {Carnero Rosell}, A. and {Davis}, T.~M. and {Hoyle}, B. and {Gruen}, D. and {Asorey}, J. and {Gschwend}, J. and {Lidman}, C. and {Kuehn}, K. and {King}, A. and {Rau}, M.~M. and {Wechsler}, R.~H. and {DeRose}, J. and {Hinton}, S.~R. and {Whiteway}, L. and {Abbott}, T.~M.~C. and {Aguena}, M. and {Allam}, S. and {Annis}, J. and {Avila}, S. and {Bernstein}, G.~M. and {Bertin}, E. and {Bridle}, S.~L. and {Brooks}, D. and {Burke}, D.~L. and {Carrasco Kind}, M. and {Carretero}, J. and {Castander}, F.~J. and {Cawthon}, R. and {Costanzi}, M. and {da Costa}, L.~N. and {Desai}, S. and {Diehl}, H.~T. and {Dietrich}, J.~P. and {Flaugher}, B. and {Fosalba}, P. and {Frieman}, J. and {Garc{\'\i}a-Bellido}, J. and {Gaztanaga}, E. and {Gerdes}, D.~W. and {Gruendl}, R.~A. and {Gutierrez}, G. and {Hollowood}, D.~L. and {Honscheid}, K. and {James}, D.~J. and {Kent}, S. and {Krause}, E. and {Kuropatkin}, N. and {Lahav}, O. and {Lima}, M. and {Maia}, M.~A.~G. and {Marshall}, J.~L. and {Melchior}, P. and {Menanteau}, F. and {Miquel}, R. and {Ogando}, R.~L.~C. and {Palmese}, A. and {Paz-Chinch{\'o}n}, F. and {Plazas}, A.~A. and {Roodman}, A. and {Rykoff}, E.~S. and {Sanchez}, E. and {Scarpine}, V. and {Schubnell}, M. and {Serrano}, S. and {Sevilla-Noarbe}, I. and {Smith}, M. and {Soares-Santos}, M. and {Suchyta}, E. and {Tarle}, G. and {Troxel}, M.~A. and {Tucker}, D.~L. and {Varga}, T.~N. and {Weller}, J. and {Wilkinson}, R.~D. and {DES Collaboration}},
        title = "{The impact of spectroscopic incompleteness in direct calibration of redshift distributions for weak lensing surveys}",
      journal = {\mnras},
         year = 2020,
        month = aug,
       volume = {496},
       number = {4},
        pages = {4769-4786},
          doi = {10.1093/mnras/staa1812},
archivePrefix = {arXiv},
       eprint = {2003.10454},
 primaryClass = {astro-ph.GA},
       adsurl = {https://ui.adsabs.harvard.edu/abs/2020MNRAS.496.4769H}
}

@ARTICLE{2020MNRAS.495.5040H,
       author = {{Hearin}, Andrew and {Korytov}, Danila and {Kovacs}, Eve and {Benson}, Andrew and {Aung}, Han and {Bradshaw}, Christopher and {Campbell}, Duncan and {LSST Dark Energy Science Collaboration}},
        title = "{Generating synthetic cosmological data with GalSampler}",
      journal = {\mnras},
         year = 2020,
        month = jul,
       volume = {495},
       number = {4},
        pages = {5040-5051},
          doi = {10.1093/mnras/staa1495},
archivePrefix = {arXiv},
       eprint = {1909.07340},
 primaryClass = {astro-ph.CO},
       adsurl = {https://ui.adsabs.harvard.edu/abs/2020MNRAS.495.5040H}
}

@ARTICLE{2021OJAp....4E...7H,
       author = {{Hearin}, Andrew P. and {Chaves-Montero}, Jon{\'a}s and {Becker}, Mathew R. and {Alarcon}, Alex},
        title = "{A Differentiable Model of the Assembly of Individual and Populations of Dark Matter Halos}",
      journal = {The Open Journal of Astrophysics},
         year = 2021,
        month = jul,
       volume = {4},
       number = {1},
          eid = {7},
        pages = {7},
          doi = {10.21105/astro.2105.05859},
archivePrefix = {arXiv},
       eprint = {2105.05859},
 primaryClass = {astro-ph.CO},
       adsurl = {https://ui.adsabs.harvard.edu/abs/2021OJAp....4E...7H}
}

@ARTICLE{2023MNRAS.521.1741H,
      author = {{Hearin}, Andrew P. and {Chaves-Montero}, Jon{\'a}s and {Alarcon}, Alex and {Becker}, Matthew R. and {Benson}, Andrew},
        title = "{DSPS: Differentiable stellar population synthesis}",
      journal = {MNRAS},
        year = 2023,
        month = may,
      volume = {521},
      number = {2},
        pages = {1741-1756},
          doi = {10.1093/mnras/stad456},
archivePrefix = {arXiv},
      eprint = {2112.06830},
primaryClass = {astro-ph.GA},
      adsurl = {https://ui.adsabs.harvard.edu/abs/2023MNRAS.521.1741H}
}

@ARTICLE{2019ApJS..245...16H,
       author = {{Heitmann}, Katrin and {Finkel}, Hal and {Pope}, Adrian and {Morozov}, Vitali and {Frontiere}, Nicholas and {Habib}, Salman and {Rangel}, Esteban and {Uram}, Thomas and {Korytov}, Danila and {Child}, Hillary and {Flender}, Samuel and {Insley}, Joe and {Rizzi}, Silvio},
        title = "{The Outer Rim Simulation: A Path to Many-core Supercomputers}",
      journal = {\apjs},
         year = 2019,
        month = nov,
       volume = {245},
       number = {1},
          eid = {16},
        pages = {16},
          doi = {10.3847/1538-4365/ab4da1},
archivePrefix = {arXiv},
       eprint = {1904.11970},
 primaryClass = {astro-ph.CO},
       adsurl = {https://ui.adsabs.harvard.edu/abs/2019ApJS..245...16H}
}

@ARTICLE{2021A&A...646A.140H,
       author = {{Heymans}, Catherine and {Tr{\"o}ster}, Tilman and {Asgari}, Marika and {Blake}, Chris and {Hildebrandt}, Hendrik and {Joachimi}, Benjamin and {Kuijken}, Konrad and {Lin}, Chieh-An and {S{\'a}nchez}, Ariel G. and {van den Busch}, Jan Luca and {Wright}, Angus H. and {Amon}, Alexandra and {Bilicki}, Maciej and {de Jong}, Jelte and {Crocce}, Martin and {Dvornik}, Andrej and {Erben}, Thomas and {Fortuna}, Maria Cristina and {Getman}, Fedor and {Giblin}, Benjamin and {Glazebrook}, Karl and {Hoekstra}, Henk and {Joudaki}, Shahab and {Kannawadi}, Arun and {K{\"o}hlinger}, Fabian and {Lidman}, Chris and {Miller}, Lance and {Napolitano}, Nicola R. and {Parkinson}, David and {Schneider}, Peter and {Shan}, HuanYuan and {Valentijn}, Edwin A. and {Verdoes Kleijn}, Gijs and {Wolf}, Christian},
        title = "{KiDS-1000 Cosmology: Multi-probe weak gravitational lensing and spectroscopic galaxy clustering constraints}",
      journal = {\aap},
         year = 2021,
        month = feb,
       volume = {646},
          eid = {A140},
        pages = {A140},
          doi = {10.1051/0004-6361/202039063},
archivePrefix = {arXiv},
       eprint = {2007.15632},
 primaryClass = {astro-ph.CO},
       adsurl = {https://ui.adsabs.harvard.edu/abs/2021A&A...646A.140H}
}

@ARTICLE{2010A&A...523A..31H,
       author = {{Hildebrandt}, H. and {Arnouts}, S. and {Capak}, P. and {Moustakas}, L.~A. and {Wolf}, C. and {Abdalla}, F.~B. and {Assef}, R.~J. and {Banerji}, M. and {Ben{\'\i}tez}, N. and {Brammer}, G.~B. and {Budav{\'a}ri}, T. and {Carliles}, S. and {Coe}, D. and {Dahlen}, T. and {Feldmann}, R. and {Gerdes}, D. and {Gillis}, B. and {Ilbert}, O. and {Kotulla}, R. and {Lahav}, O. and {Li}, I.~H. and {Miralles}, J. -M. and {Purger}, N. and {Schmidt}, S. and {Singal}, J.},
        title = "{PHAT: PHoto-z Accuracy Testing}",
      journal = {\aap},
         year = 2010,
        month = nov,
       volume = {523},
          eid = {A31},
        pages = {A31},
          doi = {10.1051/0004-6361/201014885},
archivePrefix = {arXiv},
       eprint = {1008.0658},
 primaryClass = {astro-ph.CO},
       adsurl = {https://ui.adsabs.harvard.edu/abs/2010A&A...523A..31H}
}

@ARTICLE{2007CSE.....9...90H,
       author = {{Hunter}, John D.},
        title = "{Matplotlib: A 2D Graphics Environment}",
      journal = {Computing in Science and Engineering},
         year = 2007,
        month = may,
       volume = {9},
       number = {3},
        pages = {90-95},
          doi = {10.1109/MCSE.2007.55},
       adsurl = {https://ui.adsabs.harvard.edu/abs/2007CSE.....9...90H}
}

@ARTICLE{2006A&A...457..841I,
       author = {{Ilbert}, O. and {Arnouts}, S. and {McCracken}, H.~J. and {Bolzonella}, M. and {Bertin}, E. and {Le F{\`e}vre}, O. and {Mellier}, Y. and {Zamorani}, G. and {Pell{\`o}}, R. and {Iovino}, A. and {Tresse}, L. and {Le Brun}, V. and {Bottini}, D. and {Garilli}, B. and {Maccagni}, D. and {Picat}, J.~P. and {Scaramella}, R. and {Scodeggio}, M. and {Vettolani}, G. and {Zanichelli}, A. and {Adami}, C. and {Bardelli}, S. and {Cappi}, A. and {Charlot}, S. and {Ciliegi}, P. and {Contini}, T. and {Cucciati}, O. and {Foucaud}, S. and {Franzetti}, P. and {Gavignaud}, I. and {Guzzo}, L. and {Marano}, B. and {Marinoni}, C. and {Mazure}, A. and {Meneux}, B. and {Merighi}, R. and {Paltani}, S. and {Pollo}, A. and {Pozzetti}, L. and {Radovich}, M. and {Zucca}, E. and {Bondi}, M. and {Bongiorno}, A. and {Busarello}, G. and {de La Torre}, S. and {Gregorini}, L. and {Lamareille}, F. and {Mathez}, G. and {Merluzzi}, P. and {Ripepi}, V. and {Rizzo}, D. and {Vergani}, D.},
        title = "{Accurate photometric redshifts for the CFHT legacy survey calibrated using the VIMOS VLT deep survey}",
      journal = {\aap},
         year = 2006,
        month = oct,
       volume = {457},
       number = {3},
        pages = {841-856},
          doi = {10.1051/0004-6361:20065138},
archivePrefix = {arXiv},
       eprint = {astro-ph/0603217},
 primaryClass = {astro-ph},
       adsurl = {https://ui.adsabs.harvard.edu/abs/2006A&A...457..841I}
}

@ARTICLE{2019ApJ...873..111I,
       author = {{Ivezi{\'c}}, {\v{Z}}eljko and {Kahn}, Steven M. and {Tyson}, J. Anthony and {Abel}, Bob and {Acosta}, Emily and {Allsman}, Robyn and {Alonso}, David and {AlSayyad}, Yusra and {Anderson}, Scott F. and {Andrew}, John and {Angel}, James Roger P. and {Angeli}, George Z. and {Ansari}, Reza and {Antilogus}, Pierre and {Araujo}, Constanza and {Armstrong}, Robert and {Arndt}, Kirk T. and {Astier}, Pierre and {Aubourg}, {\'E}ric and {Auza}, Nicole and {Axelrod}, Tim S. and {Bard}, Deborah J. and {Barr}, Jeff D. and {Barrau}, Aurelian and {Bartlett}, James G. and {Bauer}, Amanda E. and {Bauman}, Brian J. and {Baumont}, Sylvain and {Bechtol}, Ellen and {Bechtol}, Keith and {Becker}, Andrew C. and {Becla}, Jacek and {Beldica}, Cristina and {Bellavia}, Steve and {Bianco}, Federica B. and {Biswas}, Rahul and {Blanc}, Guillaume and {Blazek}, Jonathan and {Blandford}, Roger D. and {Bloom}, Josh S. and {Bogart}, Joanne and {Bond}, Tim W. and {Booth}, Michael T. and {Borgland}, Anders W. and {Borne}, Kirk and {Bosch}, James F. and {Boutigny}, Dominique and {Brackett}, Craig A. and {Bradshaw}, Andrew and {Brandt}, William Nielsen and {Brown}, Michael E. and {Bullock}, James S. and {Burchat}, Patricia and {Burke}, David L. and {Cagnoli}, Gianpietro and {Calabrese}, Daniel and {Callahan}, Shawn and {Callen}, Alice L. and {Carlin}, Jeffrey L. and {Carlson}, Erin L. and {Chandrasekharan}, Srinivasan and {Charles-Emerson}, Glenaver and {Chesley}, Steve and {Cheu}, Elliott C. and {Chiang}, Hsin-Fang and {Chiang}, James and {Chirino}, Carol and {Chow}, Derek and {Ciardi}, David R. and {Claver}, Charles F. and {Cohen-Tanugi}, Johann and {Cockrum}, Joseph J. and {Coles}, Rebecca and {Connolly}, Andrew J. and {Cook}, Kem H. and {Cooray}, Asantha and {Covey}, Kevin R. and {Cribbs}, Chris and {Cui}, Wei and {Cutri}, Roc and {Daly}, Philip N. and {Daniel}, Scott F. and {Daruich}, Felipe and {Daubard}, Guillaume and {Daues}, Greg and {Dawson}, William and {Delgado}, Francisco and {Dellapenna}, Alfred and {de Peyster}, Robert and {de Val-Borro}, Miguel and {Digel}, Seth W. and {Doherty}, Peter and {Dubois}, Richard and {Dubois-Felsmann}, Gregory P. and {Durech}, Josef and {Economou}, Frossie and {Eifler}, Tim and {Eracleous}, Michael and {Emmons}, Benjamin L. and {Fausti Neto}, Angelo and {Ferguson}, Henry and {Figueroa}, Enrique and {Fisher-Levine}, Merlin and {Focke}, Warren and {Foss}, Michael D. and {Frank}, James and {Freemon}, Michael D. and {Gangler}, Emmanuel and {Gawiser}, Eric and {Geary}, John C. and {Gee}, Perry and {Geha}, Marla and {Gessner}, Charles J.~B. and {Gibson}, Robert R. and {Gilmore}, D. Kirk and {Glanzman}, Thomas and {Glick}, William and {Goldina}, Tatiana and {Goldstein}, Daniel A. and {Goodenow}, Iain and {Graham}, Melissa L. and {Gressler}, William J. and {Gris}, Philippe and {Guy}, Leanne P. and {Guyonnet}, Augustin and {Haller}, Gunther and {Harris}, Ron and {Hascall}, Patrick A. and {Haupt}, Justine and {Hernandez}, Fabio and {Herrmann}, Sven and {Hileman}, Edward and {Hoblitt}, Joshua and {Hodgson}, John A. and {Hogan}, Craig and {Howard}, James D. and {Huang}, Dajun and {Huffer}, Michael E. and {Ingraham}, Patrick and {Innes}, Walter R. and {Jacoby}, Suzanne H. and {Jain}, Bhuvnesh and {Jammes}, Fabrice and {Jee}, M. James and {Jenness}, Tim and {Jernigan}, Garrett and {Jevremovi{\'c}}, Darko and {Johns}, Kenneth and {Johnson}, Anthony S. and {Johnson}, Margaret W.~G. and {Jones}, R. Lynne and {Juramy-Gilles}, Claire and {Juri{\'c}}, Mario and {Kalirai}, Jason S. and {Kallivayalil}, Nitya J. and {Kalmbach}, Bryce and {Kantor}, Jeffrey P. and {Karst}, Pierre and {Kasliwal}, Mansi M. and {Kelly}, Heather and {Kessler}, Richard and {Kinnison}, Veronica and {Kirkby}, David and {Knox}, Lloyd and {Kotov}, Ivan V. and {Krabbendam}, Victor L. and {Krughoff}, K. Simon and {Kub{\'a}nek}, Petr and {Kuczewski}, John and {Kulkarni}, Shri and {Ku}, John and {Kurita}, Nadine R. and {Lage}, Craig S. and {Lambert}, Ron and {Lange}, Travis and {Langton}, J. Brian and {Le Guillou}, Laurent and {Levine}, Deborah and {Liang}, Ming and {Lim}, Kian-Tat and {Lintott}, Chris J. and {Long}, Kevin E. and {Lopez}, Margaux and {Lotz}, Paul J. and {Lupton}, Robert H. and {Lust}, Nate B. and {MacArthur}, Lauren A. and {Mahabal}, Ashish and {Mandelbaum}, Rachel and {Markiewicz}, Thomas W. and {Marsh}, Darren S. and {Marshall}, Philip J. and {Marshall}, Stuart and {May}, Morgan and {McKercher}, Robert and {McQueen}, Michelle and {Meyers}, Joshua and {Migliore}, Myriam and {Miller}, Michelle and {Mills}, David J.},
        title = "{LSST: From Science Drivers to Reference Design and Anticipated Data Products}",
      journal = {\apj},
         year = 2019,
        month = mar,
       volume = {873},
       number = {2},
          eid = {111},
        pages = {111},
          doi = {10.3847/1538-4357/ab042c},
archivePrefix = {arXiv},
       eprint = {0805.2366},
 primaryClass = {astro-ph},
       adsurl = {https://ui.adsabs.harvard.edu/abs/2019ApJ...873..111I}
}

@article{10.1214/17-EJS1302,
       author = {{Izbicki}, Rafael and {Lee}, Ann B.},
        title = {{Converting high-dimensional regression to high-dimensional conditional density estimation}},
       volume = {11},
      journal = {Electronic Journal of Statistics},
       number = {2},
    publisher = {Institute of Mathematical Statistics and Bernoulli Society},
        pages = {2800 -- 2831},
         year = {2017},
          doi = {10.1214/17-EJS1302},
          URL = {https://doi.org/10.1214/17-EJS1302}
}

@ARTICLE{2024MNRAS.530.2688J,
       author = {{Jin}, Shoko and {Trager}, Scott C. and {Dalton}, Gavin B. and {Aguerri}, J. Alfonso L. and {Drew}, J.~E. and {Falc{\'o}n-Barroso}, Jes{\'u}s and {G{\"a}nsicke}, Boris T. and {Hill}, Vanessa and {Iovino}, Angela and {Pieri}, Matthew M. and {Poggianti}, Bianca M. and {Smith}, D.~J.~B. and {Vallenari}, Antonella and {Abrams}, Don Carlos and {Aguado}, David S. and {Antoja}, Teresa and {Arag{\'o}n-Salamanca}, Alfonso and {Ascasibar}, Yago and {Babusiaux}, Carine and {Balcells}, Marc and {Barrena}, R. and {Battaglia}, Giuseppina and {Belokurov}, Vasily and {Bensby}, Thomas and {Bonifacio}, Piercarlo and {Bragaglia}, Angela and {Carrasco}, Esperanza and {Carrera}, Ricardo and {Cornwell}, Daniel J. and {Dom{\'\i}nguez-Palmero}, Lilian and {Duncan}, Kenneth J. and {Famaey}, Benoit and {Fari{\~n}a}, Cecilia and {Gonzalez}, Oscar A. and {Guest}, Steve and {Hatch}, Nina A. and {Hess}, Kelley M. and {Hoskin}, Matthew J. and {Irwin}, Mike and {Knapen}, Johan H. and {Koposov}, Sergey E. and {Kuchner}, Ulrike and {Laigle}, Clotilde and {Lewis}, Jim and {Longhetti}, Marcella and {Lucatello}, Sara and {M{\'e}ndez-Abreu}, Jairo and {Mercurio}, Amata and {Molaeinezhad}, Alireza and {Mongui{\'o}}, Maria and {Morrison}, Sean and {Murphy}, David N.~A. and {Peralta de Arriba}, Luis and {P{\'e}rez}, Isabel and {P{\'e}rez-R{\`a}fols}, Ignasi and {Pic{\'o}}, Sergio and {Raddi}, Roberto and {Romero-G{\'o}mez}, Merc{\`e} and {Royer}, Fr{\'e}d{\'e}ric and {Siebert}, Arnaud and {Seabroke}, George M. and {Som}, Debopam and {Terrett}, David and {Thomas}, Guillaume and {Wesson}, Roger and {Worley}, C. Clare and {Alfaro}, Emilio J. and {Allende Prieto}, Carlos and {Alonso-Santiago}, Javier and {Amos}, Nicholas J. and {Ashley}, Richard P. and {Balaguer-N{\'u}{\~n}ez}, Lola and {Balbinot}, Eduardo and {Bellazzini}, Michele and {Benn}, Chris R. and {Berlanas}, Sara R. and {Bernard}, Edouard J. and {Best}, Philip and {Bettoni}, Daniela and {Bianco}, Andrea and {Bishop}, Georgia and {Blomqvist}, Michael and {Boeche}, Corrado and {Bolzonella}, Micol and {Bonoli}, Silvia and {Bosma}, Albert and {Britavskiy}, Nikolay and {Busarello}, Gianni and {Caffau}, Elisabetta and {Cantat-Gaudin}, Tristan and {Castro-Ginard}, Alfred and {Couto}, Guilherme and {Carbajo-Hijarrubia}, Juan and {Carter}, David and {Casamiquela}, Laia and {Conrado}, Ana M. and {Corcho-Caballero}, Pablo and {Costantin}, Luca and {Deason}, Alis and {de Burgos}, Abel and {De Grandi}, Sabrina and {Di Matteo}, Paola and {Dom{\'\i}nguez-G{\'o}mez}, Jes{\'u}s and {Dorda}, Ricardo and {Drake}, Alyssa and {Dutta}, Rajeshwari and {Erkal}, Denis and {Feltzing}, Sofia and {Ferr{\'e}-Mateu}, Anna and {Feuillet}, Diane and {Figueras}, Francesca and {Fossati}, Matteo and {Franciosini}, Elena and {Frasca}, Antonio and {Fumagalli}, Michele and {Gallazzi}, Anna and {Garc{\'\i}a-Benito}, Rub{\'e}n and {Gentile Fusillo}, Nicola and {Gebran}, Marwan and {Gilbert}, James and {Gledhill}, T.~M. and {Gonz{\'a}lez Delgado}, Rosa M. and {Greimel}, Robert and {Guarcello}, Mario Giuseppe and {Guerra}, Jose and {Gullieuszik}, Marco and {Haines}, Christopher P. and {Hardcastle}, Martin J. and {Harris}, Amy and {Haywood}, Misha and {Helmi}, Amina and {Hernandez}, Nauzet and {Herrero}, Artemio and {Hughes}, Sarah and {Ir{\v{s}}i{\v{c}}}, Vid and {Jablonka}, Pascale and {Jarvis}, Matt J. and {Jordi}, Carme and {Kondapally}, Rohit and {Kordopatis}, Georges and {Krogager}, Jens-Kristian and {La Barbera}, Francesco and {Lam}, Man I. and {Larsen}, S{\o}ren S. and {Lemasle}, Bertrand and {Lewis}, Ian J. and {Lhom{\'e}}, Emilie and {Lind}, Karin and {Lodi}, Marcello and {Longobardi}, Alessia and {Lonoce}, Ilaria and {Magrini}, Laura and {Ma{\'\i}z Apell{\'a}niz}, Jes{\'u}s and {Marchal}, Olivier and {Marco}, Amparo and {Martin}, Nicolas F. and {Matsuno}, Tadafumi and {Maurogordato}, Sophie and {Merluzzi}, Paola and {Miralda-Escud{\'e}}, Jordi and {Molinari}, Emilio and {Monari}, Giacomo and {Morelli}, Lorenzo and {Mottram}, Christopher J. and {Naylor}, Tim and {Negueruela}, Ignacio and {O{\~n}orbe}, Jose and {Pancino}, Elena and {Peirani}, S{\'e}bastien and {Peletier}, Reynier F. and {Pozzetti}, Lucia and {Rainer}, Monica and {Ramos}, Pau and {Read}, Shaun C. and {Rossi}, Elena Maria and {R{\"o}ttgering}, Huub J.~A. and {Rubi{\~n}o-Mart{\'\i}n}, Jose Alberto and {Sabater}, Jose and {San Juan}, Jos{\'e} and {Sanna}, Nicoletta and {Schallig}, Ellen and {Schiavon}, Ricardo P. and {Schultheis}, Mathias and {Serra}, Paolo and {Shimwell}, Timothy W. and {Sim{\'o}n-D{\'\i}az}, Sergio and {Smith}, Russell J. and {Sordo}, Rosanna and {Sorini}, Daniele and {Soubiran}, Caroline and {Starkenburg}, Else and {Steele}, Iain A. and {Stott}, John and {Stuik}, Remko and {Tolstoy}, Eline and {Tortora}, Crescenzo and {Tsantaki}, Maria and {Van der Swaelmen}, Mathieu and {van Weeren}, Reinout J. and {Vergani}, Daniela},
        title = "{The wide-field, multiplexed, spectroscopic facility WEAVE: Survey design, overview, and simulated implementation}",
      journal = {\mnras},
         year = 2024,
        month = may,
       volume = {530},
       number = {3},
        pages = {2688-2730},
          doi = {10.1093/mnras/stad557},
archivePrefix = {arXiv},
       eprint = {2212.03981},
 primaryClass = {astro-ph.IM},
       adsurl = {https://ui.adsabs.harvard.edu/abs/2024MNRAS.530.2688J}
}

@ARTICLE{2021ApJS..254...22J,
       author = {{Johnson}, Benjamin D. and {Leja}, Joel and {Conroy}, Charlie and {Speagle}, Joshua S.},
        title = "{Stellar Population Inference with Prospector}",
      journal = {\apjs},
         year = 2021,
        month = jun,
       volume = {254},
       number = {2},
          eid = {22},
        pages = {22},
          doi = {10.3847/1538-4365/abef67},
archivePrefix = {arXiv},
       eprint = {2012.01426},
 primaryClass = {astro-ph.GA},
       adsurl = {https://ui.adsabs.harvard.edu/abs/2021ApJS..254...22J}
}

@ARTICLE{2021A&A...648A..98J,
       author = {{Johnston}, Harry and {Wright}, Angus H. and {Joachimi}, Benjamin and {Bilicki}, Maciej and {Elisa Chisari}, Nora and {Dvornik}, Andrej and {Erben}, Thomas and {Giblin}, Benjamin and {Heymans}, Catherine and {Hildebrandt}, Hendrik and {Hoekstra}, Henk and {Joudaki}, Shahab and {Vakili}, Mohammadjavad},
        title = "{Organised randoms: Learning and correcting for systematic galaxy clustering patterns in KiDS using self-organising maps}",
      journal = {\aap},
         year = 2021,
        month = apr,
       volume = {648},
          eid = {A98},
        pages = {A98},
          doi = {10.1051/0004-6361/202040136},
archivePrefix = {arXiv},
       eprint = {2012.08467},
 primaryClass = {astro-ph.CO},
       adsurl = {https://ui.adsabs.harvard.edu/abs/2021A&A...648A..98J}
}

@ARTICLE{1998ApJ...498...26K,
       author = {{Kaiser}, Nick},
        title = "{Weak Lensing and Cosmology}",
      journal = {\apj},
         year = 1998,
        month = may,
       volume = {498},
       number = {1},
        pages = {26-42},
          doi = {10.1086/305515},
archivePrefix = {arXiv},
       eprint = {astro-ph/9610120},
 primaryClass = {astro-ph},
       adsurl = {https://ui.adsabs.harvard.edu/abs/1998ApJ...498...26K}
}

@ARTICLE{1998ApJ...498..541K,
       author = {{Kennicutt}, Jr., Robert C.},
        title = "{The Global Schmidt Law in Star-forming Galaxies}",
      journal = {\apj},
         year = 1998,
        month = may,
       volume = {498},
       number = {2},
        pages = {541-552},
          doi = {10.1086/305588},
archivePrefix = {arXiv},
       eprint = {astro-ph/9712213},
 primaryClass = {astro-ph},
       adsurl = {https://ui.adsabs.harvard.edu/abs/1998ApJ...498..541K}
}

@ARTICLE{2015RPPh...78h6901K,
       author = {{Kilbinger}, Martin},
        title = "{Cosmology with cosmic shear observations: a review}",
      journal = {Reports on Progress in Physics},
         year = 2015,
        month = jul,
       volume = {78},
       number = {8},
          eid = {086901},
        pages = {086901},
          doi = {10.1088/0034-4885/78/8/086901},
archivePrefix = {arXiv},
       eprint = {1411.0115},
 primaryClass = {astro-ph.CO},
       adsurl = {https://ui.adsabs.harvard.edu/abs/2015RPPh...78h6901K}
}

@ARTICLE{1985AJ.....90..418K,
       author = {{Koo}, D.~C.},
        title = "{Optical multicolors : a poor person's Z machine for galaxies.}",
      journal = {\aj},
         year = 1985,
        month = mar,
       volume = {90},
        pages = {418-440},
          doi = {10.1086/113748},
       adsurl = {https://ui.adsabs.harvard.edu/abs/1985AJ.....90..418K}
}

@ARTICLE{2013ARA&A..51..511K,
       author = {{Kormendy}, John and {Ho}, Luis C.},
        title = "{Coevolution (Or Not) of Supermassive Black Holes and Host Galaxies}",
      journal = {\araa},
         year = 2013,
        month = aug,
       volume = {51},
       number = {1},
        pages = {511-653},
          doi = {10.1146/annurev-astro-082708-101811},
archivePrefix = {arXiv},
       eprint = {1304.7762},
 primaryClass = {astro-ph.CO},
       adsurl = {https://ui.adsabs.harvard.edu/abs/2013ARA&A..51..511K}
}

@ARTICLE{2019ApJS..245...26K,
       author = {{Korytov}, Danila and {Hearin}, Andrew and {Kovacs}, Eve and {Larsen}, Patricia and {Rangel}, Esteban and {Hollowed}, Joseph and {Benson}, Andrew J. and {Heitmann}, Katrin and {Mao}, Yao-Yuan and {Bahmanyar}, Anita and {Chang}, Chihway and {Campbell}, Duncan and {DeRose}, Joseph and {Finkel}, Hal and {Frontiere}, Nicholas and {Gawiser}, Eric and {Habib}, Salman and {Joachimi}, Benjamin and {Lanusse}, Fran{\c{c}}ois and {Li}, Nan and {Mandelbaum}, Rachel and {Morrison}, Christopher and {Newman}, Jeffrey A. and {Pope}, Adrian and {Rykoff}, Eli and {Simet}, Melanie and {To}, Chun-Hao and {Vikraman}, Vinu and {Wechsler}, Risa H. and {White}, Martin and {(The LSST Dark Energy Science Collaboration}},
        title = "{CosmoDC2: A Synthetic Sky Catalog for Dark Energy Science with LSST}",
      journal = {\apjs},
         year = 2019,
        month = dec,
       volume = {245},
       number = {2},
          eid = {26},
        pages = {26},
          doi = {10.3847/1538-4365/ab510c},
archivePrefix = {arXiv},
       eprint = {1907.06530},
 primaryClass = {astro-ph.CO},
       adsurl = {https://ui.adsabs.harvard.edu/abs/2019ApJS..245...26K}
}

@ARTICLE{2022OJAp....5E...1K,
       author = {{Kovacs}, Eve and {Mao}, Yao-Yuan and {Aguena}, Michel and {Bahmanyar}, Anita and {Broussard}, Adam and {Butler}, James and {Campbell}, Duncan and {Chang}, Chihway and {Fu}, Shenming and {Heitmann}, Katrin and {Korytov}, Danila and {Lanusse}, Fran{\c{c}}ois and {Larsen}, Patricia and {Mandelbaum}, Rachel and {Morrison}, Christopher B. and {Payerne}, Constantin and {Ricci}, Marina and {Rykoff}, Eli and {S{\'a}nchez}, F. Javier and {Sevilla-Noarbe}, Ignacio and {Simet}, Melanie and {To}, Chun-Hao and {Vikraman}, Vinu and {Zhou}, Rongpu and {Avestruz}, Camille and {Benoist}, Christophe and {Benson}, Andrew J. and {Bleem}, Lindsey and {{\'C}iprianovi{\'c}}, Aleksandra and {Combet}, C{\'e}line and {Gawiser}, Eric and {He}, Shiyuan and {Joseph}, Remy and {Newman}, Jeffrey A. and {Prat}, Judit and {Schmidt}, Samuel and {Slosar}, An{\v{z}}e and {Zuntz}, Joe and {LSST Dark Energy Science Collaboration}},
        title = "{Validating Synthetic Galaxy Catalogs for Dark Energy Science in the LSST Era}",
      journal = {The Open Journal of Astrophysics},
         year = 2022,
        month = jan,
       volume = {5},
          eid = {1},
        pages = {1},
          doi = {10.21105/astro.2110.03769},
archivePrefix = {arXiv},
       eprint = {2110.03769},
 primaryClass = {astro-ph.CO},
       adsurl = {https://ui.adsabs.harvard.edu/abs/2022OJAp....5E...1K}
}

@ARTICLE{2015MNRAS.454.3500K,
       author = {{Kuijken}, Konrad and {Heymans}, Catherine and {Hildebrandt}, Hendrik and {Nakajima}, Reiko and {Erben}, Thomas and {de Jong}, Jelte T.~A. and {Viola}, Massimo and {Choi}, Ami and {Hoekstra}, Henk and {Miller}, Lance and {van Uitert}, Edo and {Amon}, Alexandra and {Blake}, Chris and {Brouwer}, Margot and {Buddendiek}, Axel and {Conti}, Ian Fenech and {Eriksen}, Martin and {Grado}, Aniello and {Harnois-D{\'e}raps}, Joachim and {Helmich}, Ewout and {Herbonnet}, Ricardo and {Irisarri}, Nancy and {Kitching}, Thomas and {Klaes}, Dominik and {La Barbera}, Francesco and {Napolitano}, Nicola and {Radovich}, Mario and {Schneider}, Peter and {Sif{\'o}n}, Crist{\'o}bal and {Sikkema}, Gert and {Simon}, Patrick and {Tudorica}, Alexandru and {Valentijn}, Edwin and {Verdoes Kleijn}, Gijs and {van Waerbeke}, Ludovic},
        title = "{Gravitational lensing analysis of the Kilo-Degree Survey}",
      journal = {\mnras},
         year = 2015,
        month = dec,
       volume = {454},
       number = {4},
        pages = {3500-3532},
          doi = {10.1093/mnras/stv2140},
archivePrefix = {arXiv},
       eprint = {1507.00738},
 primaryClass = {astro-ph.CO},
       adsurl = {https://ui.adsabs.harvard.edu/abs/2015MNRAS.454.3500K}
}

@ARTICLE{2005A&A...439..845L,
       author = {{Le F{\`e}vre}, O. and {Vettolani}, G. and {Garilli}, B. and {Tresse}, L. and {Bottini}, D. and {Le Brun}, V. and {Maccagni}, D. and {Picat}, J.~P. and {Scaramella}, R. and {Scodeggio}, M. and et al.},
        title = "{The VIMOS VLT deep survey. First epoch VVDS-deep survey: 11 564 spectra with 17.5 {\ensuremath{\leq}} IAB {\ensuremath{\leq}} 24, and the redshift distribution over 0 {\ensuremath{\leq}} z {\ensuremath{\leq}} 5}",
      journal = {\aap},
         year = 2005,
        month = sep,
       volume = {439},
       number = {3},
        pages = {845-862},
          doi = {10.1051/0004-6361:20041960},
archivePrefix = {arXiv},
       eprint = {astro-ph/0409133},
 primaryClass = {astro-ph},
       adsurl = {https://ui.adsabs.harvard.edu/abs/2005A&A...439..845L}
}

@ARTICLE{2017ApJ...838....5L,
       author = {{Leistedt}, Boris and {Hogg}, David W.},
        title = "{Data-driven, Interpretable Photometric Redshifts Trained on Heterogeneous and Unrepresentative Data}",
      journal = {\apj},
         year = 2017,
        month = mar,
       volume = {838},
       number = {1},
          eid = {5},
        pages = {5},
          doi = {10.3847/1538-4357/aa6332},
archivePrefix = {arXiv},
       eprint = {1612.00847},
 primaryClass = {astro-ph.CO},
       adsurl = {https://ui.adsabs.harvard.edu/abs/2017ApJ...838....5L}
}

\bsp
\label{lastpage}
\end{document}